\documentclass[]{fairmeta}

\usepackage[utf8]{inputenc}
\usepackage{subcaption}

\usepackage[T1]{fontenc}    %
\usepackage{hyperref}       %
\usepackage{url}            %
\usepackage{booktabs}       %
\usepackage{amsfonts}       %
\usepackage{nicefrac}       %
\usepackage{microtype}      %
\usepackage[dvipsnames]{xcolor}

\usepackage{nicematrix}
\usepackage{algorithm}
\usepackage{algpseudocode}
\usepackage{amsthm}
\usepackage[normalem]{ulem}
\usepackage{wrapfig}
\usepackage{graphicx}
\usepackage{cancel}
\usepackage{multirow}
\usepackage{multicol}
\usepackage{amssymb}
\usepackage{cleveref}

\crefformat{section}{\S#2#1#3}
\Crefformat{section}{\S#2#1#3}
\crefmultiformat{section}{\S#2#1#3}{ and \S#2#1#3}{, \S#2#1#3}{ and \S#2#1#3}
\Crefmultiformat{section}{\S#2#1#3}{ and \S#2#1#3}{, \S#2#1#3}{ and \S#2#1#3}
\crefrangeformat{section}{\S#3#1#4 to \S#5#2#6}
\Crefrangeformat{section}{\S#3#1#4 to \S#5#2#6}

\usepackage{colortbl}
\usepackage{xspace}
\usepackage{inconsolata}

\usepackage{graphicx}

\usepackage{xcolor,ifthen,marginnote}
\newboolean{showcomments}

\usepackage{etoolbox}
\newtoggle{release}
\toggletrue{release}

\usepackage{siunitx}

\newcommand\sbv{SWE-bench~Verified\xspace}
\newcommand\dmc{CodeContests\xspace}

\newcommand\swerl{SWE~RL\xspace}
\newcommand{\cwm}{\textsc{CWM}\xspace}
\newcommand{\cwms}{\textsc{CWMs}\xspace}
\newcommand{\cwmv}{\textsc{CWM}\xspace}

\newcommand{\newpara}[1]{\vspace{0.1cm}\noindent \textbf{#1}}
\newcommand\cwmrl{\textsc{CWM}-RL\xspace}

\lstdefinestyle{simple}{
  basicstyle=\ttfamily\small,     %
  columns=fullflexible,
  backgroundcolor=\color{gray!8}, %
  frame=single,                    %
  rulecolor=\color{black!30},      %
  numberstyle=\tiny\color{gray},   %
  keywordstyle=\color{blue},       %
  commentstyle=\color{green!40!black}, %
  stringstyle=\color{red!70!black},    %
  showstringspaces=false,          %
  tabsize=2,                       %
  breaklines=true,                 %
  breakatwhitespace=true,          %
  captionpos=b,                     %
  xleftmargin=3.4pt,
  xrightmargin=3.4pt
}

\definecolor{cwm_keyword_color}{HTML}{0064E0}
\definecolor{cwm_string_color}{HTML}{009B9B}
\definecolor{cwm_string_color_bright}{HTML}{C80A28}
\definecolor{cwm_comment_color}{HTML}{D75FAA}

\lstdefinestyle{simple_small}{
  basicstyle=\fontsize{8}{10}\ttfamily, %
  columns=fullflexible,
  keywordstyle=\color{cwm_keyword_color},
  commentstyle=\color{cwm_comment_color},
  stringstyle=\color{cwm_string_color},
  breaklines=true,
  frame=single,
  escapeinside={(*@}{@*)}
}

\lstdefinestyle{simple_small_small}{
  basicstyle=\fontsize{5.5}{6}\ttfamily, %
  columns=fullflexible,
  keywordstyle=\color{cwm_keyword_color},
  commentstyle=\color{cwm_comment_color},
  stringstyle=\color{cwm_string_color},
  breaklines=true,
  frame=single,
  escapeinside={(*@}{@*)}
}

\definecolor{cwm_think_color}{HTML}{AFD7FF}
\lstdefinestyle{cwm_think}{
  framerule=0pt,
  columns=fullflexible,
  basicstyle=\ttfamily\tiny,     %
  backgroundcolor=\color{cwm_think_color}, %
  showstringspaces=false,          %
  breaklines=true,                 %
  xleftmargin=3.4pt,
  xrightmargin=3.4pt,
  breakindent=0pt
}
\definecolor{cwm_prompt_color}{HTML}{A8E6CF}
\lstdefinestyle{cwm_prompt}{
  framerule=0pt,
  columns=fullflexible,
  basicstyle=\ttfamily\tiny,     %
  backgroundcolor=\color{cwm_prompt_color}, %
  showstringspaces=false,          %
  breaklines=true,                 %
  xleftmargin=3.4pt,
  xrightmargin=3.4pt,
  breakindent=0pt
}

\definecolor{cwm_act_color}{HTML}{D2D2FF}
\lstdefinestyle{cwm_act}{
  framerule=0pt,
  columns=fullflexible,
  basicstyle=\ttfamily\tiny,     %
  backgroundcolor=\color{cwm_act_color}, %
  showstringspaces=false,          %
  breaklines=true,                 %
  xleftmargin=3.4pt,
  xrightmargin=3.4pt,
  breakindent=0pt
}

\definecolor{cwm_obs_color}{HTML}{FFDCB9}
\lstdefinestyle{cwm_obs}{
  framerule=0pt,
  columns=fullflexible,
  keepspaces=true,
  basicstyle=\ttfamily\tiny,     %
  backgroundcolor=\color{cwm_obs_color}, %
  showstringspaces=false,          %
  breaklines=true,                 %
  xleftmargin=3.4pt,
  xrightmargin=3.4pt,
  breakindent=0pt
}

\usepackage{placeins}

\title{\cwm: An Open-Weights LLM for Research on Code Generation with World Models}
\author{Meta FAIR CodeGen Team}
\date{September~29,~2025}
\metadata[Inference Code]{\texttt{\href{https://github.com/facebookresearch/cwm}{github.com/facebookresearch/cwm}}}
\metadata[Model Weights]{\texttt{\href{https://ai.meta.com/resources/models-and-libraries/cwm-downloads}{ai.meta.com/resources/models-and-libraries/cwm-downloads}}, 
\newline
\texttt{\href{https://huggingface.co/facebook/cwm}{huggingface.co/facebook/cwm}},
\texttt{\href{https://huggingface.co/facebook/cwm-sft}{../cwm-sft}},
\texttt{\href{https://huggingface.co/facebook/cwm-pretrain}{../cwm-pretrain}}
}
\abstract{
We release Code World Model (\cwm), a 32-billion-parameter open-weights LLM, to advance research on code generation with world models.
To improve code understanding beyond what can be learned from training on static code alone, we mid-train \cwm on a large amount of observation-action trajectories from Python interpreter and agentic Docker environments, and perform extensive multi-task reasoning RL in verifiable coding, math, and multi-turn software engineering environments.
With \cwm, we provide a strong testbed for researchers to explore the opportunities world modeling affords for improving code generation with reasoning and planning in computational environments.
We present first steps of how world models can benefit agentic coding, enable step-by-step simulation of Python code execution, and show early results of how reasoning can benefit from the latter.
\cwm is a dense, decoder-only LLM trained with a context size of up to \SI{131}{k} tokens.
Independent of its world modeling capabilities, \cwm offers strong performance on general coding and math tasks: it reaches pass@1 scores of \SI{65.8}{\percent} on SWE-bench Verified (with test-time scaling), \SI{68.6}{\percent} on LiveCodeBench, \SI{96.6}{\percent} on Math-500, and \SI{76.0}{\percent} on AIME 2024.
To support further research on code world modeling, we release model checkpoints after mid-training, SFT, and RL. 
}

\begin{document}
\maketitle

\vspace{-0.1cm}
\section{Introduction}
\label{sec:intro}
\vspace{-0.1cm}

Software development is one of the domains where Large Language Models (LLMs) have already had a significant real-world impact~\citep{cui2024effects,bick2024rapid}.
They have quickly been adopted into the workflows of software engineers worldwide, and their capabilities are advancing fast: from only supporting programmers with small snippets of code to fixing issues or writing code bases autonomously~\citep{yeverechyahu2024impact,handa2025economic}.
However, reliably generating high-quality code remains a challenge even for the current generation of LLMs, with benchmarks consistently revealing shortcomings upon release \citep{hendrycks2021measuring,DBLP:journals/corr/abs-2107-03374,aider,DBLP:conf/iclr/JimenezYWYPPN24}.

We believe that advancing code generation with LLMs may require new training and modeling paradigms. Typically, code is treated the same as any other text data during pre-training: the model learns to predict code line by line, from left to right and top to bottom. We think this is not sufficient -- to master coding, one must understand not just what code \emph{looks like} but what it \emph{does} when executed. Such skill is instrumental to the everyday work of software engineers: at a local level, they understand how the execution of a line of code changes the state of the local variables, and, at a global level, they can make predictions about how changes to a codebase affect program outputs. Yet, teaching LLMs such \emph{code world modeling} capabilities is typically not considered before post-training.

\begin{figure}[t!]
    \centering
    \includegraphics[width=0.85\linewidth]{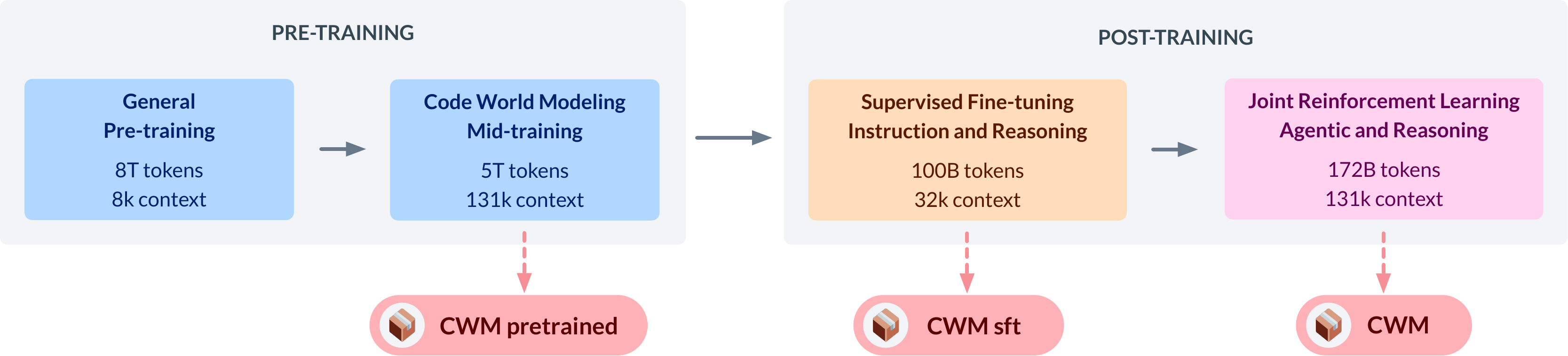}
    \caption{Overview of the \cwm training stages and the model checkpoints that we release. We generally report performance of the final CWM (instruct, RL trained) model, except where otherwise stated.}
    \label{fig:training_pipeline}
    \vspace{-0.1cm}
\end{figure}

We release Code World Model (\cwm), a new LLM for code generation and reasoning that has been trained on large amounts of code world modeling data. 
Concretely, \cwm is mid-trained on two different kinds of observation-action trajectories that capture important aspects of software development: Python code execution traces and agentic interactions with Docker environments. Mid-training on such data at scale should help improve coding performance by grounding our model's predictions in the underlying dynamical systems and provide a superior starting point for RL. 

For the Python execution data, actions are Python statements and observations contain the contents of the local variables. By training \cwm on a trajectory of observation-action pairs conditioned on the code only as context, we directly teach the model how the execution of a line of Python affects the state of the local variables. Our premise here is that teaching \cwm the semantics and not just syntax of programs should help with writing code as well as with reasoning tasks like verification, testing, and debugging.

We also train \cwm on a large-scale collection of synthetically generated agentic interactions with computational environments. These trajectories are generated with our so-called ForagerAgent, which ``forages'' for data covering agentic software engineering scenarios such as implementing missing functionality or fixing bugs from error messages. Actions here are shell-like commands or code edits generated by the agent, and observations are responses from the running environment. While it is not uncommon for recent models to include similar data, this is mostly done at smaller scale during post-training~\citep{yang2025swe}.
Data from ForagerAgent, on the other hand, is large scale and included already during mid-training, helping shape internal representations ahead of post-training.

\begin{figure}[t!]
    \centering
    \includegraphics[width=\linewidth]{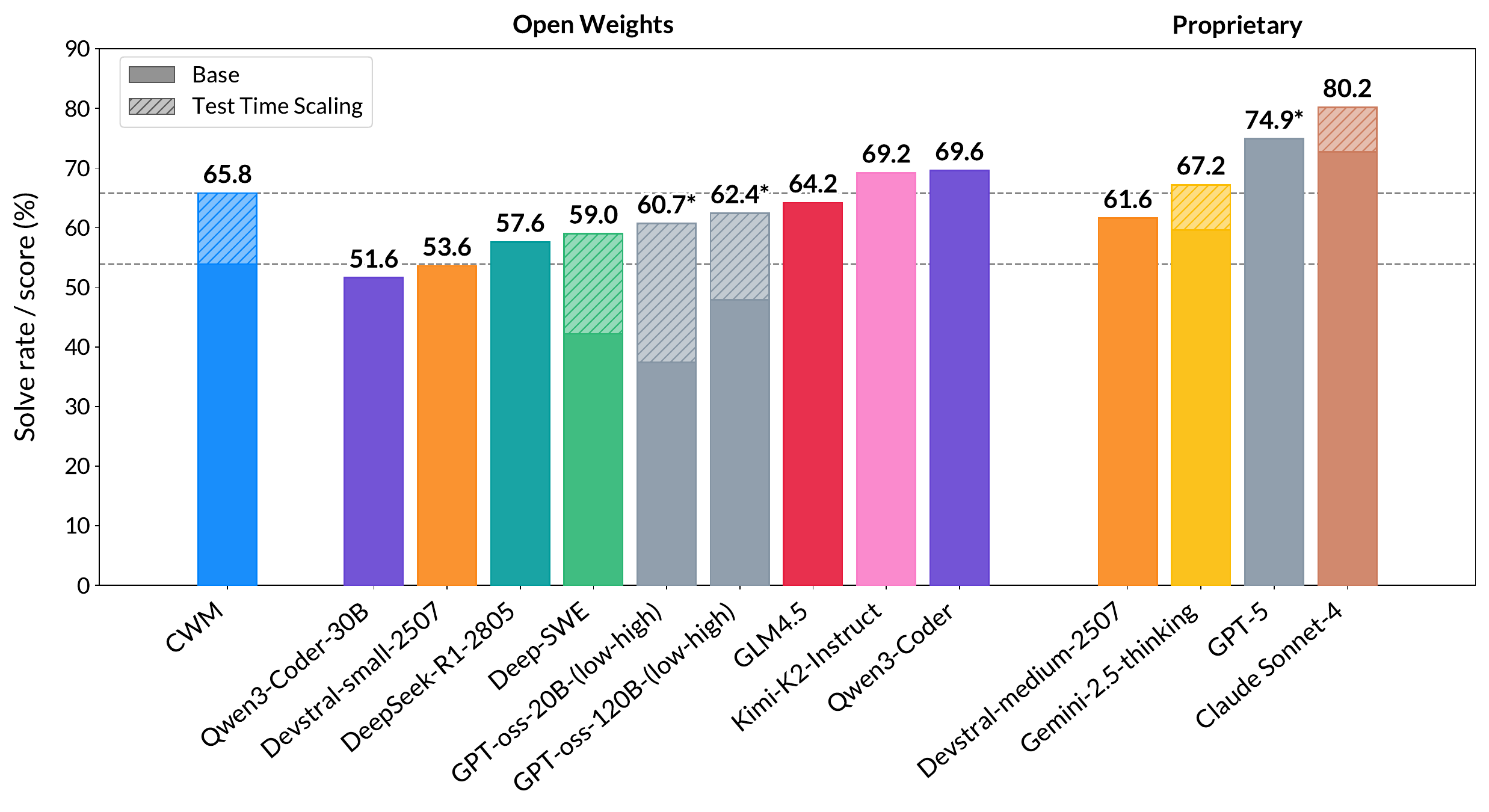}
    \vspace{-0.1cm}
    \caption{
    On \sbv, \cwm outperforms open-weight models with similar parameter counts and is even competitive with much larger or closed-weight LLMs.    
    The base score for \cwm is computed with a single attempt per instance (no retries, majority voting, or parallel candidates), averaged over multiple runs to reduce variance.
    For ``Test Time Scaling'', we generate multiple candidates in parallel and then submit one patch based on ranking.
    The ``Test Time Scaling'' score for GPT-oss models is \texttt{high} reasoning budget, while the lower score is \texttt{low}. 
    (*: GPT-5 and GPT-oss use a custom subset of \num{477} problems, while \cwm is evaluated on the full set of \num{500} problems.)    
    }
    \label{fig:sbv}
    \vspace{-0.5cm}
\end{figure}

\looseness=-1
\cwm uses a dense, decoder-only Transformer architecture~\citep{vaswani2017attention,radford2018gpt} with \SI{32}{B} parameters, interleaved sliding window attention supporting up to \SI{131}{k} tokens context size, and is trained over pre-, mid-, and post-training phases (see~\cref{fig:training_pipeline}). With quantization, inference with \cwm can be performed on a single \SI{80}{GB} NVIDIA H100. 
Beyond world modeling capabilities, \cwm achieves strong performance on general and agentic coding and reasoning tasks relative to other open-weights models of comparable size: it reaches pass@1 scores of \SI{65.8}{\percent} on SWE-bench Verified (with test-time scaling; see~\Cref{fig:sbv}), \SI{68.6}{\percent} on LiveCodeBench-v5, \SI{96.6}{\percent} on Math-500, \SI{76.0}{\percent} on AIME 2024, and \SI{94.3}{\percent} on CruxEval Output.

First and foremost, the release of \cwm is meant to enable novel research on improving code generation with world modeling.
We are excited by this vision and our report provides early supportive evidence: we give examples of how world models can benefit agentic coding, enable step-by-step simulation of Python code execution, and show early results of how reasoning can benefit from the latter. 
However, we believe the best is yet to come and hope to join forces with the open source research community to explore how world models can be used to leverage reasoning and planning to improve code generation.
To this end, we release both the final weights and intermediate checkpoints under a noncommercial research license.
Given \cwm's competitive performance, we conducted a preparedness assessment which concluded that \cwm is unlikely to increase catastrophic risks beyond those present in the current model ecosystem.\footnote{See \href{https://ai.meta.com/research/publications/cwm-preparedness}{ai.meta.com/research/publications/cwm-preparedness} and \cref{sec:discussion} for details.}

\section{Code world model datasets}
\label{sec:data}

\cwm is trained on a large variety of datasets across pre-, mid-, and post-training phases. We focus strongly on code and code world modeling data across all stages of training. We highlight two large-scale data collection efforts that empower \cwm's world modeling capabilities: Python execution traces and ForagerAgent. We refer to \Cref{sec:model} for more traditional ingredients in our datamixes.

\subsection{Executable repository images: building repositories at scale}
\label{sec:exec-envs}
\begin{figure}[t]
    \centering
   \includegraphics[width=0.75\linewidth]{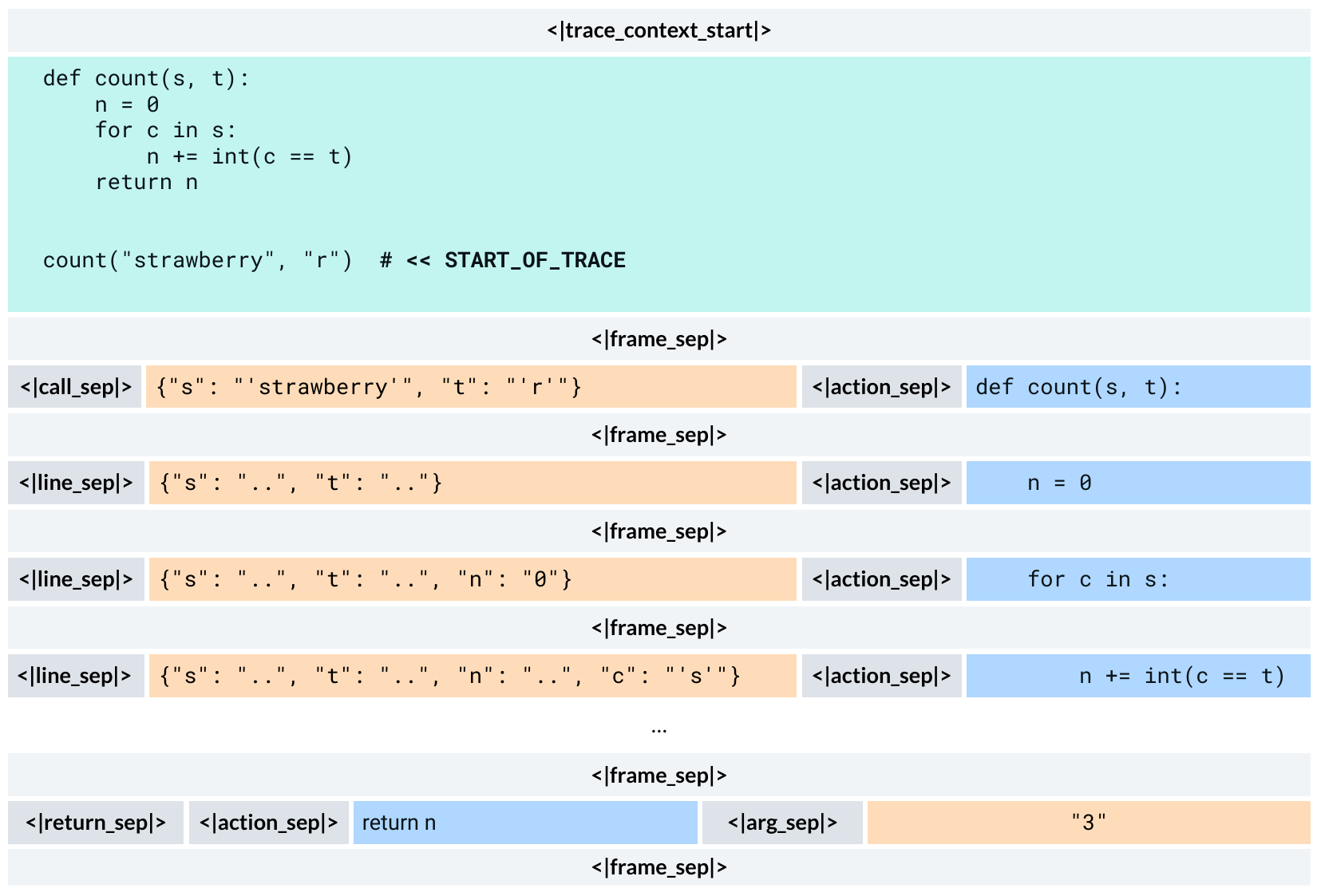}
    \caption{CWM format for Python traces. Given a source code context and a marker of the trace starting point, \cwm predicts a series of stack frames representing the Program states and the actions (executed code). }
    \label{fig:trace_format}
    \vspace{-0.5cm}
\end{figure}

\looseness=-1
A core prerequisite for capturing Python execution traces and agentic trajectories in real-world software engineering tasks is executing code in repositories at scale. For isolation and repeatability, we build these repositories as Docker containers, referred to as \textit{executable repository images}. These images contain a preconfigured environment capable of running repository code and tests without additional setup.
As manually building arbitrary GitHub repositories cannot scale to our desired dataset size, we apply both LLM- and CI-assisted methods.

For the former, an LLM-backed agent, denoted as RepoAgent, was tasked with setting up the development environment of a target repository, finding test files, and ensuring that a significant number of them could run and pass. To support its efforts, we provide RepoAgent with human-readable documentation extracted from the target repository. Although this further improves RepoAgent's success rates, human-targeted documentation can suffer from inaccuracies due to lack of verifiability and insufficient maintenance incentives. In contrast, machine-targeted instructions must remain accurate for successful builds, with platforms like GitHub immediately signaling failures. 

Therefore, we also developed the \textit{Activ} (Act in virtual) pipeline to repurpose GitHub Actions CI execution for building executable repository images. This pipeline runs the workflows locally via the \textit{act} \citep{Lee2025act} library. Since many GitHub Actions workflows are not designed for third-party execution and not limited to CI builds, we modify the target repository's source code and trigger an early exit after the completion of a single successful build.
As all GitHub Actions workflow jobs run simultaneously in individual containers and the build state is transient, we add or modify the repository's pytest configuration files, to inject a fixture that is automatically run at test-execution time. This fixture captures the build state of the container running unit tests. We then commit and push the resulting image from each repository, as further detailed in \cref{appendix:activ}. 
Running both RepoAgent and Activ methods in parallel, we created over \SI{35}{k} unique executable repository images. 

\subsection{Python tracing: neural code interpretation data}
\label{sec:tracing}

The first type of \cwm data we present is memory tracing of Python programs. This involves gathering executable functions or executable repository images, and running them using different IO pairs or CI tests, while capturing the state of the memory, chiefly the local variables, after each line is executed. This process enables us to align code and execution trace to simulate observation-action data within the computational environment. Prior work empirically shows this approach is beneficial in improving general code generation and understanding capabilities~\citep{armengol2025cannot, Zhang2025-au}. Neural code interpretation further has the potential to go beyond traditional interpreters, with applications such as tracing through unexecutable code or combining it with reasoning capabilities. 
Next, we describe the different sources from which we gather execution trace data.

\newpara{Function-level tracing.} We collect a dataset of Python functions from online sources and automatically generate input-output pairs with a combination of fuzzing and prompting Llama3-70B-Instruct. Our tracing process captures the state of the Python program (interpreter stack frames) at different intermediate execution points, corresponding to events of the Python interpreter (e.g., executed lines, return statements, exceptions). The final dataset contains over \SI{120}{M} traced Python functions. 

We post-process the raw traces to construct observation-action pairs. The observation contains local variables and stack frame metadata immediately prior to executing a line of code, the action is the specific Python line being executed, and the subsequent observation captures the resulting local variable states and additional event metadata such as return statements; we disregard global variables and external side effects. The variable values that do not change with respect to the previous step are summarized with an ellipsis. We prefix the trace data with the source code context.  

\Cref{fig:trace_format} illustrates the CWM format for Python traces. Given a Python code context and a marker of the tracing initial point, the model follows with a series of Python stack frame predictions (in the form of a JSON-formatted dictionary with the local variables) and the corresponding actions (i.e., the part of the code that is being executed). The frame, action, and argument separators, as well as the trace context start indicator, are represented using custom tokens.
We refer the reader to \Cref{sec:add_examples} for trace prediction examples and \Cref{appendix:prompting_guide} for a specification of this trace representation format.
We re-use both the tracing app and trace formatting for all other execution trace data described in the remainder of this section.

\newpara{CodeContests solutions tracing.}  We also generate tracing data for solutions to competitive programming problems. Concretely, we use Llama-3.1-70B-Instruct to generate Python solutions to training set problems in \dmc~\citep{DBLP:journals/corr/abs-2203-07814}, reusing the framework of~\citet{gehring2025rlefgroundingcodellms}. Generations are filtered to ensure a balance of incorrect and correct submissions, leading to an overall count of \SI{262}{k}. We trace these solutions with inputs from the provided unit tests and filter out long traces with more than \SI{10}{k} line events or large traces taking up more than \SI{1}{MB} disk space, leaving us with \SI{33}{k} effective code snippets and \SI{70}{k} traces.

\newpara{Repository-level tracing.} We also performed Python execution tracing for the unit tests of more than \SI{21}{k} available and traceable repository images. For a subset of these repositories, we use the repository's git log to randomly select additional commits prior to our built commit. Since the build environment is configured for the current commit's dependencies, older commits may fail to execute. We attempted tracing for up to \num{40} historical commits per repository but capped successful traces at \num{4} commits per repository to avoid over-representation of any single repository. This process resulted in around \SI{70}{k} execution-traced commits.

We post-process raw traces in two steps. First, we ``episodify'' our traces, extracting function-level traces from raw pytest traces with configurable stack depth and stochastic step-in probability. When stochastic step-in occurs, function calls are probabilistically included in their parent trace rather than a separate episode to simulate variable execution depth. In a second step, we then gather and compress the source code context from the target repository that is necessary for predicting the observation-action trajectory.
To the resulting context-trace pair, we then apply the same \cwm formatting as before.

\newpara{Natural language tracing.} Lastly, we generate a dataset of step-by-step descriptions of Python code execution in \emph{natural language} rather than our strict JSON-like format from before. Natural language explanations of code execution are closer in domain to other LLM tasks, which we hope will simplify knowledge transfer to other context such as reasoning in code generation. This less-structured format also has other advantages, such as allowing for injection of semantic context (e.g., ``this operation preserves the structure property of the max heap'') or for compressing traces by dynamically skipping less interesting parts of the trajectory (e.g., repeated logic within a for loop). We generate this data by prompting Qwen3-32B-FP8 (without thinking)~\citep{yang2025qwen3} to re-write execution traces from our function-level and CodeContests trace datasets. After removing cases where the final output prediction from Qwen diverges from the ground truth trace, we obtain \SI{75}{M} trajectories from standalone Python functions and \SI{110}{k} from CodeContests data.

\subsection{ForagerAgent: agentic midtraining data generation}
\label{sec:forager}

\looseness=-1
We mid-train \cwm on a large-scale dataset of interactions between an LLM-based software engineering agent and a computational environment. This data is generated with our so-called ForagerAgent, which collects multi-step trajectories by prompting an LLM with a software engineering task to solve in the context of a particular code repository. Exposing \cwm to such data at large scale early on should improve subsequent post-training in similar environments, as model predictions should already be grounded in environment dynamics.

The actions available to the agent are derived from the standard SWE-Agent \citep{yang2024swe} toolset: (i) create a file, (ii) edit a file, (iii) run a bash command, and (iv) view or navigate inside a file. The trajectory is concluded once the LLM, either Llama3-70B-Instruct~\citep{dubey2024llama} or Qwen3-235B-A22B (w/o thinking)~\citep{yang2025qwen3}, believes the task has been solved or the number of tokens, turns, or API costs exceed a hard limit. Like the repository-level tracing data, ForagerAgent relies on our set of executable repository images (see \cref{sec:exec-envs}) to seed problem generation. To avoid contamination, we filter out all repositories (and their forks) that are used in SWE-bench. The tasks presented to the model can be categorized into two groups: synthetic tasks and real-world tasks, which we call \emph{mutate-fix} and \emph{issue-fix}.

\newpara{Mutate-fix tasks}. For mutate-fix tasks, we start with a working codebase and then synthetically introduce a bug for the agent to fix. We begin by identifying functions (and methods -- omitted for brevity below) that can be verified using the repository test suite. 
As a first step, we filter these functions to the subset for which all unit tests pass successfully. 
We then consider the following set of mutations to synthetically introduce a bug into these functions:
\begin{itemize} 
    \item Functions: remove either a portion of the function or the entire function. 
    \item Arguments: remove arguments from the function definition or randomly re-order function call arguments.
    \item Variables: sample a pair of variables in the function and swap all their occurrences.
    \item Statements: remove an import or return statement. 
    \item Operators: replace operators (binary, unary, or boolean) in statements in the function.
\end{itemize}
We filter out mutations that cannot be applied for a given function by parsing the corresponding abstract syntax tree (AST). Lastly, we verify that applying the candidate  mutation does in fact cause the associated unit tests to fail. 
We can now use the mutation as a starting point for agentic data collection: we instruct the agent to inspect the mutated function, run its unit tests, and resolve the failing tests by fixing the bug.

\looseness=-1
\newpara{Issue-fix tasks}. For issue-fix tasks, we prompt the model to fix real issues in our set of repositories, using both issue and pull request data from GitHub.
We check out commits preceding bug-fixing PRs and task the agent with resolving failing unit tests, providing the corresponding GitHub issue descriptions for context.
We ensure unit tests are failing before the PRs and that their resolution is necessary and sufficient for addressing the issues.

\newpara{Post-processing.} To avoid overfitting to repetitive interactions, we apply a near-deduplication of trajectories foraged from the same source repository: we first represent a trajectory by the concatenation of its actions, then encode the trajectory using MinHash, and lastly drop trajectories such that the pairwise Jaccard similarity for all encoded trajectories we keep is less than \num{0.5}. Because our goal with the ForagerAgent data is to learn a comprehensive world model of agentic interactions with code environments, we do not filter trajectories based on whether they succeed at bug or issue resolution. Following the same motivation, we further train the model to predict both agent and environment turns, although we stochastically mask loss for \SI{50}{\percent} of observations as they exhibit limited diversity. Overall, we are left with \SI{3}{M} trajectories obtained from \SI{10.2}{k} images, and we refer to \Cref{tab:forager_stats} for more detailed statistics.

\begin{table}[t]
\centering
\caption{Statistics of ForagerAgent trajectories. We gather \SI{3}{M} trajectories from \SI{10.2}{k} images and \SI{3.15}{k} underlying repositories. The trajectories are split \num{55}--\num{45} between issue- and mutate-fix tasks.
}
\vspace{-0.1cm}
\label{tab:forager_stats}
\resizebox{\textwidth}{!}{
\renewcommand{\arraystretch}{1.1}
\begin{tabular}{ccc||c|ccccc}
\toprule
\textbf{Repos} & \textbf{Images} & \textbf{Trajectories} & \textbf{Issues-Fix}  & \multicolumn{5}{c}{\textbf{Mutate-Fix}}\\
\midrule
\multirow{2}{*}{3.15k} & \multirow{2}{*}{10.2k} & \multirow{2}{*}{3M} & \multirow{2}{*}{55\%} & 
Functions & Arguments & Variables & Statements & Operators \\
\cline{5-9} 
& & & & \SI{7}{\percent} & \SI{9}{\percent} & \SI{6}{\percent} & \SI{11}{\percent} & \SI{12}{\percent}
\\
\bottomrule
\end{tabular}}
\vspace{-0.4cm}
\end{table}

\section{Examples of code world modeling}
\label{sec:examples}

Before introducing the \cwm architecture and benchmark results more formally, in this section, we share a few example generations from the final model that illustrate our excitement for code world modeling.

For competitive programming, \cref{fig:example:agentic-coding} shows an example where \cwm first creates an initial solution after reasoning about the problem statement, then constructs input-output pairs to assess the correctness of its solution, before finally comparing its own prediction against the actual results of program execution.
Although we did not train \cwm for it directly, this is exactly the kind of reasoning about environment dynamics that we believe code world models can enable.
We are excited about future work that explicitly integrates reasoning about environment feedback to improve agentic code generation.

\looseness=-1
While \cwm is trained to simulate the execution of Python code line-by-line, we believe the formal integration of trace prediction into natural language reasoning is interesting future work.
\Cref{lst:reasoning_with_trace_pred} presents an early prototype of this, showing \cwm already generalizes to such behavior under light prompting.
We believe that execution trace prediction enables \emph{grounded} reasoning about code generation and execution, without requiring access to live execution environments.
This can take explicit forms, such as in our prototype, but we also envision more implicit forms of regularizing reasoning through world modeling.
We further plan to extend \cwm's trace prediction to create a ``neural debugger'' whose capabilities go beyond traditional debugging, such as jumping to future lines of code in constant time, predicting inputs to reach arbitrary states, or learning abstract representations of program states that support program verification, debugging, or generation.

\looseness=-1
Relatedly, we are excited about more direct uses of Python execution trace prediction to improve code generation.
Under appropriate prompting, \cwm can be used to predict execution traces for functions that are not given directly in the context but are instead constrained only through natural language or formal asserts.
After trace prediction, the model then constructs the function body from the generated actions, which are the executed Python statements.
We show early examples of this in \cref{lst:trace_cwm_unseen,lst:trace_cwm_unseen_reasoning}.
This method of reasoning-via-tracing for program generation is reminiscent of work by \citet{Hoare71,Dijkstra76} on using logical reasoning about program execution to guide development rather than only using it to confirm correctness post-hoc.

\begin{figure}[p]
{
\lstset{aboveskip=0pt, belowskip=8pt}
\begin{lstlisting}[style=cwm_think]
Okay, I need to solve this problem where I have to check if a given string is evenly separated. The definition says that between every pair of the same characters, there must be an even number of characters. So, for example, in "abba", each 'a' is at positions 0 and 3. The distance between them is 3-0 = 3, but the number of characters in between is 3-0-1 = 2, which is even. So that works. But in "aabba", the first two 'a's are at positions 0 and 1. The number of characters between them is 0, which is even. Let me double check. But the example says it's not correct. Hmm, wait maybe I'm misunderstanding the problem...
\end{lstlisting}%
\begin{lstlisting}[style=cwm_act]
<tool: create>
/code/solution.cpp
#include <bits/stdc++.h>
using namespace std;

bool is_evenly_separated(const string& s) {
    unordered_map<char, int> prev_index;
    ...
</tool>
\end{lstlisting}%
\begin{lstlisting}[style=cwm_obs]
File created successfully at: /code/solution.cpp (35 lines).
\end{lstlisting}
\begin{lstlisting}[style=cwm_think]
...Looking back, the code reads a string, passes it to the function. Let me create some test cases.

Test 1: 'abba' -> expected YES -> the code should return YES.

Test 2: 'aabba' -> expected NO -> because when it processes the third character, it's the third occurrence of 'a'.

Wait for 'aabba':

Indices 0:a (occurs once)

index1:a now (occurs twice. check 0 and 1 -> difference is 1 -> ??? 

But difference is -1 -> mod 2 is -1 %

Then loop reaches index 4 (a's third occurrence), which code detects count for a which is already 2 -> returns NO.

So code should output NO here.

Other test: 'aaa' (NO when processing third a).

test4: 'abcd' -> YES (each character appears once or twice in non-same parity) -> code OK.

ttest5: 'aabb' -> the two a's (index 0 and1 -> difference 1, ok). Two b's (index 2,3 difference -> 1 is ok). Code returns YES.

Now let's compile and run the code.
\end{lstlisting}%
\begin{lstlisting}[style=cwm_act]
<tool: bash>
g++ -std=c++17 -O2 solution.cpp -o solution;
echo -e "abba\n" | ./solution
echo -e "aabba\n" | ./solution
echo -e "aaa\n" | ./solution
echo -e "abcd\n" | ./solution
echo -e "aabb\n" | ./solution
</tool>
\end{lstlisting}%
\begin{lstlisting}[style=cwm_obs]
YES
NO
NO
YES
YES
\end{lstlisting}%
\begin{lstlisting}[style=cwm_think]
All the test cases match the expected results. The code works correctly for these samples and edge cases. For very large inputs, the code has O(n) time and O(1) space (since there are 26 possible characters, hash maps would use at most 26 entries). Therefore, the solution is efficient and should pass all test cases, including those up to 1e6 characters long.

The approach checks for each character occurrence and ensures that each pair of occurrences meet the parity requirement. Also, it immediately terminates upon finding a third occurrence or a pair with same parity indices, which is crucial for efficient processing.

I'm confident that there are no bugs, and the solution meets the problem's requirements.
\end{lstlisting}%
\begin{lstlisting}[style=cwm_think]
The approach involves checking each character's occurrences to ensure that no character appears more than twice, and any pair of occurrences of a character must be an odd number of positions apart (i.e., their indices have different parities).

```cpp
#include <bits/stdc++.h>
using namespace std;

bool is_evenly_separated(const string& s) {...
\end{lstlisting}
}
\caption{Example of \cwm solving a competitive programming problem in agentic fashion. The reasoning steps are shortened and some details are omitted due to space constraints. (Tool calls in purple, environment feedback in orange, and reasoning in blue.)}
\label{fig:example:agentic-coding}
\end{figure}

\begin{figure}
  \lstset{aboveskip=0pt, belowskip=3pt}
  \centering
  \begin{lstlisting}[style=cwm_prompt]
Given a python code function and an assert statement containing a specific input, provide the assertion with the exact literal output that the function returns with that input. Do not include any mathematical expressions or function calls -- only the final literal value. Your response should be solely the assertion, enclosed within [ANSWER] and [/ANSWER] tags.
You are a computational world model and can predict the program execution.
Your execution trace prediction format MUST follow this structure:
1. The execution trace prediction starts with the <|trace_context_start|> token and ends with a final <|frame_sep|> token.
2. For each code execution step:
- Begin with <|frame_sep|> followed by the event token which can be <|call_sep|>, <|line_sep|>, <|return_sep|> or <|exception_sep|>.
- After <|call_sep|> or <|line_sep|> put the local variable states as dictionary in JSON format followed by the <|action_sep|> token and the current source code line.
- After <|return_sep|>, <|exception_sep|> directly put the <|action_sep|> token and the current source code line followed by an <|arg_sep|> token and the return or exception arguments.
3. Provide the full assertion with the correct output that you obtained after <|return_sep|> in [ANSWER] and [/ANSWER] tags
Here is an example of how you would predict the output of the program using your trace prediction capability:

Python function:
def f(a,b):
    y = a
    for i in range(b):
        y += y * i
    return y
assert f(1,3) == ??
<think>
your internal reasoning
</think>
Let's verify this by putting the code into a trace context and call the function in the main() function and then trace the execution of the main function.
We indicate the entry point of the execution trace with a  # << START_OF_TRACE marker.
def f(a,b):
    y = a
    for i in range(b):
        y += y * i
    return y
def main(): # << START_OF_TRACE
    return f(1,3)
<|frame_sep|><|call_sep|>{}<|action_sep|>def main(): # << START_OF_TRACE
<|frame_sep|><|line_sep|>{}<|action_sep|>    return f(1,3)
<|frame_sep|><|call_sep|>{"a": "1", "b": "3"}<|action_sep|>def f(a,b):
<|frame_sep|><|line_sep|>{"a": "..", "b": ".."}<|action_sep|>    y = a
<|frame_sep|><|line_sep|>{"a": "..", "b": "..", "y": "1"}<|action_sep|>    for i in range(b):
<|frame_sep|><|line_sep|>{"a": "..", "b": "..", "y": "..", "i": "0"}<|action_sep|>        y += y * i
<|frame_sep|><|line_sep|>{"a": "..", "b": "..", "y": "..", "i": ".."}<|action_sep|>    for i in range(b):
<|frame_sep|><|line_sep|>{"a": "..", "b": "..", "y": "..", "i": "1"}<|action_sep|>        y += y * i
<|frame_sep|><|line_sep|>{"a": "..", "b": "..", "y": "2", "i": ".."}<|action_sep|>    for i in range(b):
<|frame_sep|><|line_sep|>{"a": "..", "b": "..", "y": "..", "i": "2"}<|action_sep|>        y += y * i
<|frame_sep|><|line_sep|>{"a": "..", "b": "..", "y": "6", "i": ".."}<|action_sep|>    for i in range(b):
<|frame_sep|><|line_sep|>{"a": "..", "b": "..", "y": "..", "i": ".."}<|action_sep|>    return y
<|frame_sep|><|return_sep|><|action_sep|>    return y
<|arg_sep|>"6"<|frame_sep|><|return_sep|><|action_sep|>    return f(1,3)
<|arg_sep|>"6"<|frame_sep|>
Now let us analyze the trace. The return argument of the function call f(1,3) in the main() function is "6" in JSON format, so the return value is 6.
[ANSWER]
assert f(1,3) == 6
[/ANSWER]

Python function:
def f(d, k):
    new_d = {}
    for key, val in d.items():
        if key < k:
            new_d[key] = val
    return new_d
assert f({1: 2, 2: 4, 3: 3}, 3) == ??
\end{lstlisting}%
\begin{lstlisting}[style=cwm_think]
Let's analyze the function call f({1: 2, 2: 4, 3: 3}, 3).
The function f takes a dictionary d and a key k. It creates a new dictionary new_d and iterates over the items in d. For each key-value pair, if the key is less thank, it adds that key-value pair to new_d.
In this case, the dictionary is {1: 2, 2: 4, 3: 3} and the key is 3. We need to check each key in the dictionary to see if it's less than 3.
- Key 1: 1 < 3 is True, so we add 1: 2 to new_d.
- Key 2: 2 < 3 is True, so we add 2: 4 to new_d.
- Key 3: 3 < 3 is False, so we skip 3: 3.
So, the resulting dictionary new_d should be {1: 2, 2: 4}.
</think>
\end{lstlisting}%
\begin{lstlisting}[style=cwm_act]
Let's verify this by putting the code into a trace context and call the function in the main() function and then trace the execution of the main function.
We indicate the entry point of the execution trace with a  # << START_OF_TRACE marker.
def f(d, k):
    new_d = {}
    for key, val in d.items():
        if key < k:
            new_d[key] = val
    return new_d
def main(): # << START_OF_TRACE
    return f({1: 2, 2: 4, 3: 3}, 3)
<|frame_sep|><|call_sep|>{}<|action_sep|>def main(): # << START_OF_TRACE
<|frame_sep|><|line_sep|>{}<|action_sep|>    return f({1: 2, 2: 4, 3: 3}, 3)
<|frame_sep|><|call_sep|>{"d": "{1: 2, 2: 4, 3: 3}", "k": "3"}<|action_sep|>def f(d, k):
<|frame_sep|><|line_sep|>{"d": "..", "k": ".."}<|action_sep|>    new_d = {}
<|frame_sep|><|line_sep|>{"d": "..", "k": "..", "new_d": "{}"}<|action_sep|>    for key, val in d.items():
<|frame_sep|><|line_sep|>{"d": "..", "k": "..", "new_d": "..", "key": "1", "val": "2"}<|action_sep|>        if key < k:
<|frame_sep|><|line_sep|>{"d": "..", "k": "..", "new_d": "..", "key": "..", "val": ".."}<|action_sep|>            new_d[key] = val
<|frame_sep|><|line_sep|>{"d": "..", "k": "..", "new_d": "{1: 2}", "key": "..", "val": ".."}<|action_sep|>    for key, val in d.items():
<|frame_sep|><|line_sep|>{"d": "..", "k": "..", "new_d": "..", "key": "2", "val": "4"}<|action_sep|>        if key < k:
<|frame_sep|><|line_sep|>{"d": "..", "k": "..", "new_d": "..", "key": "..", "val": ".."}<|action_sep|>            new_d[key] = val
<|frame_sep|><|line_sep|>{"d": "..", "k": "..", "new_d": "{1: 2, 2: 4}", "key": "..", "val": ".."}<|action_sep|>    for key, val in d.items():
<|frame_sep|><|line_sep|>{"d": "..", "k": "..", "new_d": "..", "key": "3", "val": "3"}<|action_sep|>        if key < k:
<|frame_sep|><|line_sep|>{"d": "..", "k": "..", "new_d": "..", "key": "..", "val": ".."}<|action_sep|>    for key, val in d.items():
<|frame_sep|><|line_sep|>{"d": "..", "k": "..", "new_d": "..", "key": "..", "val": ".."}<|action_sep|>    return new_d
<|frame_sep|><|return_sep|><|action_sep|>    return new_d
<|arg_sep|>"{1: 2, 2: 4}"<|frame_sep|><|return_sep|><|action_sep|>    return f({1: 2, 2: 4, 3: 3}, 3)
<|arg_sep|>"{1: 2, 2: 4}"<|frame_sep|>
Now let us analyze the trace. The return argument of the function call f({1: 2, 2: 4, 3: 3}, 3) in the main() function is "{1: 2, 2: 4}" in JSON format, so the return value is {1: 2, 2: 4}.
[ANSWER]
assert f({1: 2, 2: 4, 3: 3}, 3) == {1: 2, 2: 4}
[/ANSWER]
  \end{lstlisting}
  \caption{Example with execution trace prediction and reasoning. In this example, we add an execution trace example to the prompt. After reasoning in natural language about the code, the model uses its execution trace prediction capability to confirm the correct return value \texttt{\{1: 2, 2: 4\}}. We encode all special tokens (e.g. \texttt{<|frame\_sep|>}) as such. \label{lst:reasoning_with_trace_pred}}
\end{figure}

We refer to \cref{sec:add_examples} for additional example trajectories from \cwm, including self-correcting SWE behavior, a comparison of the advantages and disadvantages of strict trace prediction compared to natural language (and why we think their combination is worth exploring), further demonstrations of how we can adapt trace prediction to elicit code generation, and an example of using \cwm to predict program termination.

\begin{figure}[b!]
 \centering
  \begin{subfigure}{0.27\textwidth}
    \includegraphics[width=\linewidth]{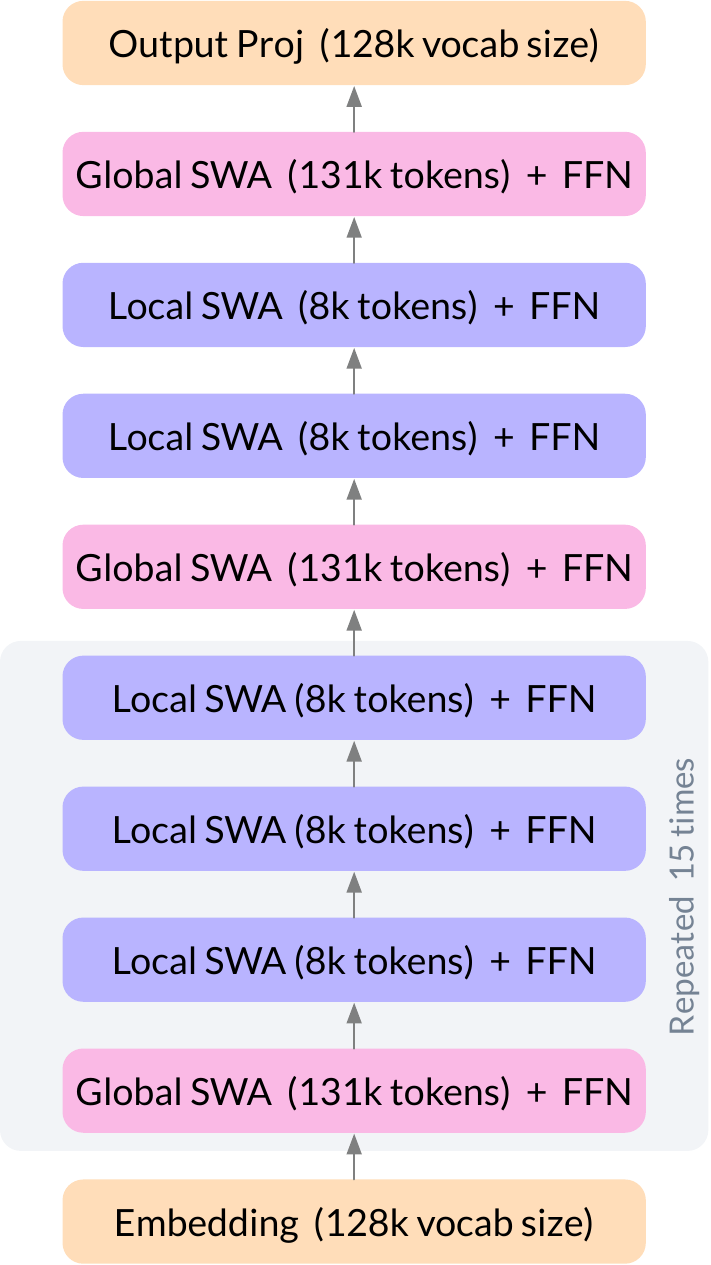}
    \caption{\cwm architecture: GQA with alternating local (\SI{8}{k}) and dynamic global (\SI{131}{k}) sliding window attention.}
    \label{fig:arch}
  \end{subfigure}\hfill
  \begin{subfigure}{0.70\textwidth}
    \includegraphics[width=\linewidth]{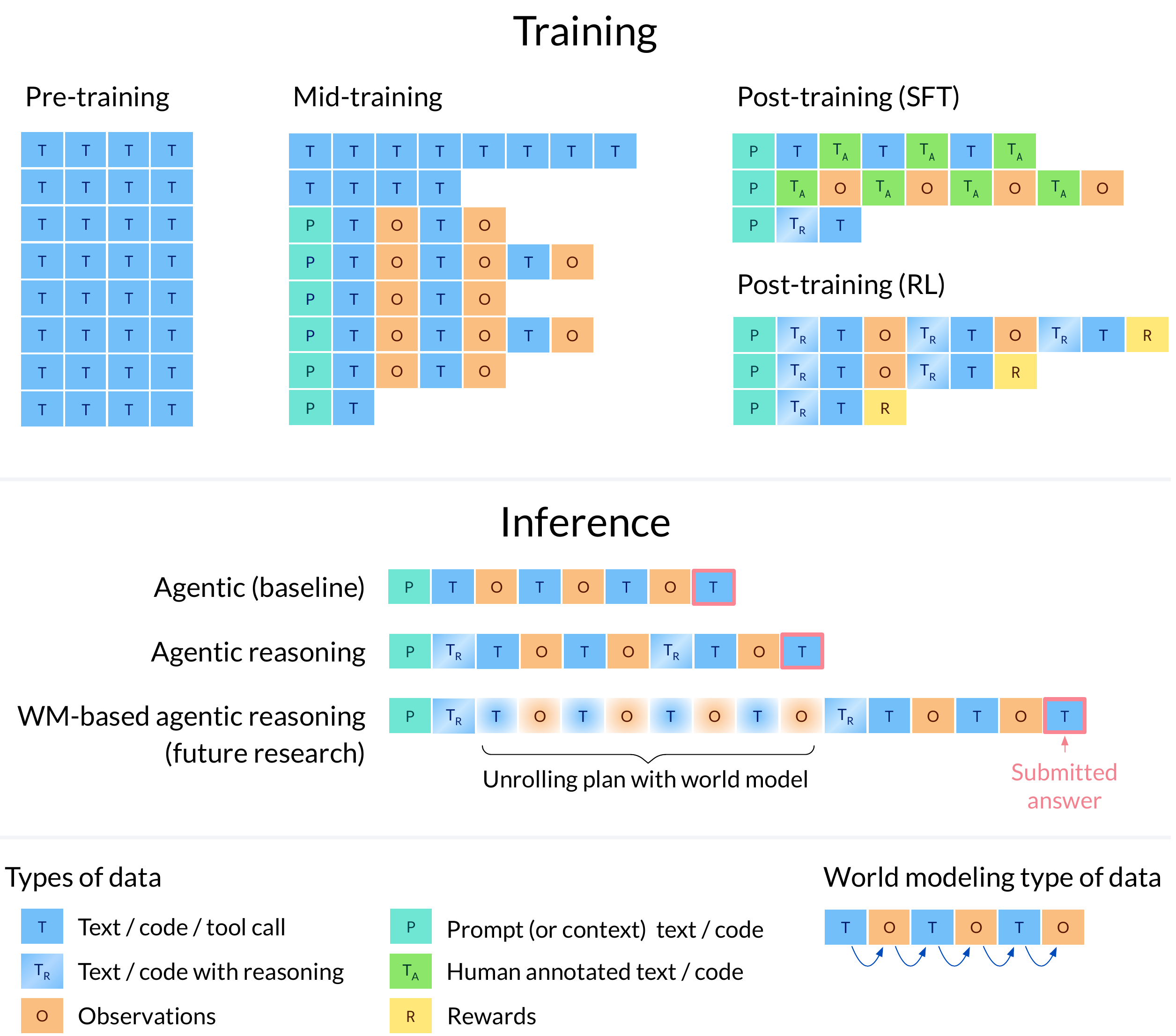}
    \caption{Overview of the \cwm training and inference types for inputs and outputs. }
    \label{fig:training_inference}
  \end{subfigure}
  \caption{Figures illustrating the \cwm Transformer architecture and the main types of data introduced in the different training steps and used at inference time. 
  }
  \label{fig:scaling}
\end{figure}

\section{\cwm: architecture, pre-training, and scaling laws}
\label{sec:model}

We next share details on the \cwm architecture, final pre-training recipe, and scaling law experiments. Specific details about our efficient training infrastructure can be found in~\Cref{sec:code_infra}. 

\subsection{Architecture and hyper-parameters}
\label{sec:architecture}

\looseness=-1
\newpara{Model architecture.} \cwm is a \num{32}-billion-parameter dense decoder-only model.
We choose a dense architecture over sparse alternatives for ease-of-use in downstream open source research. \cwm uses an alternating pattern of local and global attention blocks interleaved in a \(3\!:\!1\) ratio (see \cref{fig:arch}) with sliding window sizes of \num{8192} and \num{131072} tokens respectively. Transformer blocks use Grouped-Query-Attention \citep{ainslie2023gqa} with \num{48} query heads and \num{8} key-value heads. We use SwiGLU activation functions \citep{shazeer2020gluvariantsimprovetransformer}, RMSNorm \citep{rmsnorm} with pre-normalization, Rotary Positional Embedding (RoPE) \citep{su2021rope}, and we train with full document-causal masking. To support long-context modeling, we follow~\citet{roziere2023code, xiong2023effective} and apply Scaled RoPE with $\theta=\SI{1}{M}$ and scale factor \num{16} from mid-training onwards. We give a full overview of \cwm parameters and architecture choices in \cref{tab:modelparams}.

\newpara{Training hyper-parameters.} We train the model with the AdamW optimizer \citep{loshchilov2018decoupled} with $\beta_1=0.9$, $\beta_2=0.95$, weight decay of \num{0.1}, and gradient clipping at norm \num{1.0}.
After \num{2000} steps of linear warmup, we use a cosine decay learning rate schedule with peak learning rate \num{8e-4} and decaying by a factor of \(100\times\) over the training horizon. The cosine decay schedule is calculated for a total training duration of \SI{13}{T} tokens, with the last \SI{5}{T} tokens of the scheduler used during mid-training.
Key hyper-parameters were determined using scaling laws, which we detail in \cref{subsec:scaling_laws}.
\begin{table}[t]
\centering
\caption{Key hyper‑parameters of the 32~B \cwm.}
\label{tab:modelparams}
\begin{tabular}{ll}
\toprule
\textbf{Parameter} & \textbf{Value} \\
\midrule
Number of parameters & \SI{32}{B} \\
Layers & \num{64} \\
Hidden dimension & \num{6144} \\
Intermediate dimension & \num{21504} \\
Number of attention heads / dimension & \num{48} / \num{128} \\ 
Number of key-value heads & \num{8} \\
Local window size & \num{8192} tokens \\
Max global context & \num{131072} tokens \\
Activation function & SwiGLU \\
Normalization & RMSNorm (pre-norm) \\
Positional Encoding & Scaled RoPE ($\theta = 10^6$, scale factor $= 16$) \\
Vocabulary size & \num{128256} tokens \\
\bottomrule
\end{tabular}
\end{table}

\newpara{Tokenizer.} \cwm uses the Llama 3 tokenizer \citep{dubey2024llama} which is a fast Byte-Pair Encoding tokenizer implemented with TikToken.\footnote{See \url{https://github.com/openai/tiktoken}.} The vocabulary contains \num{128000} regular tokens as well as \num{256} reserved tokens.
We keep the control tokens from Llama 3 and leverage unused reserved tokens to support our tracing and reasoning use cases.

\subsection{Two-stage pre-training}
\label{sec:pretraining}

\cwm pre-training consists of two stages sharing learning-rate scheduler and optimizer states but differing in datamix and maximum document lengths:
\begin{enumerate}
    \item \newpara{General pre-training}: We begin with an initial pre-training phase on \SI{8}{T} tokens from a diverse range of mostly English sources, with an emphasis on coding data (making up about \SI{30}{\percent} of the mix) as well as STEM and general knowledge. We pre-train our model with a global batch size of \SI{8.4}{M} tokens and a context length of \num{8192} tokens.\footnote{Note that our ``local'' attention blocks are therefore effectively global during pre-training.}
    \item \newpara{Code world model mid-training}: We then mid-train the model for an additional \SI{5}{T} tokens. We here depart from our more generalist pre-training datamix and introduce a number of datasets in support of our code world modeling objectives. We mid-train with a global batch size of \SI{33}{M} tokens and maximum context length of \SI{131}{k} tokens.\footnote{We have observed lackluster performance when training on long-context data at smaller batch sizes. We speculate that increasing the batch size (in tokens) is beneficial for training on long-context data, as the decrease in the number of documents contained in each batch increases the variance in our gradient estimate.} 
\end{enumerate}

Mid-training is the key stage for teaching code world modeling capabilities.
Next, we discuss the changes we make to the pre-training recipe during mid-training to optimize \cwm performance.

\newpara{Mid-training datamix.} For mid-training, we introduce the ForagerAgent and Python execution tracing data, our main \cwm datasets introduced in \Cref{sec:data}, into the datamix. We additionally include code- and reasoning-related data such as datasets derived from GitHub pull requests similar to SWE-RL~\citep{wei2025swerl}, data from compiler intermediate representations \citep{cummins2024meta}, Triton PyTorch kernels similar to \citet{kernelbook2025}, and formal mathematics in Lean covering statement and proof translation, as well as world modeling (see \Cref{app:lean-data}).

\cwm-specific data makes up \SI{30}{\percent} of the overall mid-training datamix.
We further increase the fraction of general code data to \SI{40}{\percent} and keep \SI{30}{\percent} for rehearsal of the initial pre-training datamix, as this proved essential in retaining performance on standard evaluations.
Within the rehearsal fraction, we now upweight higher quality datasets such as those containing math or long context data, while making sure to avoid over-epoching.
We summarize the types of data used across \cwm training stages in \Cref{fig:training_inference}.

\newpara{Mid-training datamix ablations.} For many of the datasets introduced during mid-training we can afford to train for multiple epochs. To determine the desired number of epochs per dataset, we perform a series of scaling law experiments \citep{kaplan2020scaling} that \emph{simulate} different levels of epoching \citep{dubey2024llama}. In agreement with the literature on this \citep{muennighoff2023scaling}, we generally find that multi-epoch training improves downstream task performance, albeit at diminishing returns, before eventually leading to overfitting. By selecting target epochs such that metrics indicate little to no diminishing returns, we arrive at between \num{1} and \num{4} target epochs per dataset.
The final proportion of a dataset in the mid-training mix is then calculated such that the desired number of epochs is reached at the end of mid-training.
When estimating the number of steps per epoch, we account for both token packing (wrapping) for pre-training data and truncation for chat data.

\newpara{Long-context mid-training.} A significant amount of the mid-training data is long-context, with about \SI{30}{\percent} of documents exceeding \SI{65}{k} tokens.
This motivates our decision to increase the maximum sequence length to \SI{131}{k} tokens for all of mid-training.
Consequently, we do not need a dedicated long-context finetuning phase common in many other recipes~\citep{yang2025qwen3, agarwal2025gpt, dubey2024llama}. While our local-global pattern reduces the cost of long-context attention (see \cref{sec:architecture}), we still found that data-parallel workers with short-context documents would often wait for ranks with long-context data during distributed training. To improve iteration speeds, we ``bucketize'' all documents by sequence length, ensuring all workers draw documents from the same bucket at a given step. We choose the bucket boundaries as $(0, 16385]$,  $(16385, 65537]$, and $(65537, \infty)$ tokens and take care that the marginal probability of sampling a dataset is unchanged from bucketization. Note that, to achieve further speedups, we limit the maximum global attention size to \num{32768} in the medium bucket.

\subsection{Scaling laws}
\label{subsec:scaling_laws}
Scaling laws for LLMs that predict model performance as a function of compute, data, and model size have been studied extensively \citep{kaplan2020scaling, hoffmann2022training, bi2024deepseek}.  These empirical laws enable the estimation of the expected loss for a given compute budget, the identification of the optimal scaling strategy between model and data size, and an informed selection of training hyper-parameters. Following \citet{bi2024deepseek}, we develop scaling laws for optimal hyper-parameter prediction for the pre-training of \cwm.

We adopt the compute budget formula $C = M \cdot D$, where $M$ is the model size represented as the number of non-embedding FLOP per token and $D$ is the data scale corresponding to the total number of training tokens. For a decoder-only Transformer, the number of FLOP per token is approximated by
\begin{equation}
M = \underbrace{6N_{\text{ne}}}_{\text{linear term}} + \underbrace{6dLS}_{\text{attention term}},
\end{equation}
where \(N_{\text{ne}}\) is the number of parameters excluding embeddings, \(d\) is the model hidden dimension, \(S\) is the sequence length, and \(L\) is the number of layers. This formula explicitly accounts for the computational cost of self-attention, which constitutes a significant portion of the total compute, especially for smaller models and longer contexts where attention overhead is relatively more pronounced, as discussed in \cite{bi2024deepseek}. We refer to \cref{app:hyper_param_scaling} for further detail.

Recent LLMs are trained beyond data-optimal regimes \citep{dubey2024llama} to optimize inference costs and produce smaller yet capable models. \cite{gadre2024language} show that models scale predictably for a fixed model-to-data ratio and advocate for scaling laws that mirror the setting of the final pretraining run. Therefore, we maintain a fixed model-to-data ratio of $D/M=40$ across compute budgets, matching the target ratio of our \num{32}B parameter model pre-trained on \SI{8}{T} tokens. This ratio is roughly \num{8} times more data than would be compute optimal according to the Chinchilla paper \citep{hoffmann2022training}.

We conduct a quasi-random search over batch size and learning rate across eight increasing compute scales, ranging from \(2 \times 10^{18}\) to \(2 \times 10^{20}\) FLOP. For each scale, we keep the configurations within 1\% of the best validation loss and fit the batch size \(BS\) and learning rate \(LR\) with respect to the compute budget \(C\). Consistent with prior work, \(BS\) grows and \(LR\) declines gradually with \(C\), while near-optimal hyper-parameters span a broad range. However, likely due to our different pre-training data, our equations for learning rate and batch size diverge from \citet{bi2024deepseek}:
\begin{equation}
\begin{split}
LR(C) = 19.29 \cdot C^{-0.177},  \\
BS(C) = 30.17 \cdot C^{0.231}.
\end{split}\label{eq:label2}
\end{equation}
See \Cref{app:hyper_param_scaling} for additional details.

\section{Post-training: SFT, RL algorithms and environments}
\label{sec:posttraining}

Our post-training phase improves \cwm's ability to solve complex programming-related problems with reasoning, building on the internal code world model learned during earlier training stages. Concretely, we first perform supervised finetuning (SFT) to improve both reasoning and general instruction-following capabilities. We then carry out large-scale multi-task multi-turn reinforcement learning on coding contests, math questions, and software engineering environments. We describe the SFT stage, our RL algorithms, data and environments, and detail our joint RL training recipe. As we do not intend to develop a general-purpose chatbot we therefore deliberately omit an RLHF stage.

\subsection{SFT}
\label{sec:sft}
We perform SFT for \SI{100}{B} tokens, distributed across \SI{50}{k} steps with a global batch size of \SI{2}{M} tokens and \SI{32}{k} token sequence lengths.
We share optimization hyperparameters with pre-training but change the learning rate schedule to \SI{1}{k} steps of linear warmup followed by a constant learning rate of \num{1e-5}.
In preliminary experiments, keeping a constant learning rate achieved similar evaluation metrics to annealing with cosine schedules while enabling high learning rate training during RL. 
We further observed a performance decrease when SFT-ing at longer sequence lengths. 
We suspect this is due to the configuration of our dataloader which always sequence-packs inputs (per data-parallel rank and local batch) from a single dataset. For very small datasets and large context sizes, this reduces the amount of unique steps such datasets can be observed, which may negatively affect performance.

\looseness=-1
\newpara{Datamix.} We train on a diverse mix of internal and open-access data during SFT, including standard instruction-following datasets.
About \SI{30}{\percent} of the datamix is rehearsal from mid-training (which itself includes \SI{30}{\percent} pre-training data). This is to avoid overfitting to the SFT distribution ahead of RL and retain \cwm capabilities taught in mid-training.
Our datamix also contains agentic \swerl trajectories (see \Cref{sec:swerl}), some of which have been rejection-sampled from earlier iterations of the \cwm itself. We have generally found it useful to iteratively improve the starting point for RL by including trajectories from earlier iterations in the next SFT. Similarly, we include external datasets with reasoning traces, as we have found the performance benefit from them carries through to our final post-RL model. Specifically, we use the OpenMathReasoning~\citep{omr} and OpenCodeReasoning~\citep{ocr} datasets that rely on DeepSeek-R1~\citep{guo2025deepseek}. 

\newpara{Reasoning tokens.} For SFT training on reasoning data, we introduce \texttt{<|reasoning\_thinking\_start|>} and \texttt{<|reasoning\_thinking\_end|>} tokens that surround any reasoning text. Because we mask the loss on all \texttt{<|reasoning\_thinking\_start|>} tokens, the model does not learn to generate them.
This enables both reasoning and non-reasoning behavior for the \cwm-SFT model: non-reasoning mode is active by default and reasoning mode can be activated by injecting \texttt{<|reasoning\_thinking\_start|>} into the beginning of assistant responses.
Note that we discontinue the use of these reasoning tokens during RL as explained in \Cref{sec:rl_env_and_data}.

\subsection{RL algorithm}
\label{sec:rl_algorithm}

We use a variant of Group Relative Policy Optimization (GRPO) to train \cwm~\citep{shao2024deepseekmathpushinglimitsmathematical}.
GRPO is a policy gradient method that uses the PPO loss \citep{schulman2017proximal} in combination with Monte Carlo value estimation instead of a value model, as used in PPO. Many works have proposed improvements to and fixes of GRPO \citep{yu2025dapo, liu2025understanding, hu2025openreasonerzeroopensourceapproach,mistralai2025magistral}.
We incorporate a number of these and include further changes to support multi-turn RL and efficient asynchronous RL. A formal description of our RL algorithm can be found in~\Cref{sec:app_rl_alg}.

\newpara{Differences from GRPO.} We deviate from the original GRPO algorithm in the following ways:
\begin{itemize}
    \item \textbf{Multi-turn:} GRPO was originally developed for single turn (prompt $\rightarrow$ response) environments. Instead, we use a multi-turn variant where the sequence contains both model- and environment-generated tokens after the prompt, resulting in the need for masking via $M_{i,t}$. Furthermore, whereas GRPO used the \emph{reward} $r_i$, we use the \emph{return} $R_i$ (sum of rewards) in the advantage calculation.
    \item \textbf{Asynchronous:} Whereas GRPO uses a synchronous setup, where nodes switch between generating batches of completions and training on them, we use asynchronous RL, resulting in much higher throughput.
    \item \textbf{No $\sigma$ normalization:} GRPO calculates the advantage as the centered and scaled terminal reward $\hat{A}_{i} = (r_i - \mu) / \sigma$, where $\mu$ and $\sigma$ are mean and standard deviation of rewards in the batch. This introduces a difficulty bias \citep{liu2025understanding}, which we avoid by using the more conventional $\hat{A}_{i} = (R_i - \mu)$.
    \item \textbf{No length normalization:} As noted by \cite{liu2025understanding}, dividing the loss by the trajectory length as done in GRPO leads to a length bias, whereby the agent is incentivized to increase the length on hard problems so as to lower the average loss. To avoid this bias, we divide by the maximum number of tokens in a trajectory, which matches the maximum context size of our model $N=131072$.
    \item \textbf{Batching strategy:} We batch by a maximum token limit instead of the common approach that keeps a fixed number of trajectories per batch. This change aims at improving efficiency and to stabilize training by lowering the variance in batch size between different optimization steps -- this is important in combination with our removal of length normalization. As a result, different trajectories that belong to the same group might contribute to separate optimization steps. The proportion of groups that get split can be decreased by increasing the maximum token limit or the number of gradient accumulation steps, but in practice we observe this not to be an issue even when more than half of the groups are split.
    \item \textbf{Clip-higher:} Following ~\citep{yu2025dapo}, we use a higher upper clip value, $\varepsilon_{\text{high}} = 0.25$ and $\varepsilon_{\text{low}} = 0.2$ to prevent entropy collapse. %
    \item \textbf{No KL:} Using clip-higher to prevent entropy collapse, we found it unnecessary to use KL regularization\footnote{We recommend using the k2 estimator \citep{Schulman2020kl} whose gradient is an unbiased estimator of the forward $\operatorname{KL}(\pi_\theta, \pi_{\text{old}})$, rather than the k3 estimator used by GRPO whose gradient is an unbiased estimator of the reverse $\operatorname{KL}(\pi_{\text{old}}, \pi_\theta)$.}.
    \item \textbf{Skip zero-advantage trajectories:} The \textit{effective} batch size is the number of tokens that do not have zero advantage and contribute to the gradient. We reduce variance in the effective batch size by skipping all zero-advantage trajectories.
    \item \textbf{Skip stale trajectories:} To limit the degree of off-policyness, we skip trajectories whose most recent tokens were generated from a policy more than \num{100} training steps behind the current policy. 
    \item \textbf{Weighted mean return:} We found that longer trajectories are more likely to fail \citep{dont_over_think}, leading to the majority of \emph{tokens} having a negative advantage. To avoid biasing the token-averaged return, we compute $\mu$ as a length-weighted average.
    \item \textbf{Gibberish detection:} While gibberish typically leads to lower rewards and naturally decreases at the beginning of RL, it can increase later when some successful gibberish trajectories get reinforced, especially for agentic SWE RL. So we explicitly reject any trajectory containing any token $y_t$ that is both rare and generated with low probability: $\operatorname{id}(y_t) > 100,000$ and $\operatorname{logprob}(y_t) <  -\log(128,256) - 2$ where $128,256$ is the vocabulary size and the thresholds are tuned for high precision. Gibbrish typically consists of a window of tokens generated at low probability. BPE tokens are sorted by merge order where large $\operatorname{id}$ corresponds to rare tokens. Generating such tokens suggests that the model is generating at high entropy and over-weighing rare tokens. This method stopped any increasing gibberish generation and performed better than detectors based on logprob and position alone.    
\end{itemize}

\subsection{RL environments \& data}
\label{sec:rl_env_and_data}

We consider four types of RL tasks: Agentic software engineering (SWE) (\cref{sec:swerl}), Coding (\cref{sec:competitive-programming}), Agentic coding (\cref{sec:agentic-coding}), and Mathematics (\cref{sec:mathematics}).
Each RL task is defined by a dataset (containing prompts, a verification suite like unit tests, and additional metadata) and an environment that the agent interacts with.
We integrate these tasks into a joint RL training phase which we detail in \cref{sec:rl}.
We further refer to \Cref{sec:rl-systems} for implementation details regarding our environments and RL training infrastructure.

Our environments constitute partially observable Markov decision processes: a language model is employed as an agent, producing actions based on the preceding sequence of action-observation pairs and an initial prompt. All environments for training \cwm utilize software-based verification of outcomes, producing a single terminal reward signal per rollout.  In the remainder of this section, we describe the specification of environment, i.e., prompt and observation design, reward function, and the corresponding datasets.

During reasoning RL, we discontinue the use of SFT reasoning tokens and replace them with clear-text \texttt{<think>} \texttt{</think>} tags.
Early RL experiments on top of the SFT model showed long initial reasoning traces and slow improvements. We attribute this to our SFT reasoning data, which enhances reasoning performance but limits exploration during RL training. Switching out reasoning tags resulted in shorter responses, higher starting entropies, and significantly improved final performance. This suggests our approach leads to a best-of-both worlds scenario: the model's familiarity with reasoning responses from SFT enables rapid improvements early on in RL, while the introduction of the new tokens allows the model to develop its own reasoning that is guided -- but not restricted -- by the SFT data.

\subsubsection{Agentic SWE}
\label{sec:swerl}

\begin{figure}[t!]
    \centering
    \includegraphics[width=\linewidth]{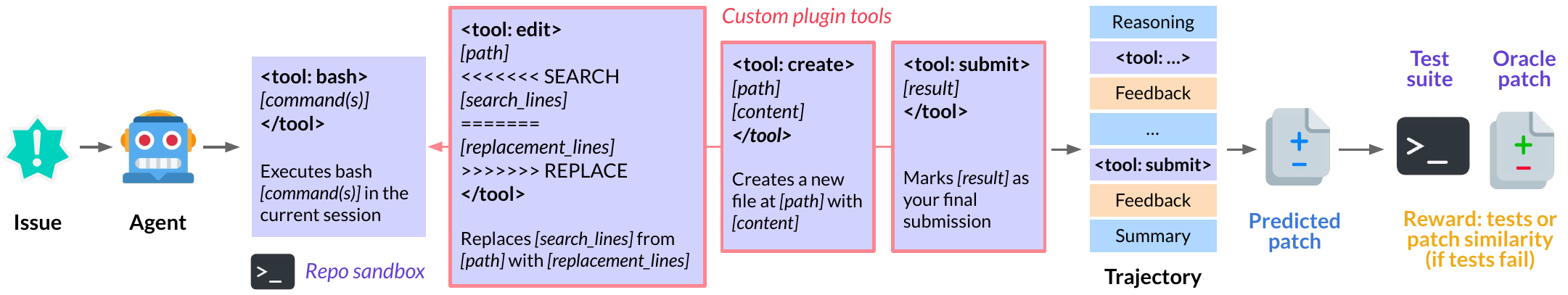}
    \caption{\swerl design. An agent solves software engineering tasks end-to-end through long-horizon agent-environment interactions via reasoning and tool use (up to 128 turns and \SI{131}{k} context size). \swerl employs a minimal toolset: \texttt{bash} as the core, with \texttt{edit}, \texttt{create}, and \texttt{submit} as lightweight \texttt{bash} plugins. The reward combines hidden test outcomes with patch similarity, where the similarity reward is applied when tests fail to provide auxiliary learning signals.}
    \label{fig:swerl:design}
\end{figure}

\newpara{Design.} Agentic \swerl substantially improves our model's software reasoning and engineering capability (e.g., on \sbv~\citep{DBLP:conf/iclr/JimenezYWYPPN24}) by enhancing the model’s agentic reasoning and tool-use skills (see \Cref{sec:swerl_ds} for the capability evolution during RL training). Its philosophy is to remain simple yet general: an LLM agent tackles a task end-to-end through reasoning and tool execution, without relying on task-specific post-processing.
The same design is applied to the agentic coding environment (see \Cref{sec:agentic-coding}). Each SWE RL trajectory has a single human user turn (besides the system prompt) containing the issue description and multiple turns of agent-environment interactions.
During training, we allow long-horizon interaction, with a maximum of \num{128} turns over a context window of \SI{131}{k} tokens.

As shown in \Cref{fig:swerl:design}, the agent is equipped with four tools to solve a given task (e.g., software issue), where we embrace a minimal tool design centered on bash and editing, inspired by Sonnet~3.5~\citep{sonnet-sbv}:
\begin{itemize}
    \item \texttt{bash}: executing commands in a stateful shell session,
    \item \texttt{edit}: modifying an existing file using the search/replace format used by Agentless~\citep{agentless} and Aider~\citep{aider},
    \item \texttt{create}: creating a new file in the sandbox, and
    \item \texttt{submit}: marking something (e.g., a file path) as the final submission according to the task requirement.
\end{itemize}
The runtime implementation of the tools follows SWE-agent~\citep{yang2024swe} and OpenHands~\citep{wang2025openhands}, where \texttt{bash} is a stateful shell session running in a persistent server process, and serves as the main component, while other customized tools are treated as plugins that can be ``de-sugared'' into simple bash commands. For example, the \texttt{edit} and \texttt{create} tools are two standalone Python scripts, and the \texttt{submit} tool, when used for file paths, reduces to \texttt{cat <path>} to retrieve the file content.

\begin{figure}[t!]
    \centering
    \includegraphics[width=\linewidth]{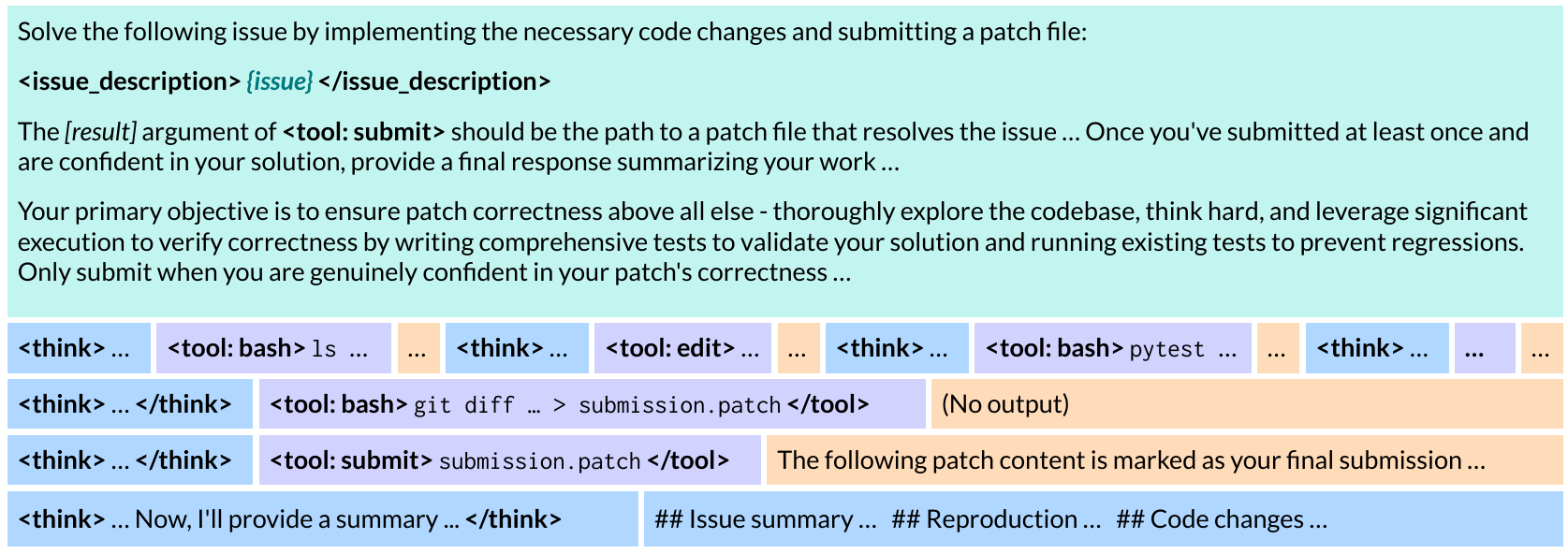}
    \caption{\swerl interaction example.     
    The agent interacts extensively with the repository sandbox through reasoning, exploration, editing, and test execution, submitting a final patch using \texttt{git diff} along with a summary.
    }
    \label{fig:swerl:example}
\end{figure}

The user prompt includes custom instructions for resolving software issues. 
For example, the prompt shown in \Cref{fig:swerl:example} asks the agent to ``...thoroughly explore the codebase, think hard, and leverage significant execution to verify correctness by writing comprehensive tests to validate your solution...'', which it follows in its subsequent actions. Notable differences from prior designs are that (1) our agent must generate the complete end-to-end patch directly via \texttt{git diff} rather than relying on task-specific post-processing, and (2) it must also produce a summary explaining how it resolves the issue, to improve clarity and usability. We also retain all reasoning turns for logical coherence.

\newpara{Reward.} We adopt a hybrid reward for \swerl. When all the hidden tests pass, the reward is $1$. If not, we adopt the patch similarity reward used in the SWE-RL paper~\citep{wei2025swerl}. Unlike the SWE-RL paper, which uses a continuous reward value, we apply a discrete and threshold-based design to improve the training stability by avoiding rewarding low-similarity patches.
In detail, when the computed similarity is above the threshold of \num{0.5}, the reward is \num{0}, otherwise, the reward is \num{-1}.
This reward shaping showed benefits in early ablation, because a higher patch similarity incentivizes the model to localize the actual bugs more precisely and to produce a closer fix to the oracle patch.
This also helps the model to gain more learning signals from difficult issues for which it cannot produce any test-passing patch.

\newpara{Data self-bootstrapping.} In \swerl, \cwm is required to solve software issues as a \emph{reasoning agent} (i.e., through both reasoning and tool use).
However, such data does not exist in the public and the format is completely new to the model before RL and different to our ForagerAgent data, so early iterations of our model struggled to interact with the software environment across long horizons without making format errors. To address this, we perform an iterative self-bootstrapping process to collect high-quality agentic reasoning traces and supply them back to the joint SFT stage so that \cwm can have a better prior distribution before RL. This process not only helps with format adherence but also significantly improves our model's software engineering capability both before and after RL.

\begin{figure}[t!]
    \centering
    \includegraphics[width=0.95\linewidth]{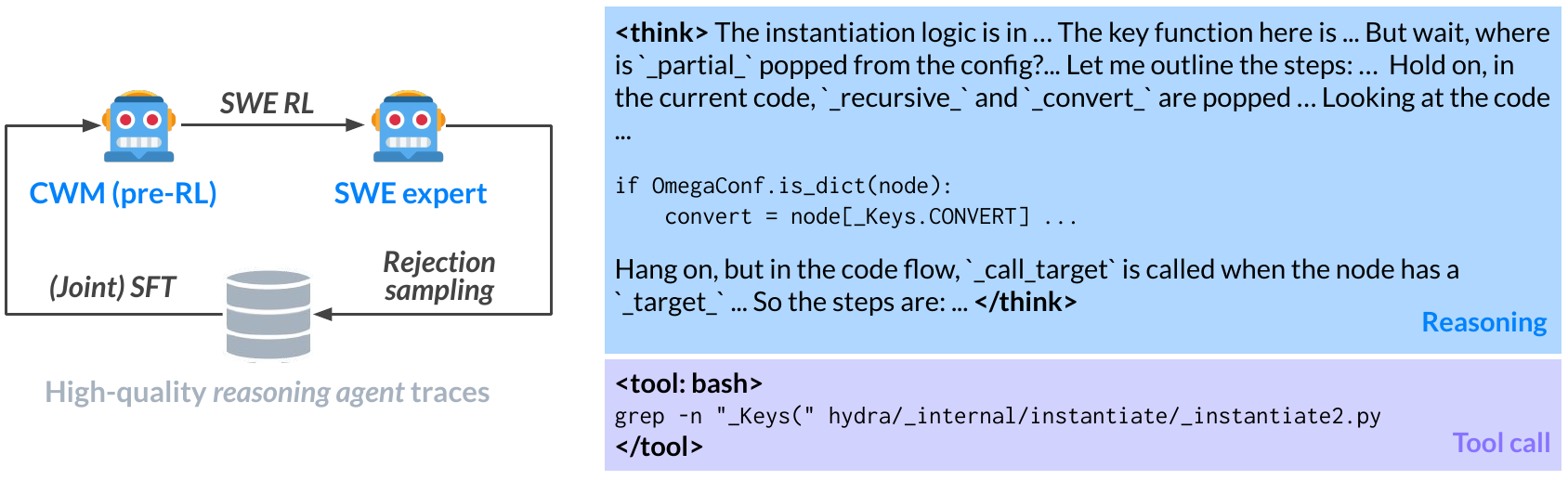}
    \caption{\swerl self-bootstrapping. Starting from a pre-RL checkpoint, we iteratively perform RL, rejection-sample high-quality reasoning traces, and feed them back into SFT. This process improves data quality and format adherence across iterations, raising success rates, and providing stronger initialization for joint RL.}
    \label{fig:swerl:self-bootstrapping}
\end{figure}

As shown in \Cref{fig:swerl:self-bootstrapping}, we start from a pre-RL \cwm checkpoint (not the final \cwm SFT) that has not been SFT-ed on any SWE trajectories in the reasoning agent format.
Over three main iterations, we perform RL and use the RL-ed model (i.e., the SWE expert in the figure) to do rejection sampling. We then select high-quality traces from the rejection samples using custom heuristics (e.g., long trajectories that pass all hidden tests without any tool use errors). Next, we perform SFT with this data on top of the original model.
This filtering helps reduce biases that RL fails to eliminate, such as the tendency to make editing mistakes. Then, iteratively, we start RL with the new SFT-ed model and collect higher-quality traces for the next round. Eventually, we include the final set of the traces into the joint SFT mix to prepare for the final joint RL.
This results in the final \cwm SFT model.
For each iteration, we redo SFT on the original midtraining checkpoint and discard old trajectories.
Importantly, we find that the bootstrapped data greatly improve the performance of the SFT checkpoint on \sbv. Without \swerl traces, the SFT model hardly resolves any issues due to format errors. With more iterations of bootstrapping, the data quality improves significantly, and the success rate increases from \num{30}\%, to \num{37}\%, and to \num{43}\% pass@1 over \sbv. During earlier iterations, we record the offline pass rate for each instance and use it as the GRPO baseline in later iterations. This lets us set the group size to 1 and speed up each epoch. We find this technique leads to faster \swerl training. In the final joint RL, we still perform online estimation of the GRPO baseline for consistency with other environments and for a higher performance ceiling.

\newpara{Data sourcing and filtering.} We reuse the executable repository images from our mid-training data generation efforts (\cref{sec:exec-envs}). Since issue solving requires additional metadata (e.g., issue text, base commit hashes, and diff patches), we join these repositories with publicly available issue and pull request metadata to create repository-issue pairs. The \texttt{git log} history enables us to create one-to-many repository-issue pairs. We also include publicly available training data such as SWE-Gym~\citep{swegym} and R2E-Gym~\citep{jain2025r2e}, further filtered by us for quality (e.g., removing non-verifiable instances whose tests cannot pass). All training data are decontaminated against \sbv at repository-level granularity, see \Cref{sec:appendix:rl-decontamination}.

We estimate the difficulty of each instance using the pass@1 score from \cwm SFT, calculated over at least \num{32} samples. Instances with a pass@1 above \num{95}\% are filtered out as easy, while those with a non-zero pass@1 are included in the primary dataset. Instances with a \num{0}\% pass@1 are placed in a secondary dataset that is sampled less frequently at the beginning. To make these hard problems solvable, we augment their prompts by adding the hidden test as a hint. This augmentation increases the pass@1 rate from \num{0}\% to approximately \num{30}\%. Later in training, we remove hints from hard instances so the model learns to solve them from scratch. Finally, this process yields \SI{12.6}{k} unique training instances: \SI{6.9}{k} in the primary set and \SI{5.7}{k} in the secondary.

\subsubsection{Coding}
\label{sec:competitive-programming}

\begin{figure*}[t!]
    \centering
    \includegraphics[width=0.7\linewidth]{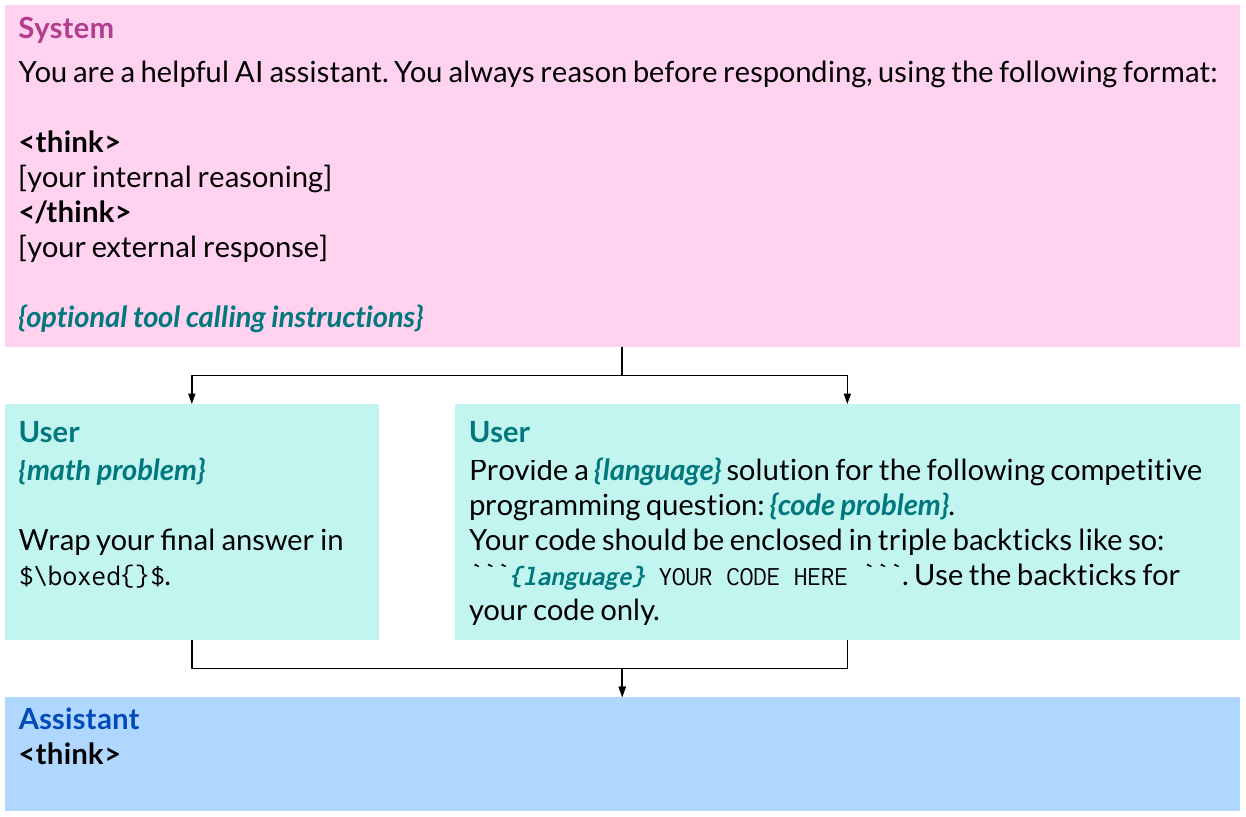}    
    \caption{Prompt template for math (left path) and competitive programming (right path) RL tasks.}    
    \label{fig:math-code-rl-prompt}
\end{figure*}

\newpara{Design.} 
RL for competitive programming aims to teach the model to write correct programs for challenging tasks and to reason about code and algorithms. Our competitive programming environment presents the problem to the agent in the first turn and optionally allows follow-up attempts, during which the environment provides execution feedback. It supports multiple programming languages and provides detailed feedback on syntax errors, timeouts, and incorrect test outputs. The environment terminates either when the maximum number of turns is reached or when the agent produces a correct solution. In the joint RL run, we limit the number of attempts to one but allow up to \SI{64}{K} tokens in responses to enable extensive reasoning.

We adapt a lightweight prompt template which is shown in~\Cref{fig:math-code-rl-prompt}. The system prompt asks for reasoning delimited by \texttt{<think>...</think>} in clear-text. The user prompt specifies the programming language and instructs the agent to put the code solution inside a markdown block.

\newpara{Reward.} We assign a reward of \num{-1} for incorrect trajectories and \num{1} for correct ones. A trajectory is correct if it meets all of the following criteria: 
\begin{itemize}
    \item Contains exactly one \texttt{</think>} tag, signaling successful reasoning completion.
    \item Contains exactly one markdown block in the model's generated answer.
    \item The code solution passes all unit tests within the specified time and memory limits. We execute the unit tests in parallel using an internal code execution service on remote machines.
\end{itemize}

\newpara{Data sourcing and filtering.}
We source coding problems from various programming contest websites. A problem typically consists of a problem description, limitations on the input and output domains, memory and time limits, and input-output examples. In addition, each problem comes with a set of tests that we use to verify the correctness of candidate solutions: a solution is considered correct only if it produces the expected output for any given test input.

We decontaminate the coding problems against test benchmarks and de-duplicate them to ensure that each training problem is unique. In both cases, we use MinHash-based similarity detection\footnote{\url{https://github.com/serega/gaoya}}, applying word- or character-based matching depending on the length of each document. This process ensures the integrity of our evaluations, which is especially important for code generation and mathematical problems, where data contamination can significantly affect performance metrics.

We use Llama-3.3-70B-Instruct to identify and remove poorly posed problems, such as those containing gibberish, missing or truncated problem statements, or lacking input/output descriptions. We do not apply any difficulty-based filtering. After decontamination (see \cref{sec:appendix:rl-decontamination}), the final code RL dataset has \SI{81}{k} prompts.

\subsubsection{Agentic coding}
\label{sec:agentic-coding}

\begin{figure}[t!]
\centering
\includegraphics[width=\linewidth]{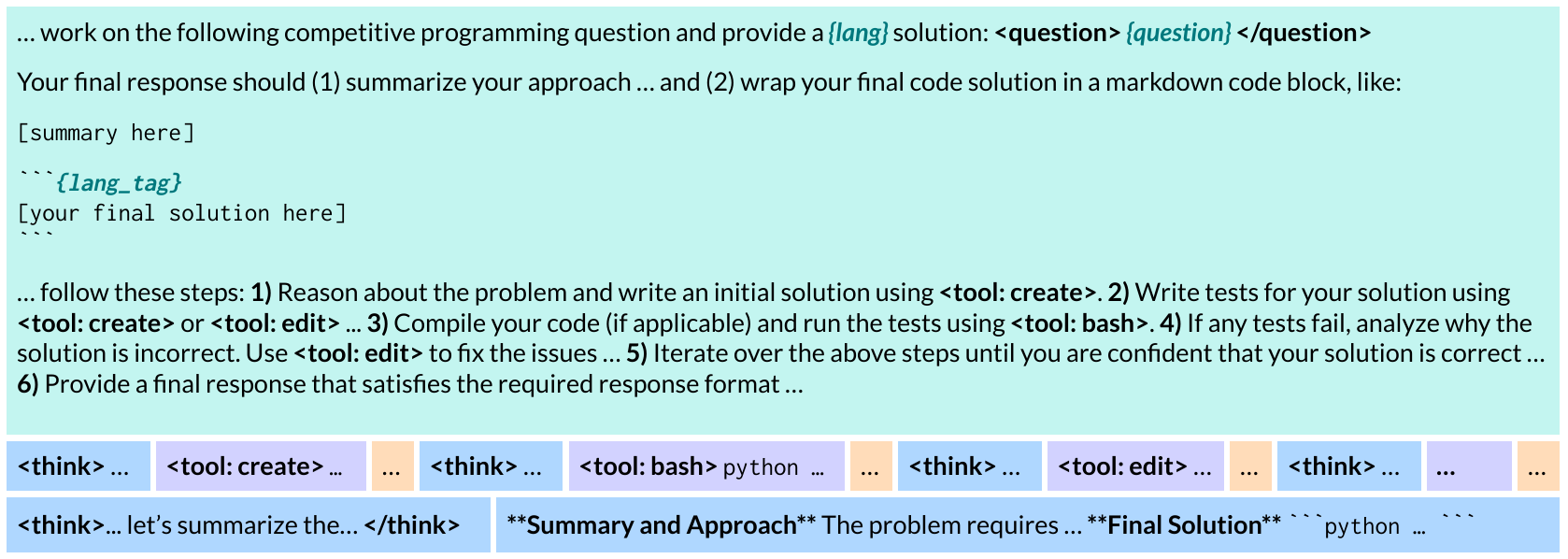}
\caption{Example interaction for the agentic coding RL environment. The agent uses reasoning and tools to solve competitive programming problems. Before generating a final solution, the agent summarizes the interaction.}
\label{fig:agentic-coding-prompt}
\end{figure}

The agentic coding environment combines the reasoning and tool use features of the \swerl environment (\Cref{sec:swerl}) with the competitive programming setup described in \Cref{sec:competitive-programming}. The user prompt is customized for solving competitive programming tasks and explicitly asks the agent to write and run tests to check and improve the solution, as illustrated in \Cref{fig:agentic-coding-prompt}. Different from \swerl, there is no \texttt{submit} tool in this environment. Instead, the agent needs to provide the solution in its final response, which is then extracted for evaluation. We consider two programming languages, Python and C++, where the images used for agentic interaction are \verb|python:3.11-slim| for Python and \verb|python:3.11-bookworm| (with \verb|gcc 12| support) for C++.
The final solution is then evaluated using the same execution infrastructure as for the competitive programming environment.

\subsubsection{Mathematics}
\label{sec:mathematics}

\newpara{Design.}
Although not the main focus on this research work, we consider mathematical reasoning as another RL task to further strengthen and generalize \cwm's reasoning capabilities. We restrict these problems to questions that have definitive and easy to verify answers. Both the questions and answers are formulated in \LaTeX, similar to much of the math content found on the web.

We also include a tool-enabled version of the math environment, adhering to the format described in \Cref{appendix:prompting_guide}. Here, the agent may invoke the Python interpreter with custom code. Standard output and error contents will form the next observation and the agent is prompted to continue solving the task. We impose a limit of \num{4} tool calls per episode and a \SI{10}{s} timeout per call.

The prompt template shares its system prompt with the coding environment but differs in the user prompt. As shown in \Cref{fig:math-code-rl-prompt}, the user prompt instructs the agent to place the final answer inside a \LaTeX{} box.

\newpara{Reward.}
Every trajectory is classified as either correct (reward = \num{1}) or incorrect (reward = \num{-1}). Correctness is defined as: 
\begin{itemize}
    \item Exactly one \texttt{</think>} tag, signaling successful reasoning completion.
    \item Exactly one \texttt{\$\textbackslash boxed\{\}\$} for the predicted answer.
    \item Our verifier emits \texttt{True} for the comparison between the predicted answer and the ground-truth answer.
\end{itemize}
Since there is no general normal form for mathematical expressions, the verifier checks whether the predicted answer is equivalent to the ground truth answer using a variety of heuristics detailed in~\cref{app:math-verify}.

\newpara{Data sourcing and filtering.} We gather math questions and answers from publicly available sources. To remove duplicates, we use the MinHash LSH algorithm to identify similar problems and verify that they have the same answer using our verifier. We also filter out problems that were solved correctly in all attempts (\num{32} out of \num{32}) by our SFT model in non-reasoning mode. This helps avoid wasting compute on easy problems and reduces the risk of reinforcing incorrect reasoning followed by a correctly memorized answer. The prompt set used for RL training contains a total of \SI{278}{k} problem-answer pairs.

\subsection{Joint RL}
\label{sec:rl}
\begin{figure}[t!]
    \centering
    \includegraphics[width=\linewidth]{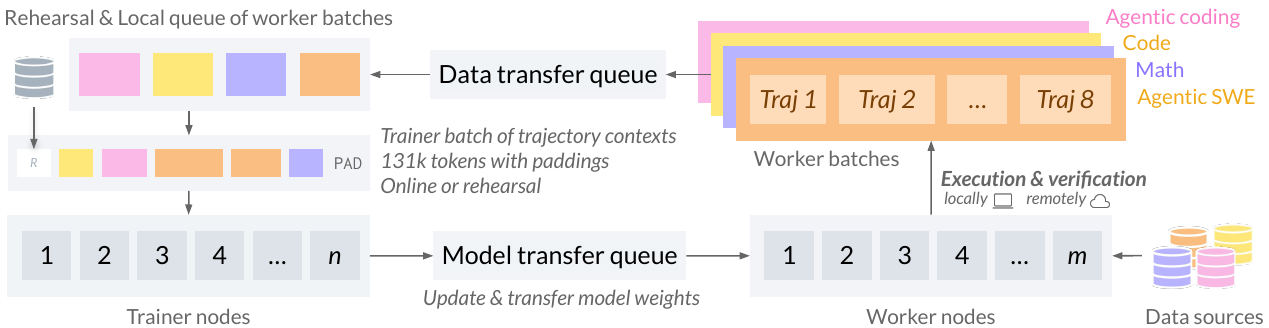}
    \caption{Async RL systems overview. Worker nodes generate trajectory batches from multiple RL environments and send them to trainer nodes via a transfer queue. Trainer nodes form training batches either from worker-provided data or the rehearsal mix, packing trajectories up to the maximum context length for a single gradient update. Environment execution and verification can occur locally on worker nodes or remotely on another cluster or in the cloud.}
    \label{fig:joint-rl}
\end{figure}

Finally, we train \cwm using all of the above-mentioned RL tasks. As shown in \Cref{fig:joint-rl}, joint RL uses our asynchronous RL infrastructure: worker nodes generate $G$ trajectories per prompt from multiple RL environments and send them to trainer nodes through the data transfer queue. The trainer nodes then form training batches either from these worker-provided batches or directly from the SFT datamix (\Cref{sec:sft}) for rehearsal. We refer to \Cref{sec:rl-systems} for further detail on engineering aspects of our asynchronous RL infrastructure.

\newpara{Data and RL environment mix.}
Worker nodes produce trajectories from three main data sources: software engineering, competitive programming, and mathematics. They use the four RL environments we describe in the previous sections, which we refer to as agentic SWE (\Cref{sec:swerl}), code (\Cref{sec:competitive-programming}), agentic coding (\Cref{sec:agentic-coding}), and 
math (\Cref{sec:mathematics}). Each data source may contain multiple datasets from different origins; however, all datasets within the same data source share a consistent format and problem domain. We sample \SI{40}{\percent} of tasks from software engineering, 
\SI{40}{\percent} from competitive programming, and \SI{20}{\percent} from mathematics. Rehearsal batches constitute $1/3$ of the training data and are integrated with a standard negative log likelihood loss, scaled by a factor of \num{0.1} to match the gradient magnitudes obtained with GRPO (\Cref{sec:rl_algorithm}).

\newpara{Three-stage training.}
We split our joint RL training into three distinct stages. Between stages, we adapt the task distribution and employ custom reward shaping techniques.

\begin{itemize}
    \item \textbf{Stage 1 -- Reasoning format bootstrapping:}
    In the initial training stage, we soft-control the length of generations in math and coding tasks with an action length reward schedule.
    For the \SI{40}{\percent} of tasks related to competitive programming, we evenly sample from four environments: code Python, code C++, agentic coding Python, and agentic coding C++ (10\% each).
    For a subset of SWE tasks identified as challenging, we include a hint in the prompts and downsample their occurence (4\% of overall tasks; \Cref{sec:swerl}).    
    \item \textbf{Stage 2 -- Increasing task diversity and data resampling:}
    After \num{14125} gradient steps, we increase the proportion of competitive programming tasks in the datamix to \SI{50}{\percent} and reduce the fraction of SWE tasks to \SI{30}{\percent}. 
    We also include additional environment variations for each task. 
    Specifically, we add Rust, Go, Java, and JavaScript versions of the code environment, which, together with Python and C++, now constitute \SI{25}{\percent} of the datamix. The other half of the competitive programming tasks use the agentic coding environment, to which we do not add new languages.
    For the SWE data, we disable plugins with a \SI{50}{\percent} chance, such that file edits require standard terminal commands.
    We further  remove hints from the challenging SWE subset and oversample it in a $4:1$ ratio when plugins are used, and reverse this ratio when plugins are disabled.
    Competitive programming and SWE datasources are filtered to include instances with a solve rate in $[0.1,0.7]$ only in order to to maximize the learning signal.
    For math tasks, we enable Python tool calling for \SI{2}{\percent} of the total datamix.

    At \num{16500} steps, we apply filtering based on solve rate with the $[0.1,0.7]$ interval to our math dataset as well.
    For SWE data, we create fine-grained subsets for each 0.1 solve-rate interval from $(0.0, 0.7]$, such as $(0.1,0.2]$ and $(0.6,0.7]$, and sample harder examples more frequently using weights inversely proportional to the interval's midpoint.
    
\end{itemize}

\newpara{Hyperparameters.}
After an initial linear warmup over 100 steps, we employ a learning rates of \num{2.5e-7} throughout training.
The maximum batch size for each gradient step is \SI{8.4}{M} tokens during the first stage and \SI{16.8}{M} tokens for the second stage.
Gradients are clipped to norm 0.1. We use $G=8$ rollouts per data point, and new model weights are broadcast to workers after 4 gradient steps. We list further GRPO-specific hyperparameters in \Cref{sec:rl_algorithm}.

\begin{figure}[t!]
    \centering\includegraphics[width=\linewidth]{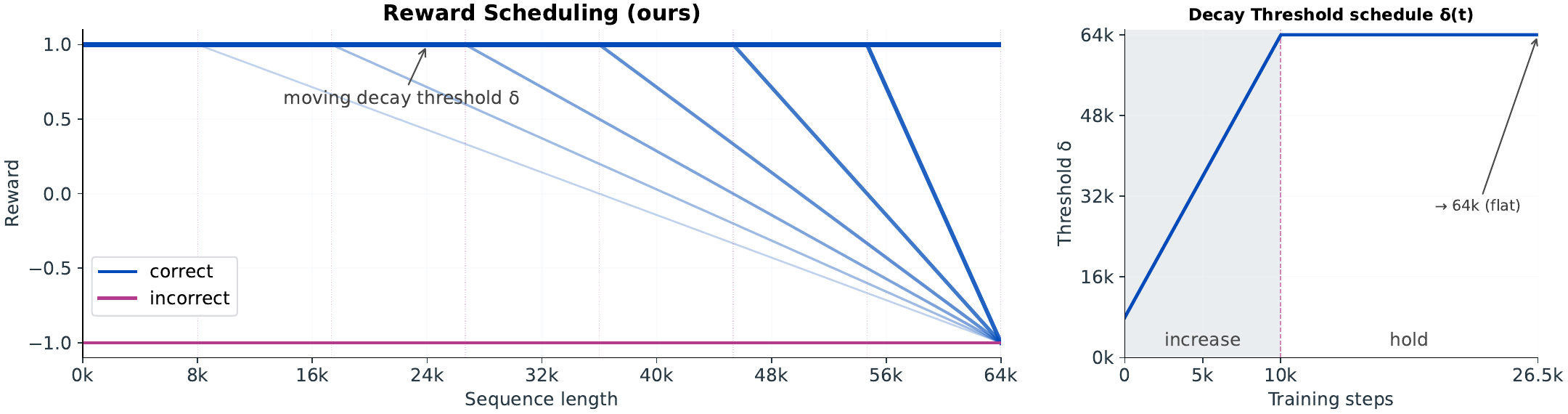}
    \caption{Length reward scheduling for RL training. The decaying threshold $\delta$ starts at \SI{8}{k} at the start of training and linearly increases to its \SI{64}{k} limit over \num{10000} steps.}
    \label{fig:reward-scheduling}
\end{figure}

\newpara{Length reward scheduling.} 
In both the code and mathematics environments, we allow context lengths of up to \SI{64}{k}. We observe that, at the start of RL training, the model rapidly increases its response length, leading to inefficient token usage. To address this, we penalize the reward for correct but overlong solutions similar to DAPO~\citep{yu2025dapo}, but gradually phase out this penalty over training. Specifically, we linearly interpolate the reward between \num{1} and \num{-1} for correct answers with a length that exceeds a soft maximum (\SI{8}{k} at the beginning of training) but is lower than the hard maximum of \SI{64}{k}. This provides a dense reward signal to the model that incentivizes it to reduce its response length, while still providing a positive signal when the answer is correct.
During training, we gradually increase the soft maximum in a continuous manner until it is equal to the hard maximum after \SI{10}{k} training steps. See \Cref{fig:reward-scheduling} for an illustration of this process. 

\vspace{-0.2cm}

\section{Code and infrastructure}
\label{sec:code_infra}
This section discusses details of our training pipeline, including efficiency-related features leveraged for \cwm training and the asynchronous RL architecture.

\subsection{Techniques for efficient training}
\label{subsec:efficient_training}

\looseness=-1
\cwm is trained on H100s using a combination of Fully-Sharded Data Parallelism (FSDP) and Tensor Parallelism (TP), see \Cref{tab:infra}. We adopt FlashAttention-3 \citep{dao2022flashattention,dao2023flashattention2} to improve training speed and reduce memory overhead. Additionally, we incorporate several optimizations towards efficient training.

\looseness=-1
\newpara{\texttt{fp8} matrix multiplication.} All linear layers in transformer blocks used \texttt{float8} low-precision mode, similar to \cite{micikevicius2022fp8}, achieving twice the nominal FLOPs of \texttt{bfloat16} on Hopper GPUs. (For RL training, fp8 precision hurt performance and hence we used \texttt{bfloat16} for linear layers in transformer blocks.) We used dynamic ``row-wise'' scaling, also called ``outer-vector'', that is, operands were scaled along their reduction dimension. We used the \texttt{e4m3} variant exclusively, and we disabled fast-accumulation throughout. The two matrix multiplication operations for the gradient computation in the backwards pass use special setups: \texttt{w.grad} is computed in \texttt{bfloat16} (which increased accuracy and precluded the need to transpose its operands to satisfy \texttt{float8} layout constraints, which make kernel fusion difficult); \texttt{in.grad} uses ``tensor-wise'' scaling for its weight operand, i.e., a single scaling factor for the whole tensor, which again makes transposition more efficient. In practice, we try to issue ``unscaled'' matrix multiplication kernels, introducing scaling in the kernel epilogue worsens performance, and perform the scaling as a manual post-processing step, which can be fused into subsequent kernels.

\newpara{Reducing communication overhead for tensor parallelism.} We reduced the communication overhead of tensor parallelism (which we implement as sequence parallelism) by more-effectively overlapping it with computation via decomposition and micropipelining, using PyTorch’s Asynchronous Tensor Parallel (Async-TP) feature\footnote{See \hyperlink{https://discuss.pytorch.org/t/distributed-w-torchtitan-introducing-async-tensor-parallelism-in-pytorch/209487}{Async tensor parallelism in PyTorch with TorchTitan}.}, derived from xFormers \citep{lefaudeux2022xformers} and originally inspired by \cite{wang2022overlap}. We implemented this optimization by-hand for the matrix multiplication needed to compute \texttt{w.grad} during the backward pass (sharded along the reduction dimension), as there is no defacto support in PyTorch.

\newpara{\texttt{fp8} with tensor parallelism.} When TP is enabled, we adapted our \texttt{fp8} recipe to further improve performance. We use ``sub-row-wise'' scaling where appropriate to align quantization boundaries with TP shards, which both avoids communication and improves accuracy. We perform all-gathers in \texttt{fp8}, which improves throughput and also enables fusing quantization into previous kernels (e.g., LayerNorm). During the backward pass, however, we sometimes all-gather the same data twice (once in \texttt{fp8} and once in \texttt{bf16}, since it will be consumed by two separate matmuls, one for each dtype). Because of Async-TP, however, this adds zero overhead and enables fusing quantization into previous kernels.

\newpara{Reducing memory consumption.} We use PyTorch’s AutoAC\footnote{Enabled by setting \texttt{torch.\_functorch.config.activation\_memory\_budget < 1}.} for activation checkpointing, which is integrated in the ``partitioner'' layer of the \texttt{torch.compile} stack, and uses an integer-linear program solver to optimize the memory-versus-recomputation tradeoff given a user-provided budget. We also leverage PyTorch’s vocab- and loss-parallel helpers to reduce memory consumption. 

\begin{table}[t!]
\caption{Summary of the training setup for the different \cwm training stages on H100 GPUs. 
 \label{tab:infra}}
\centering
\begin{tabular}{l|rr|r|rr}
\toprule
\textbf{Phase} & \textbf{Seq. Length} & \textbf{Batch size} & \textbf{\# GPUs} & \multicolumn{2}{c}{\textbf{Shards}} \\
\cline{5-6}
 & & & & \textbf{DP} & \textbf{TP} \\
\midrule
Pre-training & \SI{8}{k} & \SI{8.4}{M} & \num{2048} & \num{1024} & \num{2} \\
Mid-training & \SI{131}{k} & \SI{33.6}{M} & \num{2048} & \num{256} & \num{8} \\
\midrule
Supervised Fine-tuning  &  \SI{32}{k} & \SI{2.1}{M} & \num{256}  & 32 & 8 \\
Reinforcement Learning  & \SI{131}{k} & \SI{8.4}{M}/\SI{16.8}{M} & 2560/4608 & 64 & 8 \\
\bottomrule
\end{tabular}
\end{table}

\subsection{RL systems}
\label{sec:rl-systems}

We train our models using our own asynchronous distributed RL framework. The key distinction from the standard \texttt{llm} training lies in the data collection process: in RL, training data is gathered through rollouts where the agent interacts with an environment.

\begin{figure}[t]
    \centering
    \includegraphics[width=\linewidth]{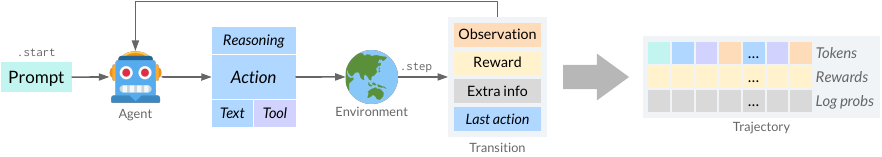}
    \caption{Overview of how agents interact with RL environments to produce trajectories.}
    \label{fig:rlenv}
\end{figure}

\newpara{Rollouts.}
As shown in \cref{fig:rlenv}, a rollout consists of a sequence of iterative agent-environment interactions. Each environment implements two methods:
\begin{itemize}
\item \texttt{start}: start a new episode by producing an initial state and an observation (prompt) based on a sample from the dataset. The state encapsulates the contents of the hidden environment along with any specific resources corresponding to the current episode.
\item \texttt{step}: takes an action (sequence of tokens) leading to a state transition. The new observation includes all information visible to the agent and necessary during training or inference, such as the latest action, observation, and reward.
\end{itemize}

All interactions between the agent and the environment are token-based. In addition, the environment can suggest context switches to erase past history or restart from scratch, allowing multi-context trajectories.
Our environments adhere to a common \emph{trajectory format}, which prescribes that a trajectory consists of a sequence of messages, whose format is detailed in \cref{appendix:prompting_guide}.

\begin{figure}[t]
 \centering
\includegraphics[width=0.7\linewidth]{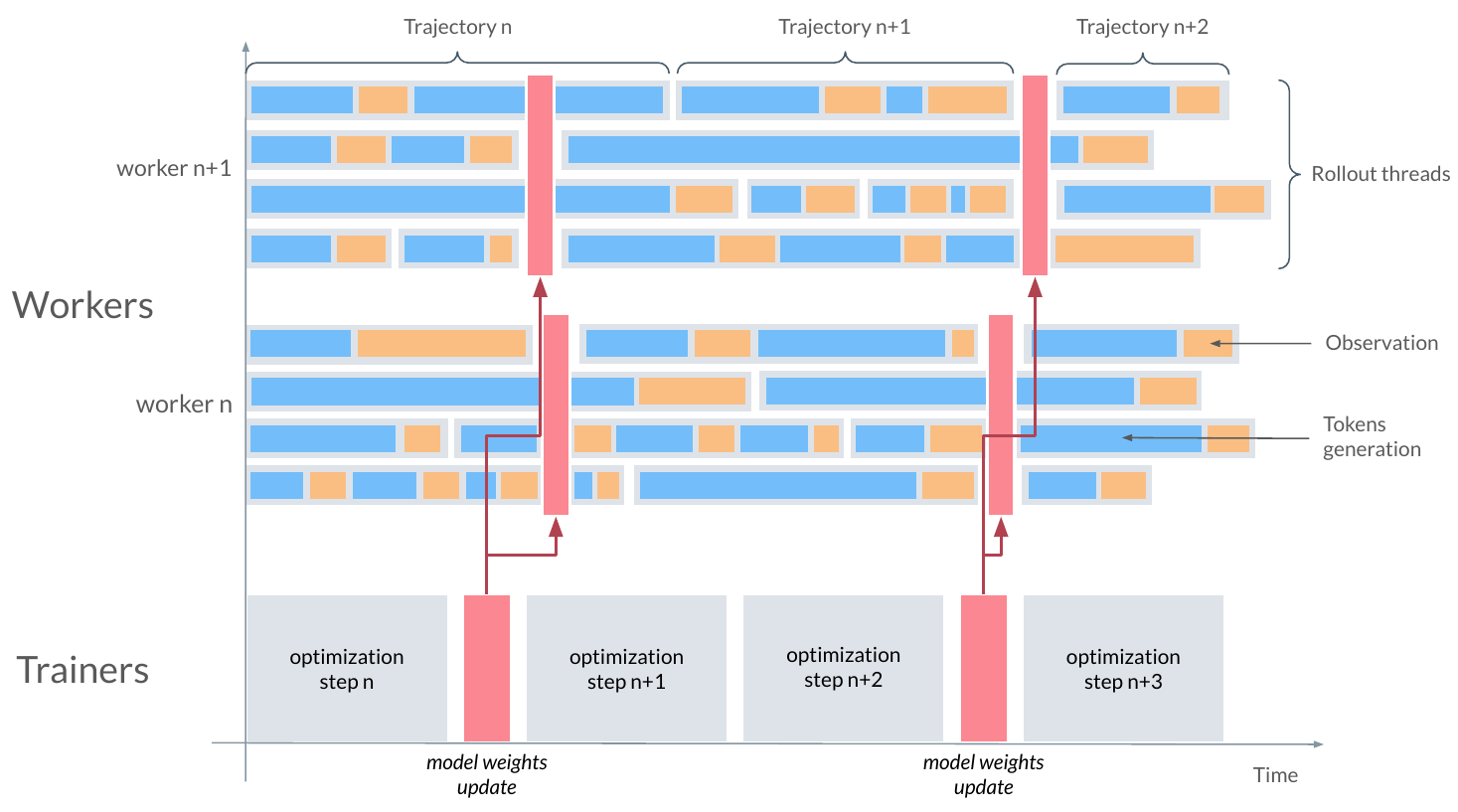}
  \caption{In \cwmrl, model weights can be updated at any time on the worker side: between trajectories, within a trajectory between steps, or even during token generation. Compared to traditional RL, this removes all synchronization overhead, maximizing worker throughput while minimizing idle time. In exchange for never blocking inference, we accept that trajectories will potentially use mixed weights, though frequent model updates ensure that generations remain reasonably on-policy. Different workers may not update their weights at the same time: the system waits for each worker to signal readiness before sending new weights to avoid memory overload.}
  \label{fig:rl_model_updates}
\end{figure}

\newpara{Training.}
GPUs are divided into a set of \emph{workers} that continuously perform rollouts and \emph{trainers} that update the current policy. Workers send batches of trajectories to the trainers as soon as they are completed and trainers send updated model weights to the workers periodically. After a model update is received and applied on a worker, the worker continues generation of partially completed trajectories using the old KV-cache. This approach ensures continuously high GPU utilization (see \Cref{fig:rl_model_updates}) and has been used in our previous work~\citep{synnaeve2019growing,gehring2025rlefgroundingcodellms, DBLP:journals/corr/abs-2503-19595, DBLP:journals/corr/abs-2503-05453} and notable RL frameworks such as PipelineRL \citep{Piche2025-nj}.

\looseness=-1
\newpara{Inference.} We use our own throughput-optimized inference backend FastGen~\citep{Carbonneaux2025fastgen}.
FastGen supports batched inference, CUDA graphs, paged attention \citep{kwon2023efficient}, chunked prefills, host-side KV-cache, tensor parallelism, and CPU/GPU profiling.
In batched inference, one generates tokens for each sequence in a batch in parallel, continuing without synchronizing CUDA streams until a block of tokens (e.g.,~\num{32}) is completed.
After each block, completed sequences are truncated at stop tokens and returned, and new sequences are added to the batch so as to keep the batch size constant.
For more details, see \cite{Carbonneaux2025fastgen}.

\newpara{Parallelism.} We support various kinds of parallelism on both trainer and worker nodes.
Trainers operate largely as in pretraining (see \Cref{sec:pretraining}), supporting FSDP and TP.
Worker nodes are grouped with TP to perform batched inference.
Whereas all trainer GPUs are synchronized, the worker groups/model replicas operate  asynchronously from each other and from the trainers.

\newpara{Model transfer.} For efficient model transfers, we use our custom PyTorch distributed backend, moodist\footnote{See \url{https://github.com/facebookresearch/moodist}.}~\citep{Mella2025moodist}. It implements efficient queues that transfer data directly between GPU and CPU memory via InfiniBand both within the same compute node and between different nodes. This facilitates transferring model weights directly from the trainer's GPU memory to the worker's CPU memory. With FSDP, each trainer has a shard of the model weights. These shards must be concatenated and sent to the workers.
Model transfer consists of three stages:
\begin{enumerate}
\item Each trainer sends their local shard from GPU memory to the CPU memory of a single worker.
\item The workers perform a distributed concatenation similar to an all-gather, such that each worker ends up with all model weights.
\item Each worker then individually applies the new weights.
\end{enumerate}
The trainers are only involved in the first stage, which minimizes the amount of time model transfer takes on the trainers. On the workers, the first and second stages both occur entirely in CPU memory, which allows them to overlap with generations. The third stage is simply a CPU to GPU memory copy, so it is reasonably fast. When TP is enabled, this process occurs individually for each data-parallel group.

\newpara{Execution infrastructure.}
Our training pipeline leverages an internal code execution service to safely execute tens of thousands of code snippets per second, in parallel across multiple programming languages and asynchronously in isolated containerized environments. This code execution service is integrated into our training loop to provide execution results including stdout, stderr, exit codes, and environment state as feedback to the LLM.

\newpara{Containerized execution for agentic RL.} We use a custom tool-based execution environment for agentic reinforcement learning, enabling agents to interact with containerized environments through structured tool calls for agentic tasks. It features a core tool execution framework based on flexible container backends (e.g., Docker execution services or Modal~\citep{modal}), implementation of remote execution servers and clients as an interface to a persistent shell session, plugins that can be defined as standalone Python scripts invoked through bash, along with evaluation infrastructure for reward calculation or benchmarking.

\section{Experimental results}
\label{sec:exp}

We begin this section by analyzing the impact of incorporating \cwm data during mid-training for a small-scale ablation. 
Next, we evaluate \cwm and compare its performance against relevant baselines, focusing on coding and mathematical reasoning tasks.
We consider agentic evaluation for coding tasks, together with additional computation-oriented evaluations covering 
(i) output prediction with execution traces and reasoning, (ii) full execution trace prediction, (iii) program termination prediction, and (iv) prediction and generation of algorithmic complexity. 
Finally, we evaluate \cwm considering established benchmarks for competitive programming, mathematical reasoning, non-reasoning evaluation, and long-context. 
Unless otherwise mentioned, we use a temperature of \num{1.0} and top-p value of \num{0.95} for all evaluations.

\subsection{The impact of \cwm data} 
To evaluate the effect of incorporating \cwm data during mid-training, we perform ablations with \SI{8}{B} parameter models trained for \SI{7}{T} tokens. We first pre-trained one model for \SI{6}{T} tokens and then studied different mid-training datamixes for the remaining \SI{1}{T} tokens, ablating the two \cwm datasets, ForagerAgent and Python execution trace data, as well as our Github PR trajectory data.\footnote{One may wonder if using the non-agentic PR data alone is sufficient for reaching strong performance on \sbv.} After mid-training, all variants underwent a fine-tuning phase comparable to our main setup for \cwm described in \cref{sec:sft} but excluding the RL phase. We report results on CruxEval-O, CruxEval-I, NLLs over \sbv (SBV) oracle patches, and NLLs over agentic SBV trajectories (truncated to \SI{32}{k} sequence length) for the models out of mid-training and pass@1 SBV numbers for the models after SFT. 

The results in \Cref{tab:cwm_ablations} show that the best performance across our set of metrics is achieved when using all datasets together. This effect carries over to our SBV evaluation of the SFT model, demonstrating how mid-training data choices can positively affect final model performance. Looking at the impact of individual datasets, we find the inclusion of the PR data helps oracle SBV NLLs and SBV pass@1 but not the agentic SBV trajectory NLLs or CruxEval. Further incorporating execution trace data significantly improves CruxEval-input and -output prediction but leaves all SBV-related metrics unaffected. Lastly, only the addition of ForagerAgent data improves agentic SBV NLLs.
The ForagerAgent data is further able to improve SBV pass@1 scores by another \SI{3.7}{\percent.}.

\begin{table}[t]
\caption{Our ablation study reveals a positive impact on performance from introducing GitHub PR trajectory, Python execution tracing, and ForagerAgent data during mid-training. We report results for CruxEval-output, CruxEval-input, NLLs on oracle \sbv (SBV) trajectories, NLLs on agentic SBV trajectories, and SBV pass@1 scores. All results are for \SI{8}{B} models, jointly pre-trained for \SI{6}{T} tokens followed by \SI{1}{T} tokens of mid-training ablation, with SBV pass@1 reported after an additional SFT phase.
\label{tab:cwm_ablations}}
\centering
\resizebox{\textwidth}{!}{
\begin{tabular}{ccc|rrrr||r}
\toprule
\textbf{PRs} & \textbf{Tracing} & \textbf{Forager} & \textbf{CruxEval-O}$\uparrow$ & \textbf{CruxEval-I}$\uparrow$ & \textbf{Oracle SBV NLL}$\downarrow$ & \textbf{Agentic SBV NLL (32k)}$\downarrow$ & \textbf{SBV}$\uparrow$ \\
\midrule
\ding{55} & \ding{55} & \ding{55} & 45.4 & 44.1 & 0.64 & 0.39 & 14.6 \\
\ding{51} & \ding{55} & \ding{55} & 44.6 & 45.8 & \textbf{0.55} & 0.37 & 18.6 \\
\ding{51} & \ding{51} & \ding{55} & 73.9 & 51.5 & \textbf{0.54} & 0.38 & 18.4 \\
\ding{51} & \ding{51} & \ding{51} & \textbf{74.5} & \textbf{54.8} & \textbf{0.54} & \textbf{0.29} & \textbf{22.1} \\
\bottomrule                  
\end{tabular}}
\end{table}

\subsection{Agentic evaluation}

\begin{figure}[p]
    \centering
    \includegraphics[width=0.9\linewidth]{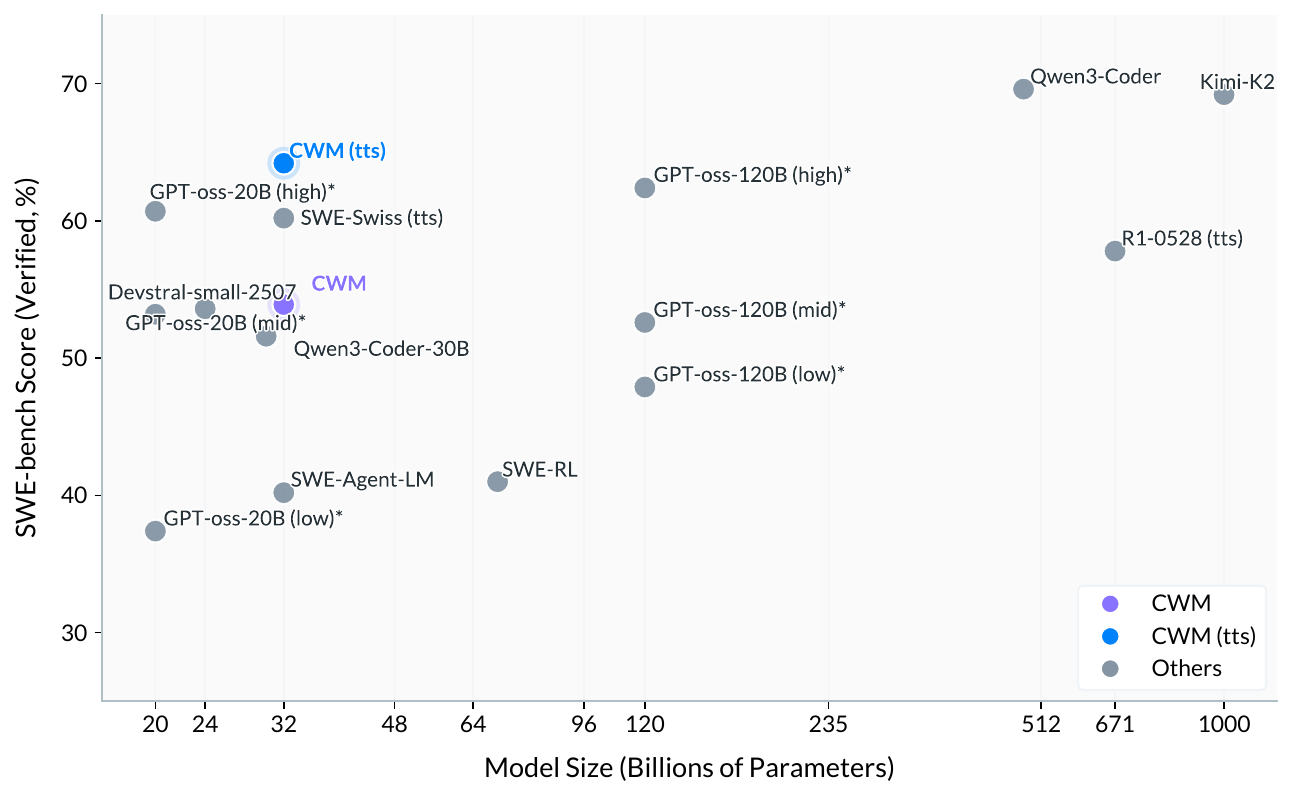}
    \caption{%
    \sbv pass@1 scores. \cwm achieves best-in-class performance with and without test-time-scaling (tts), achieving \SI{65.8}{\percent} and \SI{53.9}{\percent} respectively. Note that GPT-oss scores are computed with respect to a limited subset of \num{477} out of \num{500} problems.    
    }
    \label{fig:sbv_scatter}
\end{figure}

\newpara{\sbv.} \Cref{fig:sbv,fig:sbv_scatter} show results for \sbv. \cwm achieves pass@1 resolve rates of \SI{65.8}{\percent} with test-time-scaling and \SI{53.9}{\percent} without test-time scaling (averaged over 4 runs). With test-time scaling, \cwm outperforms open-weight models at similar size and is competitive to larger and proprietary models. The base score without test-time scaling also surpasses open-weight models with similar parameter counts and remains respectable even when comparing to much larger models such as GPT-oss-120B~\citep{agarwal2025gpt}, Qwen3-Coder~\citep{yang2025qwen3}, and Kimi K2~\citep{team2025kimi}.

For Test-Time-Scaling (TTS) on \sbv, we first generate $k$ candidate solutions as well as \num{40} \emph{novel} unit tests in parallel agentic loops for each instance.
Like Agentless~\citep{agentless}, we ask the model to generate tests that verify patch correctness \emph{and} reproduce the original bug, enabling us to filter out tests that fail to reproduce errors. Following SWE-RL~\citep{wei2025swerl}, we keep the top-5 majority tests for each instance.
Since candidate solutions are often similar in the number of \emph{existing} tests they pass, we prioritize the strongest candidates by keeping only those patches that pass the highest number of existing tests. We then execute the remaining patches on the filtered set of novel tests and select the patch with the highest pass rate for submission. In case of ties, we prioritize the majority patch, and if the tie remains, we choose the patch whose trajectory has fewer tokens. We refer to this approach as best@$k$.

\begin{figure}[p]
 \centering
  \begin{subfigure}[b]{0.48\textwidth}
    \includegraphics[width=\linewidth]{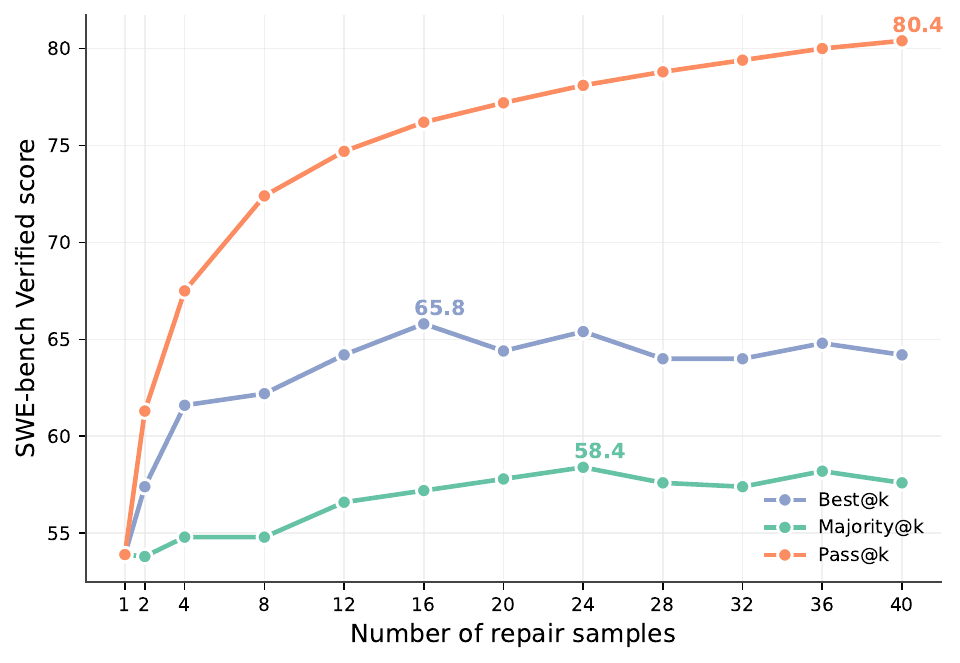}
    \caption{\label{fig:sbv_tts}}
  \end{subfigure}
  \hfill
    \begin{subfigure}[b]{0.48\textwidth}
        \centering
        \includegraphics[width=0.95\linewidth]{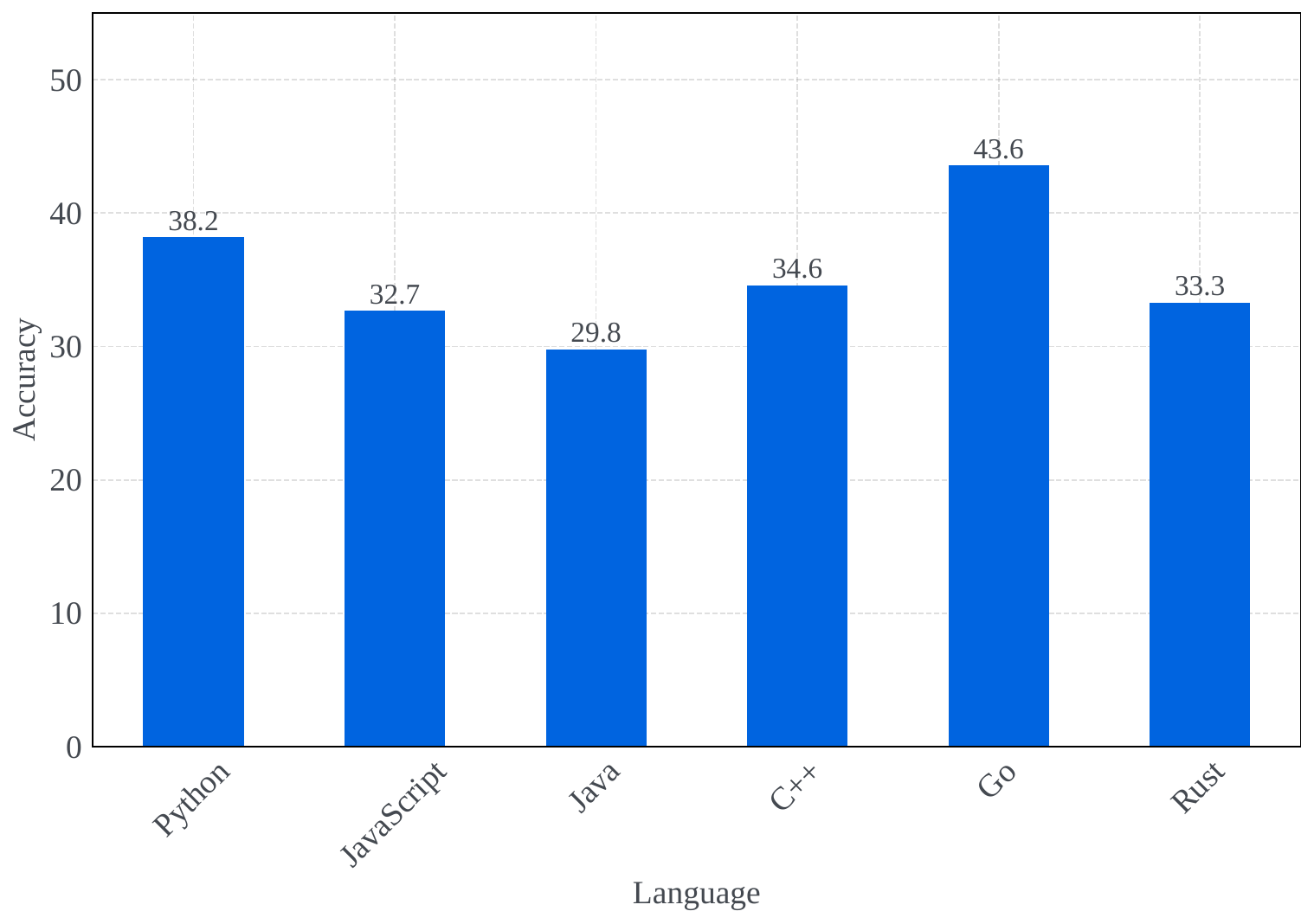}
        \caption{\label{fig:aider-lang}}        
    \end{subfigure}
  \caption{(a) Test time scaling (TTS) with both our best@$k$ method majority voting can significantly increase pass@$1$ rates for \cwm  on \sbv. (b) Accuracy of \cwm on Aider Polyglot by programming language using the whole file edit format.}
  \label{fig:tts}
\end{figure}

In \cref{fig:sbv_scatter}, we report results for best@$k$ for $k=16$, which achieves a \SI{65.8}{\percent} resolve rate. As a simple alternative to best@$k$, we found that majority voting~\citep{self_cons} of candidate patches, based on exact string matching and without any test generation or execution, leads to a pass rate of \SI{58.4}{\percent}. In \Cref{fig:sbv_tts}, we report best@$k$ and pass@$k$ across different values of $k$. 
As expected, pass@k improves monotonically with larger $k$, ultimately reaching a success rate of \SI{80.4}{\percent} at $k=40$. For best@$k$, performance improves sharply from $k=2$ before plateauing around $k=16$. For majority-voting, performance improves gradually from $k=2$ and plateaus at $k=24$.

\newpara{Alternative harnesses for \sbv.} To better understand the robustness of \cwm to the choice of evaluation harness and tool-calling implementations, we perform experiments with third-party approaches, namely Mini-SWE-Agent~\citep{yang2024swe} and OpenHands~\citep{wang2025openhands}. For both, we shortened and adapt the system prompt to better align with the \swerl prompt (see \cref{fig:agentic-coding-prompt}) and make sure to keep reasoning output as part of the message history. We configure both harnesses to use OpenAI function calling, which sends messages along with structured tool descriptions. When prompting the model, we format and append the available tools to the system prompt. When the model decides to call a tool, the call is parsed and returned in a \texttt{tool\textunderscore{}calls} field in our response. This makes sure that tools are rendered with a syntax template suitable for prompting \cwm.
For Mini-SWE-Agent, we follow the official budget of \num{250} turns. For OpenHands, we report results for \num{40}, \num{128}, and \num{500} turns. 
Additionally, we report results for our harness when limiting tool use to bash commands only. As \cref{tab:swe-bench-results-harness} shows, although resolve rates degrade when using different agents, tool implementations, or limiting tool choices, \cwm provides robust and reasonable performance across all setups.

\begin{table}[t]
    \centering
    \begin{minipage}[t]{0.45\textwidth}       
    \centering
    \caption{\sbv resolve rates for alternative agentic harnesses are lower than the \SI{53.9}{\percent} pass@1 achieved with our approach, but performance remains reasonable across the board.}
    \label{tab:swe-bench-results-harness}
    \resizebox{\textwidth}{!}{\begin{tabular}{ll|r}
        \toprule
        \textbf{Harness} & \textbf{Configuration} & \textbf{Resolve Rate (\%)} \\
        \midrule
        Mini-SWE-Agent & 250 turns & 37.6 \\  %
        OpenHands & 40 turns & 36.0\\  %
         & 128 turns & 42.6 \\ %
         & 500 turns & 40.8 \\ %
         \midrule \\
         \textbf{Ours} (bash-only) & 128 turns & 42.1 \\
         \textbf{Ours} & 128 turns & \textbf{53.9}\\
        \bottomrule
    \end{tabular}}
    \end{minipage}
    \hfill        
    \begin{minipage}[t]{0.51\textwidth}
        \centering
        \captionof{table}{Results on Aider Polyglot for \cwm and baselines from the official leaderboard.\label{tab:aider-all}}
        \resizebox{\textwidth}{!}{
        \begin{tabular}{lll}
            \toprule
            \textbf{Model} & \textbf{Format} & \textbf{Pass 1@2 (\%)} \\
            \midrule
            o3-pro (high) & Diff & \textbf{84.9} \\
            DeepSeek R1 (0528) & Diff & 71.4 \\
            Qwen3 235B A22B diff, no think & Diff & 59.6 \\
            Kimi K2 & Diff & 59.1 \\
            gpt-oss-120b (high) & Diff & 41.8 \\
            Qwen3-32B & Diff & 40.0 \\
            Gemini 2.0 Pro exp-02-05 & Whole File & 35.6 \\
            \underline{CWM} & Whole File & 35.1 \\
            Grok 3 Mini Beta (low) & Whole File & 34.7 \\
            o1-mini-2024-09-12 & Whole File & 32.9 \\
            gpt-4.1-mini & Diff & 27.1 \\
            Codestral 25.01 & Whole File & 11.1 \\
            \bottomrule
        \end{tabular}}        
    \end{minipage}
\end{table}

\looseness=-1
\newpara{Multi-lingual coding.} The Aider Polyglot benchmark~\citep{aider} measures coding ability across a diverse set of programming languages using challenging exercises from Exercism.\footnote{See \url{https://exercism.org/}.} The primary metric is the pass rate on the second attempt, allowing the model to iterate on test failures once. We make a few changes to the harness to align it with the \cwm training distribution, such as removing hard-coded assistant messages from the history, concatenating adjacent messages of the same role, removing examples from the system prompt, turning off auto-linting and stripping of reasoning traces, and reiterating in the prompt that exact matches are needed. We evaluate with reasoning, at temperature \num{0.4}, and without test-time-scaling. Although Aider Polyglot may not fully qualify as a truly agentic benchmark -- given its lack of dynamic tool use beyond code execution and limited interaction -- we include it here in light of the self-correction capabilities that it allows for.

\looseness=-1
As shown in \Cref{tab:aider-all}, CWM achieves \SI{35.1}{\percent} accuracy, comparable to other models in its class such as Qwen3-32B (\SI{40.0}{\percent}) and other models using the ``whole file'' edit format such as Gemini 2.0 Pro (\SI{35.6}{\percent}).
We also observe good generalization performance across the six languages tested in the benchmark, as shown in \Cref{fig:aider-lang}. Many top-performing models, such as o3-pro (\SI{84.9}{\percent})~\citep{o3}, DeepSeek R1 (\SI{71.4}{\percent}), and Qwen3 235B (\SI{59.6}{\percent}), achieve substantially higher scores using the ``diff'' edit format. However, \cwm was not optimized for this format and does not reach competitive performance with it.

\newpara{Terminal-Bench.} Another multi-turn agentic coding benchmark that is gaining in popularity is Terminal-Bench~\citep{tbench_2025}. In Terminal-Bench, the agent is asked to solve various complex tasks by operating directly in a tmux session.
Again, we align the prompts and response parsing of the Terminus-1\footnote{See \url{https://www.tbench.ai/terminus}.} agent provided by the benchmark with our RL training phase: we modify the system prompt to use the tools that \cwm was trained with (see \Cref{fig:swerl:example}) and parse the model output back into the format that Terminus-1 expects. We also include reasoning tokens from prior turns into the agent's history.
In this setup, \cwm achieves a \SI{26.25}{\percent} accuracy with the Terminus 1 agent following the default budget of \num{50} turns. \Cref{tab:t-bench} shows this places \cwm below o4-mini but above Gemini 2.5 Pro on the Terminal-Bench leaderboard.
\begin{table}[t]
    \centering
    \begin{minipage}[t]{0.48\textwidth}
        \centering
        \captionof{table}{Results on Terminal-Bench for \cwm and baselines from the official leaderboard.\label{tab:t-bench}}
        \resizebox{\textwidth}{!}{
        \begin{tabular}{lll}
            \toprule
            \textbf{Model} & \textbf{Agent} &  \textbf{Accuracy (\%)} \\
            \midrule
            OpenAI-Multiple & OB-1 & \textbf{59.0} \\
            GPT-5 & OB-1 &  49.0 \\
            GPT-5 & Terminus 1 & 30.0 \\
            o4-mini & Goose  & 27.5 \\
            \underline{\cwm} & Terminus 1 & 26.3 \\
            Gemini 2.5 Pro & Terminus 1 & 25.3 \\
            o4-mini & Terminus 1 & 18.5 \\
            Grok 3 Beta & Terminus 1 & 17.5 \\
            Gemini 2.5 Flash & Terminus 1 & 16.8 \\
            Qwen3-32B & TerminalAgent & 15.5 \\
            \bottomrule
        \end{tabular}}
    \end{minipage}
    \hfill        
    \begin{minipage}[t]{0.48\textwidth}
        \centering
        \captionof{table}{Execution trace prediction is competitive with reasoning for CruxEval-output pass@1 scores. For \cwm, we use temperature \num{0.6}, top-p \num{0.95}, and \num{10} generations, while for \cwm SFT we use greedy decoding. \label{tab:cruxeval_outputpred}}
        \resizebox{\textwidth}{!}{
\begin{tabular}{rl|rr}
\toprule
 \textbf{Budget} & \textbf{Mode} & \textbf{CWM SFT} & \textbf{CWM} \\
\midrule
\multirow{2}{*}{small} & Language w/o CoT & 67.8 & 66.6  \\
& Trace Step& 59.1 & 58.1 \\
\midrule
\multirow{2}{*}{large} & Language w/ CoT & 83.3 & \textbf{94.3} \\
& Trace Full& \textbf{87.3} & 87.7 \\
\bottomrule
\end{tabular}%
}
    \end{minipage}
\end{table}

\subsection{Execution trace prediction}
\label{sec:ex_tr_pred}
Next, we analyze the ability of \cwm to perform trace prediction, analyze its prediction, and explore this ability to predict program termination. 

\begin{table}
\centering
\end{table}

\newpara{CruxEval-O as execution trace prediction.} The following experiment evaluates \cwm's ability to predict Python execution traces using the format introduced in \Cref{sec:tracing}. We prompt the model with functions and input arguments from the CruxEval test set, ask it to predict the function execution trace line-by-line, and then compare its output prediction to the ground truth. To elicit trace prediction, we construct prompts following our custom trace format, with the input containing the function as the code context, the call arguments as the state, and the line containing the function definition as the first action.
In addition to this ``full'' execution trace prediction scenario, we also study a single-``step'' scenario, for which we ask the model to directly predict the return value of the function. This is achieved by replacing the \texttt{<|line\_sep|>} token with \texttt{<|return\_sep|>}.
We illustrate both formats in \Cref{fig:trace_prompts}. We compare the ``step'' scenario to classic CruxEval-output prediction, which few-shot prompts the model to directly predict outputs given function definitions and inputs. 
Consequently, we compare the ``full'' trace prediction mode to CruxEval-output with reasoning, which allows \cwm to use reasoning as introduced in \cref{sec:rl_env_and_data} before predicting the function output.

Our results in \Cref{tab:cruxeval_outputpred} show that large compute budgets, either allowing for execution trace prediction or reasoning, produce better results.
\cwm achieves a best score of \SI{94.0}{\percent} in natural language reasoning mode, while full trace prediction achieves \SI{88}{\percent}. Note that language reasoning traces are significantly more verbose, using \num{1164} tokens on average compared to \num{497} tokens for full trace prediction. We also report results for \cwm after SFT, which achieves its best result of \SI{87.3}{\percent} using full execution trace prediction.
Single-step trace prediction is not competitive with classic few-shot prompting for either \cwm model.

\newpara{Execution trace prediction analysis.} Follow previous paragraph, we present a detailed evaluation of the quality of the execution traces predicted by \cwm for validation sets of CruxEval and our function-level data. Concretely, we measure the fraction of generated traces that follow our format (Valid Trace Format) and the observation (action) exact match accuracy (Observation (Action) Exact Match), which measures the number of observations (actions) exactly matching ground truth relative to the total number of observations (actions) per execution trace. Our trace format specifies the state as a JSON dump of a dictionary containing the local variables.
We report the fraction of state predictions matching this format (Valid JSON Format). 
Additionally, Key (+Value) Match measures the average fraction of matching keys (and values) per state~prediction.

The results in \Cref{tab:trace_state_action_analysis} show that \cwm adheres to the correct trace and observation format for all data sources, achieving more than \SI{99}{\percent} format matching across the board. \cwm is able to accurately predict the execution trace as well as intermediate observations and actions, which is reflected in scores larger than \SI{96}{\percent} for Observation/Action Exact Match and larger than \SI{97}{\percent} in Key (+Value) Match.

\begin{table}[t]
\caption{Detailed analysis of execution trace prediction with \cwm and greedy decoding. We present a breakdown of the accuracy of the individual components of trace prediction for validation set inputs from CruxEval as well as our function-level data.
The CruxEval pass@$1$ score here differs from the one in \Cref{tab:cruxeval_outputpred} (\SI{87.7}{\percent}) due to greedy decoding.
Overall, we find solid accuracy across state and action prediction. \label{tab:trace_state_action_analysis}
} 
\centering
\begin{tabular}{llrr}
\toprule
 &  & CruxEval & Function-level \\
\midrule
Output & pass@1 & 88.0 & 94.4 \\
\midrule
\multirow[t]{3}{*}{Trace} & Valid Trace Format & 99.6 & 100.0 \\
 & State Exact Match & 96.9 & 96.4 \\
 & Action Exact Match & 96.5 & 98.0 \\
\midrule
\multirow[t]{3}{*}{States} & Valid JSON Format & 100.0 & 100.0 \\
 & Key Match & 99.1 & 99.0 \\
 & Key+Value Match & 98.1 & 97.9 \\
\midrule
\multirow[t]{2}{*}{Statistics} & Avg State Length (Token) & 11.7 & 18.8 \\
 & Avg Action Length (Token) & 11.2 & 10.0 \\
\bottomrule
\end{tabular}
\end{table}

\subsection{Program termination prediction}
The question of
whether a program terminates is a reasoning problem which goes beyond what can be shown by considering individual finite traces as in \cwm training: non-termination cannot be observed by executing a trace in finite time, and termination on all inputs cannot be feasibly observed by enumerating traces. \Cref{lst:trace_cwm_halteval_correct} in the Appendix illustrates termination reasoning, whereby \cwm considers several concrete inputs before generalizing to the conclusion of terminating on all inputs.

We propose HaltEval-prelim, a novel benchmark obtained by automatically translating C programs with termination annotations into Python using LLaMA-3-70B via few-shot prompting. The C programs are sourced from the International Competition on Software Verification (SVCOMP) and the Termination Problems Database (TPDB).\footnote{See \url{https://sv-comp.sosy-lab.org/} and \url{https://termination-portal.org/wiki/TPDB}.} 
Each original problem comes with termination/non-termination annotations, which we manually verify are preserved during the Python translation phase and otherwise discard. 
We obtained a balanced dataset consisting of \num{115} terminating (\texttt{T}) and \num{115} non-terminating (\texttt{NT}) Python programs.

We query LLMs to judge whether a program terminates (answer \texttt{\#T}) or diverges (answer $f(n)$ where n leads to divergence, followed by the comment \texttt{\#NT}). 
We reward a divergence claim if $f(n)$ times out after 5 seconds.  
If, however, the model predicts \texttt{\#NT}, and the ground truth is \texttt{\#T}, it is not rewarded, even if execution exceeds the timeout. For instance, if $f$'s ground truth is \texttt{\#T}  and $f(42)$ runs for 7.5 million years and then terminates, our scoring (pass@1) will not reward a \texttt{\#NT} claim for $f(42)$ even though it trips timeout.
Our use of timeout as a proxy for divergence is similar to  \cite{DBLP:conf/sigsoft/AlonD22} in judging correct non-termination claims, but different in that timeout is not used as a ground truth for termination claims. This results in an eval that is conservative in the sense that it awards scores that could be higher than that given by a perfect oracle, but never lower.
It would be worth exploring replacing the ground truths and input validation by logical proofs of termination and non-termination \citep{DBLP:journals/cacm/CookPR11,DBLP:conf/popl/GuptaHMRX08}.

\Cref{tab:HaltEval-prelim} reports results for \cwmv, Qwen3-32B, and Llama3-70B with direct prediction, prompted chain-of-thought (CoT), and reasoning (for \cwmv and Qwen3-32B only). ``Reasoning'' here means use of the \texttt{<think> ... </think>} format from RL. We report CoT prompting numbers to represent an attempt to approximate reasoning that is compatible with Llama3-70B.
As a reference, we also provide the scores of a constant classifier tagging all programs as terminating, which would obtain a pass@ of \num{0.5}.
When comparing \cwm and Qwen3, results suggest Qwen3 reaches better direct and CoT performance, however under the reasoning setup, both models significantly improved, reaching comparable performance of $\sim$\num{0.94} pass@$1$.

\begin{table}[t]
\centering
\caption{HaltEval-prelim pass@1 
scores for different LLMs in different prompting settings. For reasoning we use temperature \num{0.6}, top-p of \num{0.95}, and \num{10} generations, while for direct and CoT predictions we use greedy decoding. 
\label{tab:HaltEval-prelim}}
\begin{tabular}{l l  c ccc ccc ccc}
\toprule
 & & \multicolumn{1}{c}{\textbf{Constant}} &\multicolumn{3}{c}{\textbf{\cwm}} & \multicolumn{3}{c}{\textbf{Qwen3-32B}} & \multicolumn{3}{c}{\textbf{Llama-3-70B}} \\
\cmidrule(lr){3-3} 
\cmidrule(lr){4-6} \cmidrule(lr){7-9} \cmidrule(lr){10-11}
 & & T & Direct & CoT & Reasoning & Direct & CoT & Reasoning & Direct & CoT  \\
\midrule
\multirow{1}{*}{pass@1} 
  &  
&  0.5      & 0.37  & 0.55 & {\bf 0.94} &  0.49 & 0.68  &  {\bf 0.94} &  0.43 & 0.48  \\
\bottomrule
\end{tabular}
\end{table}

\looseness=-1
We initially designed HaltEval-prelim under the assumption that termination would be difficult to assess, given its undecidability. The strong results achieved by both \cwm and Qwen3-32B with reasoning were therefore unexpected. Still, these findings should be interpreted cautiously: the benchmark is based on small, self-contained programs and does not reflect the challenges of real-world software, where bugs must be detected in large and complex codebases. Hence, success on this preliminary dataset may not translate directly to practice. Moreover, termination in real systems is highly imbalanced -- typically with hundreds or thousands of terminating loops for every non-terminating one -- unlike the balanced distribution in our dataset \citep{vanegue2025nonterminationproving100million}.

\subsection{Algorithmic complexity prediction} 

We evaluate \cwm on two tasks from \textsc{BigO(Bench)} \citep{chambon2025bigobenchllmsgenerate}: complexity prediction, determining the Big-O time/space complexity of existing code, and complexity generation, solving coding problems while adhering to specified complexity constraints. We report all@$1$ scores, which require correct LLM output simultaneously across all possible complexity classes for a given problem. For complexity generation, we also report the pass@$1$ score with and without the complexity requirement (the solution still needs to be correct), and a best@$1$ score that corresponds to pass$@1$ on the lowest complexity class of each problem, dismissing suboptimal classes.

Results for \cwm, Qwen3-32B, Qwen3-coder-30B, and Gemma-3-27B are all presented in \Cref{tab:bigobench_extended}. To ensure the comparison with external models remains as fair as possible, we choose to re-evaluate them alongside \cwm in the same evaluation setting. 
For both tasks, we use \textsc{BigO(Bench)}'s official setup, after performing a prompt ablation that did not seem to further boost performance.
On time complexity prediction, \cwm achieves the best all@$1$ score of all compared models but fares worse for space complexity. In particular, looking at the official benchmark leaderboard,\footnote{See \url{https://facebookresearch.github.io/BigOBench/leaderboard.html} at the time of writing.} \cwm ranks second overall on time complexity prediction (all@$1$) across all reported models of all sizes.
For time complexity generation, \cwm achieves the best overall pass@$1$, best@$1$, and all@$1$ scores for our set of models, and also ranks second in general looking at the official benchmark scores.
For space complexity generation, our model ranks first for pass@$1$ on code only, and second behind Qwen3-32B in terms of the remaining metrics. We note that \cwm stands out in particular in time complexity reasoning, systematically outperforming other models across all metrics on both prediction and generation variants. Moreover, when complexity requirements are set aside, the model's performance on code-only pass@$1$ degrades far less than for other models, indicating \cwm is able to maintain focus on fundamental task requirements while effectively handling additional constraints.

\begin{table}[t]
\caption{\textsc{BigOBench} results comparing \cwm against Qwen3-32B (with reasoning), Qwen3-coder-30B, and Gemma-3-27B on complexity prediction and complexity generation, for both time and space complexity.
\cwm outperforms our set of baseline models for all metrics on time complexity prediction and generation. For space complexity generation, \cwm performs best on code-only pass@$1$ and ranks second on the remaining metrics.
We refer to the main text for details on the task and metrics.}
\label{tab:bigobench_extended}
\centering
\begin{tabular}{l|rrrrrr}
\toprule
& \textbf{CWM} & \textbf{Qwen3-32B} & \textbf{Qwen3-coder-30B} & \textbf{Gemma-3-27B} \\
\midrule
\textbf{Prediction} & & & & \\
~~ \textbf{Time Complexity} - all@$1$ & \textbf{41.3} & \underline{39.0} & 36.6 & 37.7 \\
~~ \textbf{Space Complexity} - all@$1$ & 12.3 & \textbf{15.1} & 9.1 & \underline{13.1} \\
\textbf{Generation} & & & & \\
~~ \textbf{Time Complexity} & & & & \\
~~~~ Code Only - pass@$1$ & \textbf{76.1} & \underline{70.0} & 43.8 & 34.4 \\
~~~~ Code \& Complexity - pass@$1$ & \textbf{31.3} & \underline{29.1} & 20.3 & 13.3 \\
~~~~ Code \& Complexity - best@$1$ & \textbf{48.6} & \underline{43.5} & 27.2 & 15.2 \\
~~~~ Code \& Complexity - all@$1$ & \textbf{7.6} & \underline{6.5} & 5.5 & 2.1 \\
~~ \textbf{Space Complexity} & & & & \\
~~~~ Code Only - pass@$1$ & \textbf{73.2} & \underline{65.9} & 45.1 & 36.4 \\
~~~~ Code \& Complexity - pass@$1$ & \underline{24.1} & \textbf{25.5} & 17.7 & 14.6 \\
~~~~ Code \& Complexity - best@$1$ & \underline{36.6} & \textbf{39.6} & 26.3 & 20.6 \\
~~~~ Code \& Complexity - all@$1$ & \underline{3.2} & \textbf{5.1} & 2.4 & 1.5 \\
\bottomrule
\end{tabular}
\end{table}

\subsection{Code and mathematical reasoning} 
We present results on LiveCodeBench~(LCB, \citet{DBLP:conf/iclr/JainHGLYZWSSS25}), concretely the LCBv5 and LCBv6 date ranges \texttt{01.10.2024-01.02.2025}\footnote{For LCBv5, we report results starting from October to be consistent with the numbers reported by Qwen3.}  and \texttt{01.08.2024-01.05.2025}, in \Cref{tab:rl_tasks}.\footnote{Results for Magistral were taken from the official reported numbers for 1.2 version, where no explicit dates were mentioned.}
We here compare \cwm to relevant baseline models with similar parameter counts.
Again, we observe highly competitive performance on par with Magistral-small-1.2~\citep{rastogi2025magistral}, Qwen3-32B, and gpt-oss-20B~\citep{agarwal2025gpt}.

\Cref{tab:rl_tasks} also contains pass@1 results for \cwm on Math-500~\citep{lightman2023let}, AIME24 \citep{aime}, and AIME25 -- all averaged over $n=20$ samples. \cwm performs slightly worse across the board, with notable gap compared to gpt-oss-20B (high) on AIME.

\looseness=-1
In \cref{fig:math_tts}, we additionally report test-time scaling results using majority voting and short-3@k for \cwm on AIME.
Short-m@k \citep{dont_over_think} begins sampling $k$ answers in parallel but stops sampling once the first $m$ generations are complete, and then selects the most common answer among the three.
\cwm performance on AIME24 increases by up to \SI{11}{\percent} at $k=10$ with majority voting. 
Short-3@k achieves performance comparable to majority voting, while significantly reducing computational cost for a given $k$.

\begin{table}[t]
\caption{Agentic, code, and mathematical reasoning benchmarks. We compare \cwm to baselines with roughly the same number of parameters.($\dagger$: LCB results for gpt-oss-20B (high) suffered from repeated time-outs due to repetitive reasoning, despite our prompt-tuning efforts -- which boosted gpt-oss (low/medium) scores by about \SI{10}{\percent}.)
\label{tab:rl_tasks}}
\centering
\resizebox{0.87\textwidth}{!}{
\begin{tabular}{l|cccc}
\toprule
 & \textbf{Magistral-small-1.2-24B}   & \textbf{Qwen3 -32B} & \textbf{gpt-oss-20B (low / med / high)} & \textbf{\cwm}\\
\midrule
LCBv5     & \textbf{70.0}   & 65.7  & 54.2 / 66.9 / \phantom{0}--$^\dagger$\phantom{.0}   & \underline{68.6}  \\
LCBv6     & 61.6   & 61.9     & 47.3  / \underline{62.0}  / \phantom{0}--$^\dagger$\phantom{.0} & \textbf{63.5}  \\
Math-500  & -  & \textbf{97.2}      & --                   & \underline{96.6}  \\
AIME24    & \underline{86.1}  & 81.4      & 42.1  / 80.0  / \textbf{92.1}\phantom{*} & 76.0 \\
AIME25    & \underline{77.3}  & 72.9      & 37.1  / 72.1  / \textbf{91.7}\phantom{*} & 68.2\\
\bottomrule
\end{tabular}}
\end{table}

\subsection{Non-reasoning evaluations} 
Although our main focus with \cwm is code world modeling, we also provide evaluation results of \cwm on a set of standard tasks covering code, math, and general knowledge without reasoning mode enabled. We here compare to models with similar parameter counts, such as Qwen3-32B or Gemma-3-27B, as baselines, and we use greedy generation instead of sampling at non-zero temperature.
The results in \Cref{tab:pt_tasks} show that \cwm typically performs better than Gemma-3-27B, similar to Qwen2.5-32B, but worse than Qwen3-32B. An interesting exception to this is CruxEval-O \citep{gu2024cruxeval}, where the introduction of the tracing data (see \Cref{sec:tracing}) likely helps \cwm gain an advantage. Note that we achieve even better results on CruxEval-Output when using reasoning (see \cref{sec:ex_tr_pred}).

Next, we consider two long-context evaluation benchmarks: LoCoDiff~\citep{locodiff} and RULER \citep{hsieh2024ruler}. In LoCoDiff, models are provided with the commit history of a specific file and asked to construct its final version. To succeed, models must follow the files' evolution -- from the initial commit, through diffs on multiple branches, to the resolution of merge conflicts. Performance is evaluated by the proportion of files for which the model reproduces the target version exactly. We compare \cwmv to DeepSeek-R1 0528, Claude Sonnet 4~\citep{claude}, Gemini 2.5 Pro 06-05~\citep{comanici2025gemini}, Kimi K2, GPT-5, and gpt-oss-120B. Qwen3-32B is not on the leaderboard and has a shorter native context length.

\begin{figure}[t]
    \centering
    \captionsetup[subfigure]{justification=justified}
     \begin{subfigure}[b]{0.48\textwidth}
        \includegraphics[width=\linewidth]{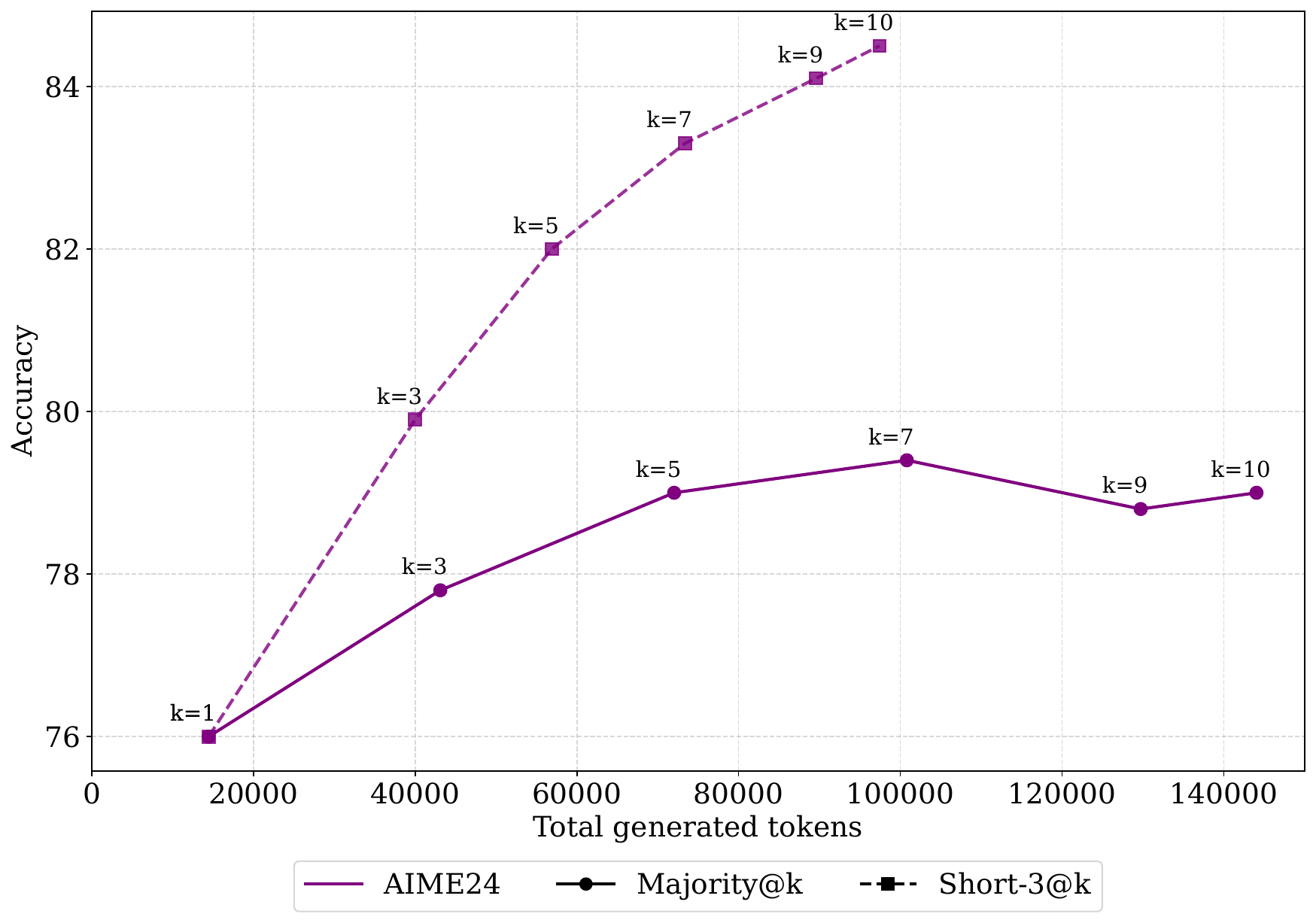}
        \caption{\label{fig:math_tts}}
      \end{subfigure}
    \hfill
    \begin{subfigure}[b]{0.48\textwidth}
        \centering
        \includegraphics[width=\linewidth]{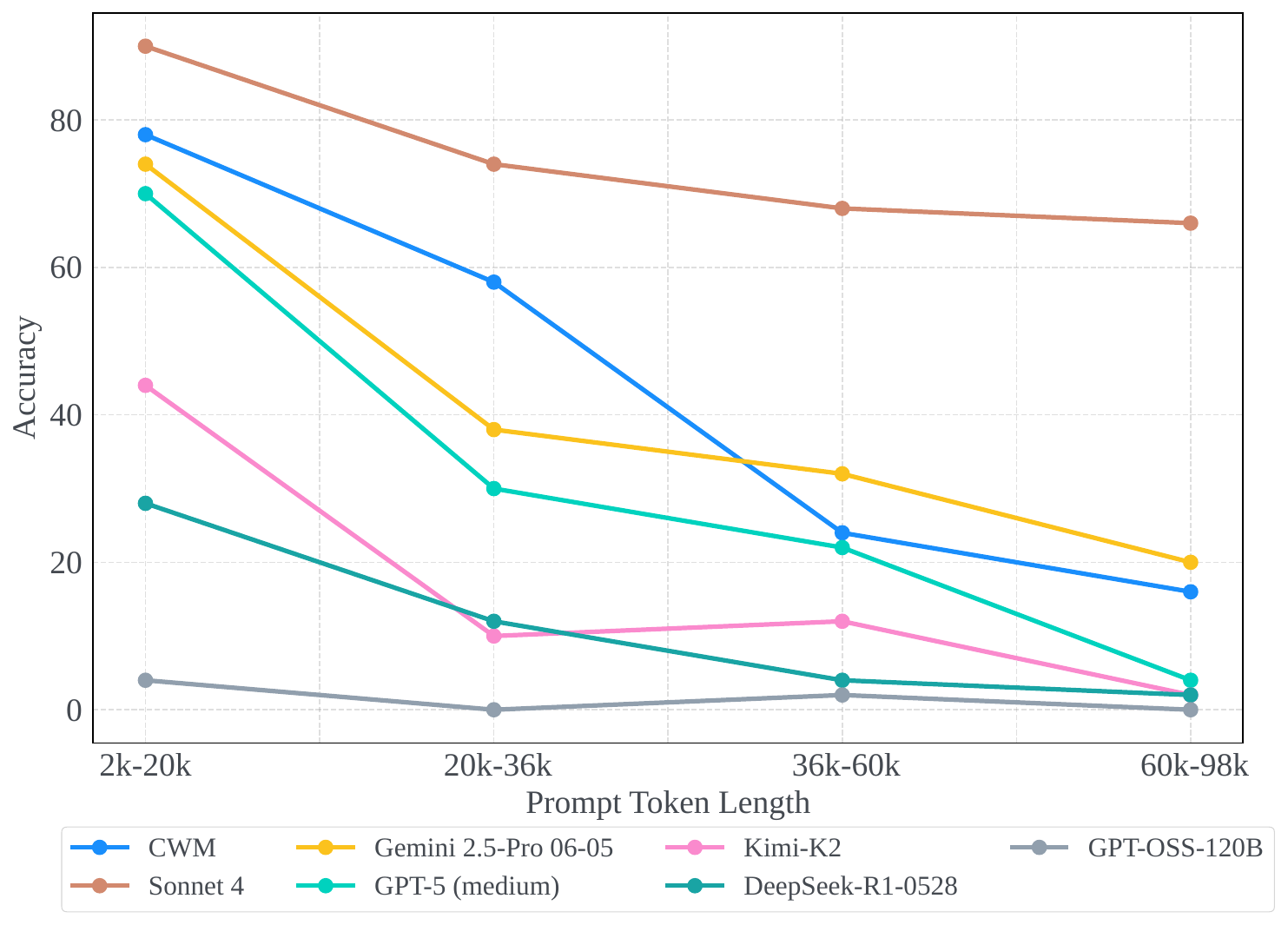}
        \caption{\label{fig:locodiff}}        
    \end{subfigure}
    \caption{(a) Test-time scaling on AIME24 with majority voting and short-3@k. See main text for details.(b) LoCoDiff results for \cwmv and baselines considering different sequence lengths buckets. }
\end{figure}

For LoCoDiff, the results in \Cref{fig:locodiff} show that, while all models suffer a degradation in performance as the sequence length increases, \cwmv provides better performance than DeepSeek-R1 0528 and gpt-oss-120B and is competitive with large scale commercial models (e.g., GPT-5 and Gemini 2.5 Pro) on both short and long sequences, with a significant gap to Claude Sonnet 4 only. We present results for RULER in \cref{sec:ruler}.

\begin{table}[t]
\caption{Performance of \cwm and $\cwm_\textup{\tiny{Mid}}$, (\cwm after mid-training), on a set of general, math, and coding tasks without any reasoning compared to a set of recent baseline models with similar parameter counts. \label{tab:pt_tasks}}
\centering
\resizebox{\textwidth}{!}{
\begin{tabular}{l|rrrrrrr}
\toprule
 & \textbf{\cwm} & \textbf{\cwm}$_\textup{\tiny{Mid}}$ &\textbf{Qwen3-32B} & \textbf{Qwen2.5-32B} & \textbf{Gemma-3-27B} & \textbf{Llama-3-70B} & \textbf{Llama-4-Scout} \\
\midrule
MMLU & 77.7 & 73.6 & \textbf{83.6} &\underline{83.3} & 78.7 & 79.3 & 78.3\\
MMLU-Pro   & \underline{60.2}  & 52.3 & \textbf{65.5} & 55.1  & 52.9 & 53.8 & 56.1\\
GPQA      & 40.6 & 31.7 & \textbf{49.5}  & \underline{48.0} & 26.3 & – & 40.4 \\
GSM8k     & \underline{93.3} & 84.7 & \textbf{93.4} & 92.9 & 81.2 & 83.7 & 85.4 \\
HumanEval-Plus   & \textbf{75.0} & 68.3 & \underline{72.1} & 66.3 & 55.8 & – & 59.9 \\
MBPP       & 73.4 & 67.8 & \textbf{78.2} & \underline{73.6} & 68.4 & 66.2 & 68.6 \\
CRUX-O     & \textbf{83.4} & \underline{78.9} & 72.5 & 67.8 & 60.0 & – & 61.9 \\
\bottomrule
\end{tabular}}
\end{table}

\FloatBarrier

\section{Transparency, Risks \& Limitations}
\label{sec:discussion}

\subsection{Transparency on external models and data}
\label{sec:ext_models}
As mentioned previously in the relevant sections, we use data from external LLMs in four contexts: (i) ForagerAgent, (ii) trace-to-natural language conversion, (iii) function tracing, and (iv) the SFT phase. For the ForagerAgent, we employ Llama3-70B-Instruct~\citep{dubey2024llama} and Qwen3-235B-A22B (without thinking)~\citep{yang2025qwen3} as base models to interact with the computational environment. For converting raw Python traces into natural language, we use Qwen3-32B-FP8 (without thinking)~\citep{yang2025qwen3}. For function tracing, we use Llama3-70B-Instruct to generate Python function inputs and to generate solutions for CodeContests data. Finally, during SFT, we incorporate trajectories from DeepSeek-R1~\citep{guo2025deepseek} through the OpenMathReasoning~\citep{omr} and OpenCodeReasoning~\citep{ocr} datasets. We used mitigated versions of the OpenCodeReasoning and OpenMathReasoning datasets, where mitigations included algorithmic bias filtering and cybersecurity protections. We applied similar mitigations when using Qwen3-32B-FP8 to generate data for training. No external LLM tokens were used beyond those explicitly mentioned in these four contexts.

\subsection{Code World Model Preparedness Report}
\label{sec:preparedness_report}
Despite its relatively small size of \SI{32}{B} parameters, \cwm outperforms open-weight models at similar size and is competitive to larger and proprietary models on verified software engineering benchmarks. To anticipate risks from this release, including potentially novel risks, we conducted an automated assessment of CWM capabilities relevant to the domains identified in our Frontier AI Framework\footnote{\label{note1}See \url{https://ai.meta.com/static-resource/meta-frontier-ai-framework}.} that could present potentially catastrophic risks, namely Cyber and Chemical \& Biological risks. As part of ongoing work to improve the robustness of our evaluations and the reliability of our models, we also include a preliminary propensity evaluation, with plans to expand this area in future assessments.

\looseness=-1
We performed this assessment by testing the relative performance of \cwm against a set popular and capable open-source models that represent a baseline of capabilities available in the open ecosystem: Qwen3-Coder-480B-A35B-Instruct~\citep{yang2025qwen3}, Llama 4 Maverick~\citep{meta_llama_llama4_2025}, and gpt-oss-120B~\citep{agarwal2025gptoss}.

Based on the results of these assessments, we believe that the open-source release of \cwm is unlikely to meaningfully increase risks related to Cybersecurity or Chemical \& Biological threats beyond the current ecosystem baseline. 
Additionally, our preliminary evaluations suggest that \cwm shows undesirable propensities at rates comparable to most open-source models though some models achieve substantially lower rates, i.e., gpt-oss-120B.

These results indicate that \cwm is within the ``moderate'' risk threshold for the catastrophic domains defined in Meta’s Frontier AI Framework.\textsuperscript{\ref{note1}} We share the details in the Code World Model Preparedness Report.\footnote{Code World Model Preparedness Report, available at \url{https://ai.meta.com/research/publications/cwm-preparedness}.}

\subsection{Limitations \& future research}
\label{sec:limitations}
We explicitly release CWM as a research model under a noncommercial research license for the community to explore the opportunities afforded by world modeling and reasoning in computational environments. As such, our models come with a number of limitations which we outline below to help the research community make the most of \cwm, while being aware of its shortcomings and avoiding accidental misuse.

As these are research-only models, they are not suitable for  production use cases. Although we have performed some limited evaluations, we have not conducted a full range of possible evaluations for these models.  The performance of \cwm in production and real-world scenarios has not been evaluated by Meta.  These models have not been fully evaluated or trained for user-facing interactions and they are not intended for such use. Researchers are recommended to exercise caution  when deploying or using these models.

Similarly, \cwm should not be used as a general-purpose assistant or chat model. While it was exposed to some level of instruction-following data during SFT, \cwm has not undergone any thorough optimization for general chat-bot use, such as RLHF \citep{ouyang2022training}. General chat use is not an intended use of CWM and generations may diverge from expectations and/or be inappropriate or inaccurate. Further, \cwm training focuses strongly on code generation and reasoning with code. Thus, our models may be lacking in other domains such as factual knowledge or classic natural language tasks.

\cwm is not trained for use as a general-purpose assistant or chat model and has not been aligned on, or fully evaluated for, content risks. We make available system level protections -- like Llama Guard, Prompt Guard, and Code Shield -- as a solution to help manage content generation in research environments.\footnote{See \url{https://www.llama.com/llama-protections}.} However, these system level protections alone are unlikely to be sufficient to enable production uses of \cwm and further evaluations and fine-tuning may be required. \cwm is intended to be used in English only. It is not multilingual and performance in other languages has not been evaluated or optimized.

Lastly, while we are excited about the opportunities that world modeling affords, these are only our first steps in this direction. Our code world modeling dataset collection efforts focus on explicit Python execution, and expanding this set to include other programming languages or symbolic execution is left for future work. Robust ways to leverage world model knowledge to improve performance across a variety of tasks via prompting or fine tuning is a ripe area for research. Similarly, planning with code world models, either using formal inference frameworks or informally during reasoning, is an exciting direction for research and core to our motivation for building \cwm in the first place. In some sense, one might compare the current state of \cwms to LLMs before CoT \citep{wei2023chainofthoughtpromptingelicitsreasoning}: the capabilities are there, we just need to find out how to make the most of them.

\section{Conclusion}
\label{sec:con}
Our vision is for Code World Models to bridge the gap between language‑level reasoning and executable semantics. We believe that coding and agentic use cases of LLMs will benefit from having a world model, a learned transition function between states conditioned on actions.
With the release of \cwm, we present the first steps of this vision.
Our ablations already show that world modeling data, Python execution traces, and executable Docker environments can be directly beneficial for downstream task performance.
More broadly though, \cwm provides a strong test-bed for future research in zero-shot planning, grounded chain-of-thought, and reinforcement learning with sparse, verifiable rewards.
Similar to our early results with execution trace prediction, we believe that the Python tracing world model enables research on reasoning about code generation, execution, correctness, and verification.
World models should improve reinforcement learning because agents that are already familiar with the dynamics of the environment can focus on learning which actions lead to rewards.
More research is needed to consistently leverage the benefits of incorporating world models into LLMs during pre-training across tasks.
Ultimately, models that can reason about the consequences of their actions should be much more efficient in their interactions with the environment which should allow for scaling the complexity of the tasks they perform.

\FloatBarrier

\newpage
\section*{Authors: Meta FAIR CodeGen Team}

Alphabetic order for core contributors, from second author onward and excluding senior authors, and contributors.

\subsection*{Core contributors}
\begin{multicols}{3}
Jade Copet\\
Quentin Carbonneaux\\
Gal Cohen\\
Jonas Gehring\\
Jacob Kahn\\
Jannik Kossen\\
Felix Kreuk\\
Emily McMilin\\
Michel Meyer\\
Yuxiang Wei\\
David Zhang\\
Kunhao Zheng
\end{multicols}

\subsection*{Contributors}
\begin{multicols}{3}
Jordi Armengol-Estapé\\
Pedram Bashiri\\
Maximilian Beck\\
Pierre Chambon\\
Abhishek Charnalia\\
Chris Cummins\\
Juliette Decugis\\
Zacharias V. Fisches\\
François Fleuret\\
Fabian Gloeckle\\
Alex Gu\\
Michael Hassid\\
Daniel Haziza\\
Badr Youbi Idrissi\\
Christian Keller\\
Rahul Kindi\\
Hugh Leather\\
Gallil Maimon\\
Aram Markosyan\\
Francisco Massa\\
Pierre-Emmanuel Mazaré\\
Vegard Mella\\
Naila Murray\\
Keyur Muzumdar\\
Peter O'Hearn\\
Matteo Pagliardini\\
Dmitrii Pedchenko\\
Tal Remez\\
Volker Seeker\\
Marco Selvi\\
Oren Sultan\\
Sida Wang\\
Luca Wehrstedt\\
Ori Yoran\\
Lingming Zhang
\end{multicols}

\subsection*{Senior core contributors}
\begin{multicols}{3}
Taco Cohen\\
Yossi Adi\\
Gabriel Synnaeve
\end{multicols}
\newpage

\bibliographystyle{assets/plainnat}
\bibliography{main}

\begin{thebibliography}{115}
\providecommand{\natexlab}[1]{#1}
\providecommand{\url}[1]{\texttt{#1}}
\expandafter\ifx\csname urlstyle\endcsname\relax
  \providecommand{\doi}[1]{doi: #1}\else
  \providecommand{\doi}{doi: \begingroup \urlstyle{rm}\Url}\fi

\bibitem[Agarwal et~al.(2025)Agarwal, Ahmad, Ai, Altman, Applebaum, Arbus,
  Arora, Bai, Baker, Bao, et~al.]{agarwal2025gpt}
Sandhini Agarwal, Lama Ahmad, Jason Ai, Sam Altman, Andy Applebaum, Edwin
  Arbus, Rahul~K Arora, Yu~Bai, Bowen Baker, Haiming Bao, et~al.
\newblock gpt-oss-120b \& gpt-oss-20b model card.
\newblock \emph{arXiv preprint arXiv:2508.10925}, 2025.

\bibitem[Ahmad et~al.(2025)Ahmad, Narenthiran, Majumdar, Ficek, Jain, Huang,
  Noroozi, and Ginsburg]{ocr}
Wasi~Uddin Ahmad, Sean Narenthiran, Somshubra Majumdar, Aleksander Ficek,
  Siddhartha Jain, Jocelyn Huang, Vahid Noroozi, and Boris Ginsburg.
\newblock Opencodereasoning: Advancing data distillation for competitive
  coding.
\newblock \emph{arXiv preprint arXiv:2504.01943}, 2025.

\bibitem[{Aider Team}(2025)]{aider}
{Aider Team}.
\newblock aider, 2025.
\newblock \url{https://github.com/Aider-AI/aider}.
\newblock GitHub repository; accessed 2025-08-18.

\bibitem[Ainslie et~al.(2023)Ainslie, Lee-Thorp, De~Jong, Zemlyanskiy,
  Lebr{\'o}n, and Sanghai]{ainslie2023gqa}
Joshua Ainslie, James Lee-Thorp, Michiel De~Jong, Yury Zemlyanskiy, Federico
  Lebr{\'o}n, and Sumit Sanghai.
\newblock Gqa: Training generalized multi-query transformer models from
  multi-head checkpoints.
\newblock \emph{arXiv preprint arXiv:2305.13245}, 2023.

\bibitem[Alon and David(2022)]{DBLP:conf/sigsoft/AlonD22}
Yoav Alon and Cristina David.
\newblock Using graph neural networks for program termination.
\newblock In Abhik Roychoudhury, Cristian Cadar, and Miryung Kim, editors,
  \emph{Proceedings of the 30th {ACM} Joint European Software Engineering
  Conference and Symposium on the Foundations of Software Engineering,
  {ESEC/FSE} 2022, Singapore, Singapore, November 14-18, 2022}, pages 910--921.
  {ACM}, 2022.
\newblock \doi{10.1145/3540250.3549095}.
\newblock \url{https://doi.org/10.1145/3540250.3549095}.

\bibitem[Anthropic(2025)]{claude}
Anthropic.
\newblock Claude 3.7 sonnet and claude code, February 2025.
\newblock \url{https://www.anthropic.com/news/claude-3-7-sonnet}.

\bibitem[{Anthropic}(2025)]{sonnet-sbv}
{Anthropic}.
\newblock Raising the bar on swe-bench verified with claude 3.5 sonnet.
\newblock 2025.
\newblock \url{https://www.anthropic.com/engineering/swe-bench-sonnet}.
\newblock Accessed 2025-08-18.

\bibitem[Aram H.~Markosyan(2024)]{ahm2025leanuniverse}
Hugh~Leather Aram H.~Markosyan, Gabriel~Synnaeve.
\newblock Leanuniverse: A library for consistent and scalable lean4 dataset
  management, 2024.

\bibitem[Armengol-Estap{\'e} et~al.(2025)Armengol-Estap{\'e}, Carbonneaux,
  Zhang, Markosyan, Seeker, Cummins, Kambadur, O'Boyle, Wang, Synnaeve,
  et~al.]{armengol2025cannot}
Jordi Armengol-Estap{\'e}, Quentin Carbonneaux, Tianjun Zhang, Aram~H
  Markosyan, Volker Seeker, Chris Cummins, Melanie Kambadur, Michael~FP
  O'Boyle, Sida Wang, Gabriel Synnaeve, et~al.
\newblock What i cannot execute, i do not understand: Training and evaluating
  llms on program execution traces.
\newblock \emph{arXiv preprint arXiv:2503.05703}, 2025.

\bibitem[Austin et~al.(2021)Austin, Odena, Nye, Bosma, Michalewski, Dohan,
  Jiang, Cai, Terry, Le, and Sutton]{DBLP:journals/corr/abs-2108-07732}
Jacob Austin, Augustus Odena, Maxwell~I. Nye, Maarten Bosma, Henryk
  Michalewski, David Dohan, Ellen Jiang, Carrie~J. Cai, Michael Terry, Quoc~V.
  Le, and Charles Sutton.
\newblock Program synthesis with large language models.
\newblock \emph{CoRR}, abs/2108.07732, 2021.
\newblock \url{https://arxiv.org/abs/2108.07732}.

\bibitem[Azerbayev et~al.(2023)Azerbayev, Piotrowski, Schoelkopf, Ayers, Radev,
  and Avigad]{azerbayev2023proofnet}
Zhangir Azerbayev, Bartosz Piotrowski, Hailey Schoelkopf, Edward~W. Ayers,
  Dragomir Radev, and Jeremy Avigad.
\newblock Proofnet: Autoformalizing and formally proving undergraduate-level
  mathematics, 2023.
\newblock \url{https://arxiv.org/abs/2302.12433}.

\bibitem[Bi et~al.(2024)Bi, Chen, Chen, Chen, Dai, Deng, Ding, Dong, Du, Fu,
  et~al.]{bi2024deepseek}
Xiao Bi, Deli Chen, Guanting Chen, Shanhuang Chen, Damai Dai, Chengqi Deng,
  Honghui Ding, Kai Dong, Qiushi Du, Zhe Fu, et~al.
\newblock Deepseek llm: Scaling open-source language models with longtermism.
\newblock \emph{arXiv preprint arXiv:2401.02954}, 2024.

\bibitem[Bick et~al.(2024)Bick, Blandin, and Deming]{bick2024rapid}
Alexander Bick, Adam Blandin, and David~J Deming.
\newblock The rapid adoption of generative ai.
\newblock Technical report, National Bureau of Economic Research, 2024.

\bibitem[Bisk et~al.(2020)Bisk, Zellers, Bras, Gao, and
  Choi]{DBLP:conf/aaai/BiskZLGC20}
Yonatan Bisk, Rowan Zellers, Ronan~Le Bras, Jianfeng Gao, and Yejin Choi.
\newblock {PIQA:} reasoning about physical commonsense in natural language.
\newblock In \emph{The Thirty-Fourth {AAAI} Conference on Artificial
  Intelligence, {AAAI} 2020, The Thirty-Second Innovative Applications of
  Artificial Intelligence Conference, {IAAI} 2020, The Tenth {AAAI} Symposium
  on Educational Advances in Artificial Intelligence, {EAAI} 2020, New York,
  NY, USA, February 7-12, 2020}, pages 7432--7439. {AAAI} Press, 2020.
\newblock \doi{10.1609/AAAI.V34I05.6239}.
\newblock \url{https://doi.org/10.1609/aaai.v34i05.6239}.

\bibitem[Carbonneaux(2025)]{Carbonneaux2025fastgen}
Quentin Carbonneaux.
\newblock Fastgen, 2025.
\newblock \url{https://github.com/facebookresearch/fastgen}.

\bibitem[Chambon et~al.(2025)Chambon, Roziere, Sagot, and
  Synnaeve]{chambon2025bigobenchllmsgenerate}
Pierre Chambon, Baptiste Roziere, Benoit Sagot, and Gabriel Synnaeve.
\newblock Bigo(bench) -- can llms generate code with controlled time and space
  complexity?, 2025.
\newblock \url{https://arxiv.org/abs/2503.15242}.

\bibitem[Chen et~al.(2021)Chen, Tworek, Jun, Yuan, de~Oliveira~Pinto, Kaplan,
  Edwards, Burda, Joseph, Brockman, Ray, Puri, Krueger, Petrov, Khlaaf, Sastry,
  Mishkin, Chan, Gray, Ryder, Pavlov, Power, Kaiser, Bavarian, Winter, Tillet,
  Such, Cummings, Plappert, Chantzis, Barnes, Herbert{-}Voss, Guss, Nichol,
  Paino, Tezak, Tang, Babuschkin, Balaji, Jain, Saunders, Hesse, Carr, Leike,
  Achiam, Misra, Morikawa, Radford, Knight, Brundage, Murati, Mayer, Welinder,
  McGrew, Amodei, McCandlish, Sutskever, and
  Zaremba]{DBLP:journals/corr/abs-2107-03374}
Mark Chen, Jerry Tworek, Heewoo Jun, Qiming Yuan, Henrique~Pond{\'{e}}
  de~Oliveira~Pinto, Jared Kaplan, Harri Edwards, Yuri Burda, Nicholas Joseph,
  Greg Brockman, Alex Ray, Raul Puri, Gretchen Krueger, Michael Petrov, Heidy
  Khlaaf, Girish Sastry, Pamela Mishkin, Brooke Chan, Scott Gray, Nick Ryder,
  Mikhail Pavlov, Alethea Power, Lukasz Kaiser, Mohammad Bavarian, Clemens
  Winter, Philippe Tillet, Felipe~Petroski Such, Dave Cummings, Matthias
  Plappert, Fotios Chantzis, Elizabeth Barnes, Ariel Herbert{-}Voss,
  William~Hebgen Guss, Alex Nichol, Alex Paino, Nikolas Tezak, Jie Tang, Igor
  Babuschkin, Suchir Balaji, Shantanu Jain, William Saunders, Christopher
  Hesse, Andrew~N. Carr, Jan Leike, Joshua Achiam, Vedant Misra, Evan Morikawa,
  Alec Radford, Matthew Knight, Miles Brundage, Mira Murati, Katie Mayer, Peter
  Welinder, Bob McGrew, Dario Amodei, Sam McCandlish, Ilya Sutskever, and
  Wojciech Zaremba.
\newblock Evaluating large language models trained on code.
\newblock \emph{CoRR}, abs/2107.03374, 2021.
\newblock \url{https://arxiv.org/abs/2107.03374}.

\bibitem[Clark et~al.(2018)Clark, Cowhey, Etzioni, Khot, Sabharwal, Schoenick,
  and Tafjord]{DBLP:journals/corr/abs-1803-05457}
Peter Clark, Isaac Cowhey, Oren Etzioni, Tushar Khot, Ashish Sabharwal, Carissa
  Schoenick, and Oyvind Tafjord.
\newblock Think you have solved question answering? try arc, the {AI2}
  reasoning challenge.
\newblock \emph{CoRR}, abs/1803.05457, 2018.
\newblock \url{http://arxiv.org/abs/1803.05457}.

\bibitem[Cobbe et~al.(2021)Cobbe, Kosaraju, Bavarian, Chen, Jun, Kaiser,
  Plappert, Tworek, Hilton, Nakano, Hesse, and
  Schulman]{DBLP:journals/corr/abs-2110-14168}
Karl Cobbe, Vineet Kosaraju, Mohammad Bavarian, Mark Chen, Heewoo Jun, Lukasz
  Kaiser, Matthias Plappert, Jerry Tworek, Jacob Hilton, Reiichiro Nakano,
  Christopher Hesse, and John Schulman.
\newblock Training verifiers to solve math word problems.
\newblock \emph{CoRR}, abs/2110.14168, 2021.
\newblock \url{https://arxiv.org/abs/2110.14168}.

\bibitem[Cohen et~al.(2025)Cohen, Zhang, Zheng, Tang, Munos, and
  Synnaeve]{DBLP:journals/corr/abs-2503-05453}
Taco Cohen, David~W. Zhang, Kunhao Zheng, Yunhao Tang, R{\'{e}}mi Munos, and
  Gabriel Synnaeve.
\newblock Soft policy optimization: Online off-policy {RL} for sequence models.
\newblock \emph{CoRR}, abs/2503.05453, 2025.
\newblock \doi{10.48550/ARXIV.2503.05453}.
\newblock \url{https://doi.org/10.48550/arXiv.2503.05453}.

\bibitem[Comanici et~al.(2025)Comanici, Bieber, Schaekermann, Pasupat,
  Sachdeva, Dhillon, Blistein, Ram, Zhang, Rosen, et~al.]{comanici2025gemini}
Gheorghe Comanici, Eric Bieber, Mike Schaekermann, Ice Pasupat, Noveen
  Sachdeva, Inderjit Dhillon, Marcel Blistein, Ori Ram, Dan Zhang, Evan Rosen,
  et~al.
\newblock Gemini 2.5: Pushing the frontier with advanced reasoning,
  multimodality, long context, and next generation agentic capabilities.
\newblock \emph{arXiv preprint arXiv:2507.06261}, 2025.

\bibitem[{CompFiles authors}(2025)]{compfiles}
{CompFiles authors}.
\newblock Compfiles.
\newblock \url{https://github.com/dwrensha/compfiles}, 2025.

\bibitem[Cook et~al.(2011)Cook, Podelski, and
  Rybalchenko]{DBLP:journals/cacm/CookPR11}
Byron Cook, Andreas Podelski, and Andrey Rybalchenko.
\newblock Proving program termination.
\newblock \emph{Commun. {ACM}}, 54\penalty0 (5):\penalty0 88--98, 2011.
\newblock \doi{10.1145/1941487.1941509}.
\newblock \url{https://doi.org/10.1145/1941487.1941509}.

\bibitem[Cui et~al.(2024)Cui, Demirer, Jaffe, Musolff, Peng, and
  Salz]{cui2024effects}
Zheyuan Cui, Mert Demirer, Sonia Jaffe, Leon Musolff, Sida Peng, and Tobias
  Salz.
\newblock {The Effects of Generative AI on High Skilled Work: Evidence from
  Three Field Experiments with Software Developers}.
\newblock \emph{SSRN eLibrary}, 2024.
\newblock \doi{10.2139/ssrn.4945566}.

\bibitem[Cummins et~al.(2024)Cummins, Seeker, Grubisic, Roziere, Gehring,
  Synnaeve, and Leather]{cummins2024meta}
Chris Cummins, Volker Seeker, Dejan Grubisic, Baptiste Roziere, Jonas Gehring,
  Gabriel Synnaeve, and Hugh Leather.
\newblock Meta large language model compiler: Foundation models of compiler
  optimization.
\newblock \emph{arXiv preprint arXiv:2407.02524}, 2024.

\bibitem[Dao(2024)]{dao2023flashattention2}
Tri Dao.
\newblock Flash{A}ttention-2: Faster attention with better parallelism and work
  partitioning.
\newblock In \emph{International Conference on Learning Representations
  (ICLR)}, 2024.

\bibitem[Dao et~al.(2022)Dao, Fu, Ermon, Rudra, and
  R{\'e}]{dao2022flashattention}
Tri Dao, Daniel~Y. Fu, Stefano Ermon, Atri Rudra, and Christopher R{\'e}.
\newblock Flash{A}ttention: Fast and memory-efficient exact attention with
  {IO}-awareness.
\newblock In \emph{Advances in Neural Information Processing Systems
  (NeurIPS)}, 2022.

\bibitem[Dijkstra(1976)]{Dijkstra76}
Edsger~W. Dijkstra.
\newblock \emph{A Discipline of Programming}.
\newblock Prentice-Hall, 1976.
\newblock ISBN 013215871X.
\newblock \url{https://www.worldcat.org/oclc/01958445}.

\bibitem[Dua et~al.(2019)Dua, Wang, Dasigi, Stanovsky, Singh, and
  Gardner]{DBLP:conf/naacl/DuaWDSS019}
Dheeru Dua, Yizhong Wang, Pradeep Dasigi, Gabriel Stanovsky, Sameer Singh, and
  Matt Gardner.
\newblock {DROP:} {A} reading comprehension benchmark requiring discrete
  reasoning over paragraphs.
\newblock In Jill Burstein, Christy Doran, and Thamar Solorio, editors,
  \emph{Proceedings of the 2019 Conference of the North American Chapter of the
  Association for Computational Linguistics: Human Language Technologies,
  {NAACL-HLT} 2019, Minneapolis, MN, USA, June 2-7, 2019, Volume 1 (Long and
  Short Papers)}, pages 2368--2378. Association for Computational Linguistics,
  2019.
\newblock \doi{10.18653/V1/N19-1246}.
\newblock \url{https://doi.org/10.18653/v1/n19-1246}.

\bibitem[Dubey et~al.(2024)Dubey, Jauhri, Pandey, Kadian, Al-Dahle, Letman,
  Mathur, Schelten, Yang, Fan, et~al.]{dubey2024llama}
Abhimanyu Dubey, Abhinav Jauhri, Abhinav Pandey, Abhishek Kadian, Ahmad
  Al-Dahle, Aiesha Letman, Akhil Mathur, Alan Schelten, Amy Yang, Angela Fan,
  et~al.
\newblock The llama 3 herd of models.
\newblock \emph{arXiv e-prints}, pages arXiv--2407, 2024.

\bibitem[Gadre et~al.(2024)Gadre, Smyrnis, Shankar, Gururangan, Wortsman, Shao,
  Mercat, Fang, Li, Keh, et~al.]{gadre2024language}
Samir~Yitzhak Gadre, Georgios Smyrnis, Vaishaal Shankar, Suchin Gururangan,
  Mitchell Wortsman, Rulin Shao, Jean Mercat, Alex Fang, Jeffrey Li, Sedrick
  Keh, et~al.
\newblock Language models scale reliably with over-training and on downstream
  tasks.
\newblock \emph{arXiv preprint arXiv:2403.08540}, 2024.

\bibitem[Gao et~al.(2025)Gao, Song, Yang, Cai, Miao, Dong, Li, Ma, Chen, Xu,
  Tang, Wang, Zan, Quan, Zhang, Sha, Zhang, Ren, Liu, and
  Chang]{DBLP:conf/iclr/Gao0YCMDLMCXTWZ25}
Bofei Gao, Feifan Song, Zhe Yang, Zefan Cai, Yibo Miao, Qingxiu Dong, Lei Li,
  Chenghao Ma, Liang Chen, Runxin Xu, Zhengyang Tang, Benyou Wang, Daoguang
  Zan, Shanghaoran Quan, Ge~Zhang, Lei Sha, Yichang Zhang, Xuancheng Ren,
  Tianyu Liu, and Baobao Chang.
\newblock Omni-math: {A} universal olympiad level mathematic benchmark for
  large language models.
\newblock In \emph{The Thirteenth International Conference on Learning
  Representations, {ICLR} 2025, Singapore, April 24-28, 2025}. OpenReview.net,
  2025.
\newblock \url{https://openreview.net/forum?id=yaqPf0KAlN}.

\bibitem[Gehring et~al.(2025)Gehring, Zheng, Copet, Mella, Carbonneaux, Cohen,
  and Synnaeve]{gehring2025rlefgroundingcodellms}
Jonas Gehring, Kunhao Zheng, Jade Copet, Vegard Mella, Quentin Carbonneaux,
  Taco Cohen, and Gabriel Synnaeve.
\newblock Rlef: Grounding code llms in execution feedback with reinforcement
  learning, 2025.
\newblock \url{https://arxiv.org/abs/2410.02089}.

\bibitem[Gu et~al.(2024)Gu, Rozière, Leather, Solar-Lezama, Synnaeve, and
  Wang]{gu2024cruxeval}
Alex Gu, Baptiste Rozière, Hugh Leather, Armando Solar-Lezama, Gabriel
  Synnaeve, and Sida~I. Wang.
\newblock Cruxeval: A benchmark for code reasoning, understanding and
  execution.
\newblock \emph{arXiv preprint arXiv:2401.03065}, 2024.

\bibitem[Guo et~al.(2025)Guo, Yang, Zhang, Song, Zhang, Xu, Zhu, Ma, Wang, Bi,
  et~al.]{guo2025deepseek}
Daya Guo, Dejian Yang, Haowei Zhang, Junxiao Song, Ruoyu Zhang, Runxin Xu,
  Qihao Zhu, Shirong Ma, Peiyi Wang, Xiao Bi, et~al.
\newblock Deepseek-r1: Incentivizing reasoning capability in llms via
  reinforcement learning.
\newblock \emph{arXiv preprint arXiv:2501.12948}, 2025.

\bibitem[Gupta et~al.(2008)Gupta, Henzinger, Majumdar, Rybalchenko, and
  Xu]{DBLP:conf/popl/GuptaHMRX08}
Ashutosh Gupta, Thomas~A. Henzinger, Rupak Majumdar, Andrey Rybalchenko, and
  Ru{-}Gang Xu.
\newblock Proving non-termination.
\newblock In George~C. Necula and Philip Wadler, editors, \emph{Proceedings of
  the 35th {ACM} {SIGPLAN-SIGACT} Symposium on Principles of Programming
  Languages, {POPL} 2008, San Francisco, California, USA, January 7-12, 2008},
  pages 147--158. {ACM}, 2008.
\newblock \doi{10.1145/1328438.1328459}.
\newblock \url{https://doi.org/10.1145/1328438.1328459}.

\bibitem[Handa et~al.(2025)Handa, Tamkin, McCain, Huang, Durmus, Heck, Mueller,
  Hong, Ritchie, Belonax, et~al.]{handa2025economic}
Kunal Handa, Alex Tamkin, Miles McCain, Saffron Huang, Esin Durmus, Sarah Heck,
  Jared Mueller, Jerry Hong, Stuart Ritchie, Tim Belonax, et~al.
\newblock Which economic tasks are performed with ai? evidence from millions of
  claude conversations.
\newblock \emph{arXiv preprint arXiv:2503.04761}, 2025.

\bibitem[Hassid et~al.(2025)Hassid, Synnaeve, Adi, and
  Schwartz]{dont_over_think}
Michael Hassid, Gabriel Synnaeve, Yossi Adi, and Roy Schwartz.
\newblock Don't overthink it. preferring shorter thinking chains for improved
  llm reasoning.
\newblock \emph{arXiv preprint arXiv:2505.17813}, 2025.

\bibitem[Hendrycks et~al.(2021{\natexlab{a}})Hendrycks, Basart, Kadavath,
  Mazeika, Arora, Guo, Burns, Puranik, He, Song,
  et~al.]{hendrycks2021measuring}
Dan Hendrycks, Steven Basart, Saurav Kadavath, Mantas Mazeika, Akul Arora,
  Ethan Guo, Collin Burns, Samir Puranik, Horace He, Dawn Song, et~al.
\newblock Measuring coding challenge competence with apps.
\newblock \emph{arXiv preprint arXiv:2105.09938}, 2021{\natexlab{a}}.

\bibitem[Hendrycks et~al.(2021{\natexlab{b}})Hendrycks, Burns, Kadavath, Arora,
  Basart, Tang, Song, and Steinhardt]{DBLP:conf/nips/HendrycksBKABTS21}
Dan Hendrycks, Collin Burns, Saurav Kadavath, Akul Arora, Steven Basart, Eric
  Tang, Dawn Song, and Jacob Steinhardt.
\newblock Measuring mathematical problem solving with the {MATH} dataset.
\newblock In Joaquin Vanschoren and Sai{-}Kit Yeung, editors, \emph{Proceedings
  of the Neural Information Processing Systems Track on Datasets and Benchmarks
  1, NeurIPS Datasets and Benchmarks 2021, December 2021, virtual},
  2021{\natexlab{b}}.
\newblock
  \url{https://datasets-benchmarks-proceedings.neurips.cc/paper/2021/hash/be83ab3ecd0db773eb2dc1b0a17836a1-Abstract-round2.html}.

\bibitem[Hoare(1971)]{Hoare71}
C.~A.~R. Hoare.
\newblock Proof of a program: {FIND}.
\newblock \emph{Commun. {ACM}}, 14\penalty0 (1):\penalty0 39--45, 1971.
\newblock \doi{10.1145/362452.362489}.
\newblock \url{https://doi.org/10.1145/362452.362489}.

\bibitem[Hoffmann et~al.(2022)Hoffmann, Borgeaud, Mensch, Buchatskaya, Cai,
  Rutherford, Casas, Hendricks, Welbl, Clark, et~al.]{hoffmann2022training}
Jordan Hoffmann, Sebastian Borgeaud, Arthur Mensch, Elena Buchatskaya, Trevor
  Cai, Eliza Rutherford, Diego de~Las Casas, Lisa~Anne Hendricks, Johannes
  Welbl, Aidan Clark, et~al.
\newblock Training compute-optimal large language models.
\newblock \emph{arXiv preprint arXiv:2203.15556}, 2022.

\bibitem[Hsieh et~al.(2024)Hsieh, Sun, Kriman, Acharya, Rekesh, Jia, Zhang, and
  Ginsburg]{hsieh2024ruler}
Cheng-Ping Hsieh, Simeng Sun, Samuel Kriman, Shantanu Acharya, Dima Rekesh, Fei
  Jia, Yang Zhang, and Boris Ginsburg.
\newblock Ruler: What's the real context size of your long-context language
  models?
\newblock \emph{arXiv preprint arXiv:2404.06654}, 2024.

\bibitem[Hu et~al.(2025)Hu, Zhang, Han, Jiang, Zhang, and
  Shum]{hu2025openreasonerzeroopensourceapproach}
Jingcheng Hu, Yinmin Zhang, Qi~Han, Daxin Jiang, Xiangyu Zhang, and Heung-Yeung
  Shum.
\newblock Open-reasoner-zero: An open source approach to scaling up
  reinforcement learning on the base model, 2025.
\newblock \url{https://arxiv.org/abs/2503.24290}.

\bibitem[Jain et~al.(2025{\natexlab{a}})Jain, Han, Gu, Li, Yan, Zhang, Wang,
  Solar{-}Lezama, Sen, and Stoica]{DBLP:conf/iclr/JainHGLYZWSSS25}
Naman Jain, King Han, Alex Gu, Wen{-}Ding Li, Fanjia Yan, Tianjun Zhang, Sida
  Wang, Armando Solar{-}Lezama, Koushik Sen, and Ion Stoica.
\newblock Livecodebench: Holistic and contamination free evaluation of large
  language models for code.
\newblock In \emph{The Thirteenth International Conference on Learning
  Representations, {ICLR} 2025, Singapore, April 24-28, 2025}. OpenReview.net,
  2025{\natexlab{a}}.
\newblock \url{https://openreview.net/forum?id=chfJJYC3iL}.

\bibitem[Jain et~al.(2025{\natexlab{b}})Jain, Singh, Shetty, Zheng, Sen, and
  Stoica]{jain2025r2e}
Naman Jain, Jaskirat Singh, Manish Shetty, Liang Zheng, Koushik Sen, and Ion
  Stoica.
\newblock R2e-gym: Procedural environments and hybrid verifiers for scaling
  open-weights swe agents.
\newblock \emph{arXiv preprint arXiv:2504.07164}, 2025{\natexlab{b}}.

\bibitem[Jimenez et~al.(2024)Jimenez, Yang, Wettig, Yao, Pei, Press, and
  Narasimhan]{DBLP:conf/iclr/JimenezYWYPPN24}
Carlos~E. Jimenez, John Yang, Alexander Wettig, Shunyu Yao, Kexin Pei, Ofir
  Press, and Karthik~R. Narasimhan.
\newblock Swe-bench: Can language models resolve real-world github issues?
\newblock In \emph{The Twelfth International Conference on Learning
  Representations, {ICLR} 2024, Vienna, Austria, May 7-11, 2024}.
  OpenReview.net, 2024.
\newblock \url{https://openreview.net/forum?id=VTF8yNQM66}.

\bibitem[Kaplan et~al.(2020)Kaplan, McCandlish, Henighan, Brown, Chess, Child,
  Gray, Radford, Wu, and Amodei]{kaplan2020scaling}
Jared Kaplan, Sam McCandlish, Tom Henighan, Tom~B Brown, Benjamin Chess, Rewon
  Child, Scott Gray, Alec Radford, Jeffrey Wu, and Dario Amodei.
\newblock Scaling laws for neural language models.
\newblock \emph{arXiv preprint arXiv:2001.08361}, 2020.

\bibitem[{Kimi Team} et~al.(2025){Kimi Team}, Bai, Bao, Chen, Chen, Chen, Chen,
  Chen, Chen, Chen, et~al.]{team2025kimi}
{ } {Kimi Team}, Yifan Bai, Yiping Bao, Guanduo Chen, Jiahao Chen, Ningxin
  Chen, Ruijue Chen, Yanru Chen, Yuankun Chen, Yutian Chen, et~al.
\newblock Kimi k2: Open agentic intelligence.
\newblock \emph{arXiv preprint arXiv:2507.20534}, 2025.

\bibitem[Kwon et~al.(2023)Kwon, Li, Zhuang, Sheng, Zheng, Yu, Gonzalez, Zhang,
  and Stoica]{kwon2023efficient}
Woosuk Kwon, Zhuohan Li, Siyuan Zhuang, Ying Sheng, Lianmin Zheng, Cody~Hao Yu,
  Joseph~E. Gonzalez, Hao Zhang, and Ion Stoica.
\newblock Efficient memory management for large language model serving with
  pagedattention.
\newblock In \emph{Proceedings of the ACM SIGOPS 29th Symposium on Operating
  Systems Principles}, 2023.

\bibitem[Kydlicek et~al.(2025)Kydlicek, Lozovskaya, Habib, and
  Fourrier]{math_verify}
Hynek Kydlicek, Alina Lozovskaya, Nathan Habib, and Clémentine Fourrier.
\newblock Math-verify, 2025.
\newblock \url{https://github.com/huggingface/Math-Verify}.

\bibitem[Lee(2019)]{Lee2025act}
Casey Lee.
\newblock act, 2019.
\newblock \url{https://github.com/nektos/act}.

\bibitem[Lefaudeux et~al.(2022)Lefaudeux, Massa, Liskovich, Xiong, Caggiano,
  Naren, Xu, Hu, Tintore, Zhang, et~al.]{lefaudeux2022xformers}
Benjamin Lefaudeux, Francisco Massa, Diana Liskovich, Wenhan Xiong, Vittorio
  Caggiano, Sean Naren, Min Xu, Jieru Hu, Marta Tintore, Susan Zhang, et~al.
\newblock xformers: A modular and hackable transformer modelling library, 2022.

\bibitem[Li et~al.(2024)Li, Chiang, Frick, Dunlap, Wu, Zhu, Gonzalez, and
  Stoica]{DBLP:journals/corr/abs-2406-11939}
Tianle Li, Wei{-}Lin Chiang, Evan Frick, Lisa Dunlap, Tianhao Wu, Banghua Zhu,
  Joseph~E. Gonzalez, and Ion Stoica.
\newblock From crowdsourced data to high-quality benchmarks: Arena-hard and
  benchbuilder pipeline.
\newblock \emph{CoRR}, abs/2406.11939, 2024.
\newblock \doi{10.48550/ARXIV.2406.11939}.
\newblock \url{https://doi.org/10.48550/arXiv.2406.11939}.

\bibitem[Li et~al.(2022)Li, Choi, Chung, Kushman, Schrittwieser, Leblond,
  Eccles, Keeling, Gimeno, Lago, Hubert, Choy, de~Masson~d'Autume, Babuschkin,
  Chen, Huang, Welbl, Gowal, Cherepanov, Molloy, Mankowitz, Robson, Kohli,
  de~Freitas, Kavukcuoglu, and Vinyals]{DBLP:journals/corr/abs-2203-07814}
Yujia Li, David~H. Choi, Junyoung Chung, Nate Kushman, Julian Schrittwieser,
  R{\'{e}}mi Leblond, Tom Eccles, James Keeling, Felix Gimeno, Agustin~Dal
  Lago, Thomas Hubert, Peter Choy, Cyprien de~Masson~d'Autume, Igor Babuschkin,
  Xinyun Chen, Po{-}Sen Huang, Johannes Welbl, Sven Gowal, Alexey Cherepanov,
  James Molloy, Daniel~J. Mankowitz, Esme~Sutherland Robson, Pushmeet Kohli,
  Nando de~Freitas, Koray Kavukcuoglu, and Oriol Vinyals.
\newblock Competition-level code generation with alphacode.
\newblock \emph{CoRR}, abs/2203.07814, 2022.
\newblock \doi{10.48550/ARXIV.2203.07814}.
\newblock \url{https://doi.org/10.48550/arXiv.2203.07814}.

\bibitem[Lightman et~al.(2023)Lightman, Kosaraju, Burda, Edwards, Baker, Lee,
  Leike, Schulman, Sutskever, and Cobbe]{lightman2023let}
Hunter Lightman, Vineet Kosaraju, Yuri Burda, Harrison Edwards, Bowen Baker,
  Teddy Lee, Jan Leike, John Schulman, Ilya Sutskever, and Karl Cobbe.
\newblock Let's verify step by step.
\newblock In \emph{The Twelfth International Conference on Learning
  Representations}, 2023.

\bibitem[Lin et~al.(2025)Lin, Tang, Lyu, Wu, Lin, Yang, Li, Xia, Chen, Arora,
  and Jin]{lin2025goedelprover}
Yong Lin, Shange Tang, Bohan Lyu, Jiayun Wu, Hongzhou Lin, Kaiyu Yang, Jia Li,
  Mengzhou Xia, Danqi Chen, Sanjeev Arora, and Chi Jin.
\newblock Goedel-prover: A frontier model for open-source automated theorem
  proving, 2025.
\newblock \url{https://arxiv.org/abs/2502.07640}.

\bibitem[Liu et~al.(2025)Liu, Chen, Li, Qi, Pang, Du, Lee, and
  Lin]{liu2025understanding}
Zichen Liu, Changyu Chen, Wenjun Li, Penghui Qi, Tianyu Pang, Chao Du, Wee~Sun
  Lee, and Min Lin.
\newblock Understanding r1-zero-like training: A critical perspective.
\newblock \emph{arXiv preprint arXiv:2503.20783}, 2025.

\bibitem[Loshchilov and Hutter(2019)]{loshchilov2018decoupled}
Ilya Loshchilov and Frank Hutter.
\newblock Decoupled weight decay regularization.
\newblock In \emph{International Conference on Learning Representations}, 2019.
\newblock \url{https://openreview.net/forum?id=Bkg6RiCqY7}.

\bibitem[mathlib Community(2020)]{leanmathlib2020}
The mathlib Community.
\newblock The lean mathematical library.
\newblock In \emph{Proceedings of the 9th ACM SIGPLAN International Conference
  on Certified Programs and Proofs}, CPP 2020, page 367–381, New York, NY,
  USA, 2020. Association for Computing Machinery.
\newblock ISBN 9781450370974.
\newblock \doi{10.1145/3372885.3373824}.
\newblock \url{https://doi.org/10.1145/3372885.3373824}.

\bibitem[Mella(2025)]{Mella2025moodist}
Vegard Mella.
\newblock Moodist, 2025.
\newblock \url{https://github.com/facebookresearch/moodist}.

\bibitem[{Mentat AI Team}(2025)]{locodiff}
{Mentat AI Team}.
\newblock Locodiff-bench: Natural long-context code benchmark, 2025.
\newblock \url{https://github.com/AbanteAI/LoCoDiff-bench}.

\bibitem[{Meta AI}(2025)]{meta_llama_llama4_2025}
{Meta AI}.
\newblock Llama 4 model card.
\newblock
  \url{https://github.com/meta-llama/llama-models/blob/main/models/llama4/MODEL_CARD.md},
  2025.
\newblock Accessed: 2025-09-18.

\bibitem[Meurer et~al.(2017)Meurer, Smith, Paprocki, \v{C}ert\'{i}k, Kirpichev,
  Rocklin, Kumar, Ivanov, Moore, Singh, Rathnayake, Vig, Granger, Muller,
  Bonazzi, Gupta, Vats, Johansson, Pedregosa, Curry, Terrel, Rou\v{c}ka, Saboo,
  Fernando, Kulal, Cimrman, and Scopatz]{sympy}
Aaron Meurer, Christopher~P. Smith, Mateusz Paprocki, Ond\v{r}ej
  \v{C}ert\'{i}k, Sergey~B. Kirpichev, Matthew Rocklin, AMiT Kumar, Sergiu
  Ivanov, Jason~K. Moore, Sartaj Singh, Thilina Rathnayake, Sean Vig, Brian~E.
  Granger, Richard~P. Muller, Francesco Bonazzi, Harsh Gupta, Shivam Vats,
  Fredrik Johansson, Fabian Pedregosa, Matthew~J. Curry, Andy~R. Terrel,
  \v{S}t\v{e}p\'{a}n Rou\v{c}ka, Ashutosh Saboo, Isuru Fernando, Sumith Kulal,
  Robert Cimrman, and Anthony Scopatz.
\newblock Sympy: symbolic computing in python.
\newblock \emph{PeerJ Computer Science}, 3:\penalty0 e103, January 2017.
\newblock ISSN 2376-5992.
\newblock \doi{10.7717/peerj-cs.103}.
\newblock \url{https://doi.org/10.7717/peerj-cs.103}.

\bibitem[Micikevicius et~al.(2022)Micikevicius, Stosic, Burgess, Cornea, Dubey,
  Grisenthwaite, Ha, Heinecke, Judd, Kamalu, et~al.]{micikevicius2022fp8}
Paulius Micikevicius, Dusan Stosic, Neil Burgess, Marius Cornea, Pradeep Dubey,
  Richard Grisenthwaite, Sangwon Ha, Alexander Heinecke, Patrick Judd, John
  Kamalu, et~al.
\newblock Fp8 formats for deep learning.
\newblock \emph{arXiv preprint arXiv:2209.05433}, 2022.

\bibitem[Mihaylov et~al.(2018)Mihaylov, Clark, Khot, and
  Sabharwal]{DBLP:conf/emnlp/MihaylovCKS18}
Todor Mihaylov, Peter Clark, Tushar Khot, and Ashish Sabharwal.
\newblock Can a suit of armor conduct electricity? {A} new dataset for open
  book question answering.
\newblock In Ellen Riloff, David Chiang, Julia Hockenmaier, and Jun'ichi
  Tsujii, editors, \emph{Proceedings of the 2018 Conference on Empirical
  Methods in Natural Language Processing, Brussels, Belgium, October 31 -
  November 4, 2018}, pages 2381--2391. Association for Computational
  Linguistics, 2018.
\newblock \doi{10.18653/V1/D18-1260}.
\newblock \url{https://doi.org/10.18653/v1/d18-1260}.

\bibitem[Mistral-AI et~al.(2025)Mistral-AI, :, Rastogi, Jiang, Lo, Berrada,
  Lample, Rute, Barmentlo, Yadav, Khandelwal, Chandu, Blier, Saulnier, Dinot,
  Darrin, Gupta, Soletskyi, Vaze, Scao, Wang, Yang, Liu, Sablayrolles, Héliou,
  Martin, Ehrenberg, Agarwal, Roux, Darcet, Mensch, Bout, Rozière, Monicault,
  Bamford, Wallenwein, Renaudin, Lanfranchi, Dabert, Mizelle, de~las Casas,
  Chane-Sane, Fugier, Hanna, Delerce, Guinet, Novikov, Martin, Jaju,
  Ludziejewski, Chabran, Delignon, Studnia, Amar, Roberts, Denize, Saxena,
  Jain, Zhao, Martin, Gao, Lavaud, Pellat, Guillaumin, Felardos, Augustin,
  Seznec, Raghuraman, Duchenne, Wang, von Platen, Saffer, Jacob, Wambergue,
  Kurylowicz, Muddireddy, Chagniot, Stock, Agrawal, Sauvestre, Delacourt,
  Gandhi, Subramanian, Dalal, Gandhi, Ghosh, Mishra, Aithal, Antoniak,
  Schueller, Lavril, Robert, Wang, Lacroix, Nemychnikova, Paltz, Richard, Li,
  Marshall, Zhang, and Tang]{mistralai2025magistral}
Mistral-AI, :, Abhinav Rastogi, Albert~Q. Jiang, Andy Lo, Gabrielle Berrada,
  Guillaume Lample, Jason Rute, Joep Barmentlo, Karmesh Yadav, Kartik
  Khandelwal, Khyathi~Raghavi Chandu, Léonard Blier, Lucile Saulnier, Matthieu
  Dinot, Maxime Darrin, Neha Gupta, Roman Soletskyi, Sagar Vaze, Teven~Le Scao,
  Yihan Wang, Adam Yang, Alexander~H. Liu, Alexandre Sablayrolles, Amélie
  Héliou, Amélie Martin, Andy Ehrenberg, Anmol Agarwal, Antoine Roux, Arthur
  Darcet, Arthur Mensch, Baptiste Bout, Baptiste Rozière, Baudouin~De
  Monicault, Chris Bamford, Christian Wallenwein, Christophe Renaudin,
  Clémence Lanfranchi, Darius Dabert, Devon Mizelle, Diego de~las Casas,
  Elliot Chane-Sane, Emilien Fugier, Emma~Bou Hanna, Gauthier Delerce, Gauthier
  Guinet, Georgii Novikov, Guillaume Martin, Himanshu Jaju, Jan Ludziejewski,
  Jean-Hadrien Chabran, Jean-Malo Delignon, Joachim Studnia, Jonas Amar,
  Josselin~Somerville Roberts, Julien Denize, Karan Saxena, Kush Jain, Lingxiao
  Zhao, Louis Martin, Luyu Gao, Lélio~Renard Lavaud, Marie Pellat, Mathilde
  Guillaumin, Mathis Felardos, Maximilian Augustin, Mickaël Seznec, Nikhil
  Raghuraman, Olivier Duchenne, Patricia Wang, Patrick von Platen, Patryk
  Saffer, Paul Jacob, Paul Wambergue, Paula Kurylowicz, Pavankumar~Reddy
  Muddireddy, Philomène Chagniot, Pierre Stock, Pravesh Agrawal, Romain
  Sauvestre, Rémi Delacourt, Sanchit Gandhi, Sandeep Subramanian, Shashwat
  Dalal, Siddharth Gandhi, Soham Ghosh, Srijan Mishra, Sumukh Aithal, Szymon
  Antoniak, Thibault Schueller, Thibaut Lavril, Thomas Robert, Thomas Wang,
  Timothée Lacroix, Valeriia Nemychnikova, Victor Paltz, Virgile Richard,
  Wen-Ding Li, William Marshall, Xuanyu Zhang, and Yunhao Tang.
\newblock Magistral, 2025.
\newblock \url{https://arxiv.org/abs/2506.10910}.

\bibitem[{Modal Team}()]{modal}
{Modal Team}.
\newblock Modal: High-performance ai infrastructure.
\newblock \url{https://modal.com/docs}.
\newblock Accessed 2025-08-18.

\bibitem[Moshkov et~al.(2025)Moshkov, Hanley, Sorokin, Toshniwal, Henkel,
  Schifferer, Du, and Gitman]{omr}
Ivan Moshkov, Darragh Hanley, Ivan Sorokin, Shubham Toshniwal, Christof Henkel,
  Benedikt Schifferer, Wei Du, and Igor Gitman.
\newblock Aimo-2 winning solution: Building state-of-the-art mathematical
  reasoning models with openmathreasoning dataset.
\newblock \emph{arXiv preprint arXiv:2504.16891}, 2025.

\bibitem[Moura and Ullrich(2021)]{moura2021lean}
Leonardo~de Moura and Sebastian Ullrich.
\newblock The lean 4 theorem prover and programming language.
\newblock In Andr{\'e} Platzer and Geoff Sutcliffe, editors, \emph{Automated
  Deduction -- CADE 28}, pages 625--635, Cham, 2021. Springer International
  Publishing.
\newblock ISBN 978-3-030-79876-5.

\bibitem[Muennighoff et~al.(2023)Muennighoff, Rush, Barak, Le~Scao, Tazi,
  Piktus, Pyysalo, Wolf, and Raffel]{muennighoff2023scaling}
Niklas Muennighoff, Alexander Rush, Boaz Barak, Teven Le~Scao, Nouamane Tazi,
  Aleksandra Piktus, Sampo Pyysalo, Thomas Wolf, and Colin~A Raffel.
\newblock Scaling data-constrained language models.
\newblock \emph{Advances in Neural Information Processing Systems},
  36:\penalty0 50358--50376, 2023.

\bibitem[OpenAI(2024)]{aime}
OpenAI.
\newblock Learning to reason with llms, September 2024.
\newblock \url{https://openai.com/index/learning-to-reason-with-llms/}.

\bibitem[OpenAI(2025{\natexlab{a}})]{agarwal2025gptoss}
OpenAI.
\newblock gpt-oss-120b \& gpt-oss-20b model card.
\newblock \emph{arXiv preprint arXiv:2508.10925}, 2025{\natexlab{a}}.

\bibitem[OpenAI(2025{\natexlab{b}})]{o3}
OpenAI.
\newblock Claude 3.7 sonnet and claude code, April 2025{\natexlab{b}}.
\newblock \url{https://openai.com/index/introducing-o3-and-o4-mini/}.

\bibitem[Ouyang et~al.(2022)Ouyang, Wu, Jiang, Almeida, Wainwright, Mishkin,
  Zhang, Agarwal, Slama, Ray, et~al.]{ouyang2022training}
Long Ouyang, Jeffrey Wu, Xu~Jiang, Diogo Almeida, Carroll Wainwright, Pamela
  Mishkin, Chong Zhang, Sandhini Agarwal, Katarina Slama, Alex Ray, et~al.
\newblock Training language models to follow instructions with human feedback.
\newblock \emph{Advances in neural information processing systems},
  35:\penalty0 27730--27744, 2022.

\bibitem[Paliskara and Saroufim(2025)]{kernelbook2025}
Sahan Paliskara and Mark Saroufim.
\newblock Kernelbook, 5 2025.
\newblock \url{https://huggingface.co/datasets/GPUMODE/KernelBook}.

\bibitem[Pan et~al.(2025)Pan, Wang, Neubig, Jaitly, Ji, Suhr, and
  Zhang]{swegym}
Jiayi Pan, Xingyao Wang, Graham Neubig, Navdeep Jaitly, Heng Ji, Alane Suhr,
  and Yizhe Zhang.
\newblock Training software engineering agents and verifiers with swe‑gym.
\newblock In \emph{Proceedings of the 42nd International Conference on Machine
  Learning (ICML 2025)}, 2025.
\newblock \url{https://arxiv.org/abs/2412.21139}.
\newblock arXiv:2412.21139, accepted at ICML 2025.

\bibitem[Piche et~al.(2025)Piche, Pardinas, Kamalloo, and
  Bahdanau]{Piche2025-nj}
Alex Piche, Rafael Pardinas, Ehsan Kamalloo, and Dzmitry Bahdanau.
\newblock {Pipeline RL: fast LLM agent training}, 2025.
\newblock \url{https://huggingface.co/blog/ServiceNow/pipelinerl}.

\bibitem[Radford et~al.(2018)Radford, Narasimhan, Salimans, and
  Sutskever]{radford2018gpt}
Alec Radford, Karthik Narasimhan, Tim Salimans, and Ilya Sutskever.
\newblock Improving language understanding by generative pre-training.
\newblock Technical report, OpenAI, 2018.
\newblock
  \url{https://cdn.openai.com/research-covers/language-unsupervised/language_understanding_paper.pdf}.

\bibitem[Rastogi et~al.(2025)Rastogi, Jiang, Lo, Berrada, Lample, Rute,
  Barmentlo, Yadav, Khandelwal, Chandu, et~al.]{rastogi2025magistral}
Abhinav Rastogi, Albert~Q Jiang, Andy Lo, Gabrielle Berrada, Guillaume Lample,
  Jason Rute, Joep Barmentlo, Karmesh Yadav, Kartik Khandelwal, Khyathi~Raghavi
  Chandu, et~al.
\newblock Magistral.
\newblock \emph{arXiv preprint arXiv:2506.10910}, 2025.

\bibitem[Rein et~al.(2023)Rein, Hou, Stickland, Petty, Pang, Dirani, Michael,
  and Bowman]{DBLP:journals/corr/abs-2311-12022}
David Rein, Betty~Li Hou, Asa~Cooper Stickland, Jackson Petty, Richard~Yuanzhe
  Pang, Julien Dirani, Julian Michael, and Samuel~R. Bowman.
\newblock {GPQA:} {A} graduate-level google-proof q{\&}a benchmark.
\newblock \emph{CoRR}, abs/2311.12022, 2023.
\newblock \doi{10.48550/ARXIV.2311.12022}.
\newblock \url{https://doi.org/10.48550/arXiv.2311.12022}.

\bibitem[Roziere et~al.(2023)Roziere, Gehring, Gloeckle, Sootla, Gat, Tan, Adi,
  Liu, Sauvestre, Remez, et~al.]{roziere2023code}
Baptiste Roziere, Jonas Gehring, Fabian Gloeckle, Sten Sootla, Itai Gat,
  Xiaoqing~Ellen Tan, Yossi Adi, Jingyu Liu, Romain Sauvestre, Tal Remez,
  et~al.
\newblock Code llama: Open foundation models for code.
\newblock \emph{arXiv preprint arXiv:2308.12950}, 2023.

\bibitem[Saavedra et~al.(2024)Saavedra, Silva, and Monperrus]{saavedra2024}
Nuno Saavedra, Andr\'{e} Silva, and Martin Monperrus.
\newblock Gitbug-actions: Building reproducible bug-fix benchmarks with github
  actions.
\newblock In \emph{Proceedings of the 2024 IEEE/ACM 46th International
  Conference on Software Engineering: Companion Proceedings}, ICSE-Companion
  '24, page 1–5, New York, NY, USA, 2024. Association for Computing
  Machinery.
\newblock ISBN 9798400705021.
\newblock \doi{10.1145/3639478.3640023}.
\newblock \url{https://doi.org/10.1145/3639478.3640023}.

\bibitem[Sakaguchi et~al.(2020)Sakaguchi, Bras, Bhagavatula, and
  Choi]{DBLP:conf/aaai/SakaguchiBBC20}
Keisuke Sakaguchi, Ronan~Le Bras, Chandra Bhagavatula, and Yejin Choi.
\newblock Winogrande: An adversarial winograd schema challenge at scale.
\newblock In \emph{The Thirty-Fourth {AAAI} Conference on Artificial
  Intelligence, {AAAI} 2020, The Thirty-Second Innovative Applications of
  Artificial Intelligence Conference, {IAAI} 2020, The Tenth {AAAI} Symposium
  on Educational Advances in Artificial Intelligence, {EAAI} 2020, New York,
  NY, USA, February 7-12, 2020}, pages 8732--8740. {AAAI} Press, 2020.
\newblock \doi{10.1609/AAAI.V34I05.6399}.
\newblock \url{https://doi.org/10.1609/aaai.v34i05.6399}.

\bibitem[Schulman(2020)]{Schulman2020kl}
John Schulman.
\newblock Approximating kl divergence, 2020.
\newblock \url{https://joschu.net/blog/kl-approx.html}.

\bibitem[Schulman et~al.(2017)Schulman, Wolski, Dhariwal, Radford, and
  Klimov]{schulman2017proximal}
John Schulman, Filip Wolski, Prafulla Dhariwal, Alec Radford, and Oleg Klimov.
\newblock Proximal policy optimization algorithms.
\newblock \emph{arXiv preprint arXiv:1707.06347}, 2017.

\bibitem[Shao et~al.(2024)Shao, Wang, Zhu, Xu, Song, Bi, Zhang, Zhang, Li, Wu,
  and Guo]{shao2024deepseekmathpushinglimitsmathematical}
Zhihong Shao, Peiyi Wang, Qihao Zhu, Runxin Xu, Junxiao Song, Xiao Bi, Haowei
  Zhang, Mingchuan Zhang, Y.~K. Li, Y.~Wu, and Daya Guo.
\newblock Deepseekmath: Pushing the limits of mathematical reasoning in open
  language models, 2024.
\newblock \url{https://arxiv.org/abs/2402.03300}.

\bibitem[Shazeer(2020)]{shazeer2020gluvariantsimprovetransformer}
Noam Shazeer.
\newblock Glu variants improve transformer, 2020.
\newblock \url{https://arxiv.org/abs/2002.05202}.

\bibitem[Su et~al.(2021)Su, Lu, Pan, Murtadha, Wen, and Liu]{su2021rope}
Jianlin Su, Yu~Lu, Shengfeng Pan, Ahmed Murtadha, Bo~Wen, and Yunfeng Liu.
\newblock Roformer: Enhanced transformer with rotary position embedding, 2021.
\newblock \url{https://arxiv.org/abs/2104.09864}.

\bibitem[Synnaeve et~al.(2019)Synnaeve, Gehring, Lin, Haziza, Usunier,
  Rothermel, Mella, Ju, Carion, Gustafson, et~al.]{synnaeve2019growing}
Gabriel Synnaeve, Jonas Gehring, Zeming Lin, Daniel Haziza, Nicolas Usunier,
  Danielle Rothermel, Vegard Mella, Da~Ju, Nicolas Carion, Laura Gustafson,
  et~al.
\newblock Growing up together: Structured exploration for large action spaces.
\newblock 2019.

\bibitem[Talmor et~al.(2019)Talmor, Herzig, Lourie, and
  Berant]{DBLP:conf/naacl/TalmorHLB19}
Alon Talmor, Jonathan Herzig, Nicholas Lourie, and Jonathan Berant.
\newblock Commonsenseqa: {A} question answering challenge targeting commonsense
  knowledge.
\newblock In Jill Burstein, Christy Doran, and Thamar Solorio, editors,
  \emph{Proceedings of the 2019 Conference of the North American Chapter of the
  Association for Computational Linguistics: Human Language Technologies,
  {NAACL-HLT} 2019, Minneapolis, MN, USA, June 2-7, 2019, Volume 1 (Long and
  Short Papers)}, pages 4149--4158. Association for Computational Linguistics,
  2019.
\newblock \doi{10.18653/V1/N19-1421}.
\newblock \url{https://doi.org/10.18653/v1/n19-1421}.

\bibitem[Tang et~al.(2025)Tang, Zheng, Synnaeve, and
  Munos]{DBLP:journals/corr/abs-2503-19595}
Yunhao Tang, Kunhao Zheng, Gabriel Synnaeve, and R{\'{e}}mi Munos.
\newblock Optimizing language models for inference time objectives using
  reinforcement learning.
\newblock \emph{CoRR}, abs/2503.19595, 2025.
\newblock \doi{10.48550/ARXIV.2503.19595}.
\newblock \url{https://doi.org/10.48550/arXiv.2503.19595}.

\bibitem[{The Terminal-Bench Team}(2025)]{tbench_2025}
{The Terminal-Bench Team}.
\newblock Terminal-bench: A benchmark for ai agents in terminal environments,
  Apr 2025.
\newblock \url{https://github.com/laude-institute/terminal-bench}.

\bibitem[Vanegue et~al.(2025)Vanegue, Villard, O'Hearn, and
  Raad]{vanegue2025nonterminationproving100million}
Julien Vanegue, Jules Villard, Peter O'Hearn, and Azalea Raad.
\newblock Non-termination proving: 100 million loc and beyond, 2025.
\newblock \url{https://arxiv.org/abs/2509.05293}.

\bibitem[Vaswani et~al.(2017)Vaswani, Shazeer, Parmar, Uszkoreit, Jones, Gomez,
  Kaiser, and Polosukhin]{vaswani2017attention}
Ashish Vaswani, Noam Shazeer, Niki Parmar, Jakob Uszkoreit, Llion Jones,
  Aidan~N Gomez, {\L}ukasz Kaiser, and Illia Polosukhin.
\newblock Attention is all you need.
\newblock \emph{Advances in neural information processing systems}, 30, 2017.

\bibitem[Wang et~al.(2022{\natexlab{a}})Wang, Wei, Sabne, Davis, Ilbeyi,
  Hechtman, Chen, Murthy, Maggioni, Zhang, et~al.]{wang2022overlap}
Shibo Wang, Jinliang Wei, Amit Sabne, Andy Davis, Berkin Ilbeyi, Blake
  Hechtman, Dehao Chen, Karthik~Srinivasa Murthy, Marcello Maggioni, Qiao
  Zhang, et~al.
\newblock Overlap communication with dependent computation via decomposition in
  large deep learning models.
\newblock In \emph{Proceedings of the 28th ACM International Conference on
  Architectural Support for Programming Languages and Operating Systems, Volume
  1}, pages 93--106, 2022{\natexlab{a}}.

\bibitem[Wang et~al.(2025)Wang, Li, Song, Xu, Tang, Zhuge, Pan, Song, Li,
  Singh, Tran, Li, Ma, Zheng, Qian, Shao, Muennighoff, Zhang, Hui, Lin,
  Brennan, Peng, Ji, and Neubig]{wang2025openhands}
Xingyao Wang, Boxuan Li, Yufan Song, Frank~F. Xu, Xianu Tang, Mingchen Zhuge,
  Jiayi Pan, Yueqi Song, Bowen Li, Jaskirat Singh, Hoang~H. Tran, Fuqiang Li,
  Ren Ma, Mingzhang Zheng, Bill Qian, Yanjun Shao, Niklas Muennighoff, Yizhe
  Zhang, Binyuan Hui, Junyang Lin, Robert Brennan, Hao Peng, Heng Ji, and
  Graham Neubig.
\newblock Openhands: An open platform for {AI} software developers as
  generalist agents.
\newblock In \emph{The Thirteenth International Conference on Learning
  Representations}, 2025.
\newblock \url{https://openreview.net/forum?id=OJd3ayDDoF}.

\bibitem[Wang et~al.(2022{\natexlab{b}})Wang, Wei, Schuurmans, Le, Chi, Narang,
  Chowdhery, and Zhou]{self_cons}
Xuezhi Wang, Jason Wei, Dale Schuurmans, Quoc Le, Ed~Chi, Sharan Narang,
  Aakanksha Chowdhery, and Denny Zhou.
\newblock Self-consistency improves chain of thought reasoning in language
  models.
\newblock \emph{arXiv preprint arXiv:2203.11171}, 2022{\natexlab{b}}.

\bibitem[Wei et~al.(2023)Wei, Wang, Schuurmans, Bosma, Ichter, Xia, Chi, Le,
  and Zhou]{wei2023chainofthoughtpromptingelicitsreasoning}
Jason Wei, Xuezhi Wang, Dale Schuurmans, Maarten Bosma, Brian Ichter, Fei Xia,
  Ed~Chi, Quoc Le, and Denny Zhou.
\newblock Chain-of-thought prompting elicits reasoning in large language
  models, 2023.
\newblock \url{https://arxiv.org/abs/2201.11903}.

\bibitem[Wei et~al.(2024)Wei, Karina, Chung, Jiao, Papay, Glaese, Schulman, and
  Fedus]{DBLP:journals/corr/abs-2411-04368}
Jason Wei, Nguyen Karina, Hyung~Won Chung, Yunxin~Joy Jiao, Spencer Papay,
  Amelia Glaese, John Schulman, and William Fedus.
\newblock Measuring short-form factuality in large language models.
\newblock \emph{CoRR}, abs/2411.04368, 2024.
\newblock \doi{10.48550/ARXIV.2411.04368}.
\newblock \url{https://doi.org/10.48550/arXiv.2411.04368}.

\bibitem[Wei et~al.(2025)Wei, Duchenne, Copet, Carbonneaux, Zhang, Fried,
  Synnaeve, Singh, and Wang]{wei2025swerl}
Yuxiang Wei, Olivier Duchenne, Jade Copet, Quentin Carbonneaux, Lingming Zhang,
  Daniel Fried, Gabriel Synnaeve, Rishabh Singh, and Sida~I. Wang.
\newblock Swe-rl: Advancing llm reasoning via reinforcement learning on open
  software evolution.
\newblock \emph{arXiv preprint arXiv:2502.18449}, 2025.

\bibitem[Xia et~al.(2024)Xia, Deng, Dunn, and Zhang]{agentless}
Chunqiu~Steven Xia, Yinlin Deng, Soren Dunn, and Lingming Zhang.
\newblock Agentless: Demystifying llm-based software engineering agents.
\newblock \emph{arXiv preprint}, 2024.

\bibitem[Xiong et~al.(2023)Xiong, Liu, Molybog, Zhang, Bhargava, Hou, Martin,
  Rungta, Sankararaman, Oguz, et~al.]{xiong2023effective}
Wenhan Xiong, Jingyu Liu, Igor Molybog, Hejia Zhang, Prajjwal Bhargava, Rui
  Hou, Louis Martin, Rashi Rungta, Karthik~Abinav Sankararaman, Barlas Oguz,
  et~al.
\newblock Effective long-context scaling of foundation models.
\newblock \emph{arXiv preprint arXiv:2309.16039}, 2023.

\bibitem[Yang et~al.(2025{\natexlab{a}})Yang, Li, Yang, Zhang, Hui, Zheng, Yu,
  Gao, Huang, Lv, et~al.]{yang2025qwen3}
An~Yang, Anfeng Li, Baosong Yang, Beichen Zhang, Binyuan Hui, Bo~Zheng, Bowen
  Yu, Chang Gao, Chengen Huang, Chenxu Lv, et~al.
\newblock Qwen3 technical report.
\newblock \emph{arXiv preprint arXiv:2505.09388}, 2025{\natexlab{a}}.

\bibitem[Yang et~al.(2024)Yang, Jimenez, Wettig, Lieret, Yao, Narasimhan, and
  Press]{yang2024swe}
John Yang, Carlos~E Jimenez, Alexander Wettig, Kilian Lieret, Shunyu Yao,
  Karthik Narasimhan, and Ofir Press.
\newblock Swe-agent: Agent-computer interfaces enable automated software
  engineering.
\newblock \emph{Advances in Neural Information Processing Systems},
  37:\penalty0 50528--50652, 2024.

\bibitem[Yang et~al.(2025{\natexlab{b}})Yang, Leret, Jimenez, Wettig, Khandpur,
  Zhang, Hui, Press, Schmidt, and Yang]{yang2025swe}
John Yang, Kilian Leret, Carlos~E Jimenez, Alexander Wettig, Kabir Khandpur,
  Yanzhe Zhang, Binyuan Hui, Ofir Press, Ludwig Schmidt, and Diyi Yang.
\newblock Swe-smith: Scaling data for software engineering agents.
\newblock \emph{arXiv preprint arXiv:2504.21798}, 2025{\natexlab{b}}.

\bibitem[Yeverechyahu et~al.(2024)Yeverechyahu, Mayya, and
  Oestreicher-Singer]{yeverechyahu2024impact}
Doron Yeverechyahu, Raveesh Mayya, and Gal Oestreicher-Singer.
\newblock The impact of large language models on open-source innovation:
  Evidence from github copilot.
\newblock \emph{arXiv preprint arXiv:2409.08379}, 2024.

\bibitem[Ying et~al.(2025)Ying, Wu, Geng, Yuan, Lin, and
  Chen]{ying2025leanworkbook}
Huaiyuan Ying, Zijian Wu, Yihan Geng, Zheng Yuan, Dahua Lin, and Kai Chen.
\newblock Lean workbook: A large-scale lean problem set formalized from natural
  language math problems, 2025.
\newblock \url{https://arxiv.org/abs/2406.03847}.

\bibitem[Yu et~al.(2025)Yu, Zhang, Zhu, Yuan, Zuo, Yue, Fan, Liu, Liu, Liu,
  et~al.]{yu2025dapo}
Qiying Yu, Zheng Zhang, Ruofei Zhu, Yufeng Yuan, Xiaochen Zuo, Yu~Yue, Tiantian
  Fan, Gaohong Liu, Lingjun Liu, Xin Liu, et~al.
\newblock Dapo: An open-source llm reinforcement learning system at scale.
\newblock \emph{arXiv preprint arXiv:2503.14476}, 2025.

\bibitem[Yue et~al.(2024)Yue, Madaan, Moskovitz, Strouse, and
  Singh]{DBLP:journals/corr/abs-2412-08819}
Albert~S. Yue, Lovish Madaan, Ted Moskovitz, DJ~Strouse, and Aaditya~K. Singh.
\newblock {HARP:} {A} challenging human-annotated math reasoning benchmark.
\newblock \emph{CoRR}, abs/2412.08819, 2024.
\newblock \doi{10.48550/ARXIV.2412.08819}.
\newblock \url{https://doi.org/10.48550/arXiv.2412.08819}.

\bibitem[Zellers et~al.(2019)Zellers, Holtzman, Bisk, Farhadi, and
  Choi]{DBLP:conf/acl/ZellersHBFC19}
Rowan Zellers, Ari Holtzman, Yonatan Bisk, Ali Farhadi, and Yejin Choi.
\newblock Hellaswag: Can a machine really finish your sentence?
\newblock In Anna Korhonen, David~R. Traum, and Llu{\'{\i}}s M{\`{a}}rquez,
  editors, \emph{Proceedings of the 57th Conference of the Association for
  Computational Linguistics, {ACL} 2019, Florence, Italy, July 28- August 2,
  2019, Volume 1: Long Papers}, pages 4791--4800. Association for Computational
  Linguistics, 2019.
\newblock \doi{10.18653/V1/P19-1472}.
\newblock \url{https://doi.org/10.18653/v1/p19-1472}.

\bibitem[Zhang and Sennrich(2019)]{rmsnorm}
Biao Zhang and Rico Sennrich.
\newblock \emph{Root mean square layer normalization}.
\newblock Curran Associates Inc., Red Hook, NY, USA, 2019.

\bibitem[Zhang et~al.()Zhang, Defferrard, Rainone, and Memisevic]{Zhang2025-au}
David~W Zhang, Michaël Defferrard, Corrado Rainone, and Roland Memisevic.
\newblock {Grounding code understanding in step-by-step execution}.
\newblock \url{https://openreview.net/forum?id=MUr7Fl93QS}.

\bibitem[Zheng et~al.(2022)Zheng, Han, and Polu]{zheng2022minif2f}
Kunhao Zheng, Jesse~Michael Han, and Stanislas Polu.
\newblock Minif2f: a cross-system benchmark for formal olympiad-level
  mathematics, 2022.
\newblock \url{https://arxiv.org/abs/2109.00110}.

\bibitem[Zheng et~al.(2023)Zheng, Chiang, Sheng, Zhuang, Wu, Zhuang, Lin, Li,
  Li, Xing, Zhang, Gonzalez, and Stoica]{DBLP:conf/nips/ZhengC00WZL0LXZ23}
Lianmin Zheng, Wei{-}Lin Chiang, Ying Sheng, Siyuan Zhuang, Zhanghao Wu,
  Yonghao Zhuang, Zi~Lin, Zhuohan Li, Dacheng Li, Eric~P. Xing, Hao Zhang,
  Joseph~E. Gonzalez, and Ion Stoica.
\newblock Judging llm-as-a-judge with mt-bench and chatbot arena.
\newblock In Alice Oh, Tristan Naumann, Amir Globerson, Kate Saenko, Moritz
  Hardt, and Sergey Levine, editors, \emph{Advances in Neural Information
  Processing Systems 36: Annual Conference on Neural Information Processing
  Systems 2023, NeurIPS 2023, New Orleans, LA, USA, December 10 - 16, 2023},
  2023.
\newblock
  \url{http://papers.nips.cc/paper\_files/paper/2023/hash/91f18a1287b398d378ef22505bf41832-Abstract-Datasets\_and\_Benchmarks.html}.

\end{thebibliography}

\newpage
\beginappendix

\renewcommand\thefigure{\thesection.\arabic{figure}}    
\section{Acknowledgments}
The authors thank Ariel Stolerman, Ayelet Regev Dabah, Dani Shames, Tamir Meyer and Nadav Azaria for support in building executable repository images at scale; Jeff Yang, Yonatan Komornik and Tarun Anand for support in curating GitHub PR and Issue metadata; Qian Liang, Meng Zhang, Hanwen Zha, Ananya Saxena, Emily Dinan, Melanie Kambadur for the support in data preparation; Yining Yang, Sten Sootla, Chris Waterson and Michael Jiang for support in the development of RepoAgent and additional repository images; 
Eslam Elnikety, Jamie Cahill, Christine Wang, Don Landrum, Sadman Fahmid, Andrew Hamiel, Ned Newton, Andrii Golovei, Rashmi Narasimha, Zack Leman, Mehrdad Mahdavi, Leon Yang, Joshua Fink, Sargun Dillon, Jeff Hanson and Zach Wentz for the internal sandboxing platform and the code execution and Docker execution services built atop it, enabling secure and massively parallel execution of untrusted code; Mathurin Videau, Leonid Shamis, Jeremy Reizenstein, Maria Lomeli, Lucca Bertoncini, Vivien Cabannes, Charles Arnal and Pascal Kesseli for their contributions to the CWM research codebase and training and evaluation infrastructure; Julien Vanegue for advice on practical aspects of the halting problem; 
Daniel Fried and Rishabh Singh for support in designing and developing Agentic SWE RL;
the Modal team -- especially Jonathon Belotti, Matthew Saltz, Colin Weld, Peyton Walters, Deven Navani, Michael Waskom, Advay Pal, Akshat Bubna, Alec Powell, Lucy Zhang, and Eric Zhang -- for extensive support with remote execution, infrastructure, and platform stability;
Lovish Madaan, Binh Tang, Viktor Kerkez, Rishabh Agarwal, Alan Schelten, Xuewei Wang and Jeremy Fu for support with mathematical expression comparison code.

\section{\cwm Examples}
\label{sec:add_examples}

Extending \cref{sec:examples}, we here present additional examples of using \cwm for SWE reasoning, trace prediction, and a combination of the two.

\newpara{Reasoning agent.}  \Cref{fig:example:swerl-bash} shows an example of \cwm solving an \sbv problem in a bash-only environment, which is more challenging than environments that provide dedicated tools for common tasks such as file editing. In this example, the model makes incorrect edits in the initial turns but realizes its error and restores the original file state using \texttt{git checkout}, followed by producing a correct edit with \texttt{sed}.

\Cref{fig:example:swerl-submit} demonstrates that \cwm can leverage test execution to verify patch correctness before submission. In this specific example, the agent makes sure that the changes it makes do not break any existing functionality. Only after this verification, the agent submits the patch and generates a summary.

Lastly, \cref{fig:example:swerl-edit} shows the default \swerl setting where \cwm is paired with the \texttt{edit} tool. In this example, \cwm performs extensive reasoning before making the edit. The \texttt{edit} tool then provides agent-friendly feedback showing the surrounding code after the change.

\newpara{Python execution trace prediction.} \Cref{lst:trace_cwm_correct_qwen_wrong,lst:trace_cwm_wrong_qwen_correct} showcase Python execution trace prediction at inference time and compare it to reasoning about program execution in natural language.
For \cref{lst:trace_cwm_correct_qwen_wrong}, a Python list is modified while iterating over it. 
In execution trace prediction mode, the model tracks all list modifications and predicts the output correctly.
With natural language reasoning, the model fails to predict the correct return value -- even though it appears to recognize the list modification during reasoning.
Conversely, \cref{lst:trace_cwm_wrong_qwen_correct} presents an example requiring the evaluation of a complex Python statement.
Execution trace prediction fails to correctly predict the outcome of the statement in a single prediction step, but natural language reasoning breaks down the complex statement into simpler expressions and then combines those into a  correct result.
We believe that combining the groundedness of trace prediction with the flexibility of natural language reasoning makes for interesting future research.

\Cref{lst:trace_debugger_demo} demonstrates how \cwm's execution trace prediction capabilities allow it to function as a neural Python debugger. 
We think that equipping \cwm with debugging capabilities that are not available with traditional debuggers, such as skipping loops in constant time, jumping to arbitrary lines of code, or predicting inputs to reach arbitrary states is highly interesting future work.

\Cref{fig:trace_prompts} shows how we execution trace prediction for CruxEval output prediction in our experiments in \cref{sec:ex_tr_pred}.

\Cref{lst:trace_cwm_unseen} demonstrates how \cwm's Python execution trace prediction capability can be used for code generation. 
By specifying a set of \texttt{asserts} consistent with the desired behavior and simply pretending to \texttt{import} the desired function, without actually giving a function definition, \cwm starts to generate actions consistent with the desired function.
It is possible that the model acquired this capability of jointly tracing and generating code because for some tracing data we do not include the source context of third-party libraries. 

We expand on this in \cref{lst:trace_cwm_unseen_reasoning}, the example of mixing tracing and code generation discussed earlier in Section \ref{sec:examples}.
Future work could build on \cwm's capabilities here and explore how execution trace prediction can be used to improve code generation.

\newpara{Program Termination.} \Cref{lst:trace_cwm_halteval_correct} llustrates termination reasoning, whereby CWM considers several concrete inputs before generalizing to the conclusion of terminating on all inputs.

\begin{figure}[htbp!]
{
\lstset{aboveskip=0pt, belowskip=5pt}
\begin{lstlisting}[style=cwm_think]
The issue persists due to an indentation error. I need to correct the indentation in the sky_coordinate.py file. However, since the 'edit' tool is not available, I'll try a different approach.

Let me check the current state of the __getattr__ method in sky_coordinate.py.
\end{lstlisting}%
\begin{lstlisting}[style=cwm_act]
<tool: bash>
sed -n 890,910p /testbed/astropy/coordinates/sky_coordinate.py
</tool>
\end{lstlisting}%
\begin{lstlisting}[style=cwm_obs]
<output>
                return getattr(self._sky_coord_frame, attr)

            # Try to interpret as a new frame for transforming.
            frame_cls = frame_transform_graph.lookup_name(attr)
            if frame_cls is not None and self.frame.is_transformable_to(frame_cls):
                return self.transform_to(attr)

        # Fail
        return self.__getattribute__(attr)
            f"'{self.__class__.__name__}' object has no attribute '{attr}'"
        )

    def __setattr__(self, attr, val):
        # This is to make anything available through __getattr__ immutable
        if "_sky_coord_frame" in self.__dict__:
            if self._is_name(attr):
                raise AttributeError(f"'{attr}' is immutable")

            if not attr.startswith("_") and hasattr(self._sky_coord_frame, attr):
                setattr(self._sky_coord_frame, attr, val)
                return
</output>
<budget>
* Remaining turns: 82
* Remaining tokens: 85000
</budget>
\end{lstlisting}%
\begin{lstlisting}[style=cwm_think]
The issue is that the line `return self.__getattribute__(attr)` is not properly replacing the original `raise AttributeError` line, and there's an extra line with `f"'{self.__class__.__name__}' object has no attribute '{attr}'"` that's causing a syntax error.

I need to correctly replace the entire block. Let me try to fix this by restoring the original file and applying the change again carefully.
\end{lstlisting}%
\begin{lstlisting}[style=cwm_act]
<tool: bash>
git checkout /testbed/astropy/coordinates/sky_coordinate.py
</tool>
\end{lstlisting}%
\begin{lstlisting}[style=cwm_obs]
<output>
Updated 1 path from the index
</output>
<budget>
* Remaining turns: 81
* Remaining tokens: 84837
</budget>
\end{lstlisting}%
\begin{lstlisting}[style=cwm_think]
Now that the sky_coordinate.py file has been restored, I'll apply the correct change using sed, making sure to replace the entire raise AttributeError block with the correct line.

The original code at line 898 is:

        # Fail
        raise AttributeError(
            f"'{self.__class__.__name__}' object has no attribute '{attr}'"
        )

I need to replace this with:

        # Fail
        return self.__getattribute__(attr)

Let me use sed to do this correctly.
\end{lstlisting}%
\begin{lstlisting}[style=cwm_act]
<tool: bash>
sed -i '898,900c\        # Fail\n        return self.__getattribute__(attr)' /testbed/astropy/coordinates/sky_coordinate.py
</tool>
\end{lstlisting}
}
\caption{Example intermediate turns in \swerl (bash-only). The agent learns to correct indentation mistakes from wrong edits by restoring the original file state using \texttt{git checkout}, followed by applying a correct edit using \texttt{sed}.}
\label{fig:example:swerl-bash}
\end{figure}

\begin{figure}[htbp!]
{
\lstset{aboveskip=0pt, belowskip=5pt}
\begin{lstlisting}[style=cwm_think]
The code ran without errors, and `model.dual_coef_` was printed, which should be an empty CSR matrix. However, the output is empty, which suggests that `model.dual_coef_` is indeed an empty matrix. 

Now, I need to verify that this fix doesn't break any other functionality. To do that, I'll run the tests for the SVM module in sklearn to ensure that all tests pass.
\end{lstlisting}%
\begin{lstlisting}[style=cwm_act]
<tool: bash>
python -m pytest sklearn/svm/tests/
</tool>
\end{lstlisting}%
\begin{lstlisting}[style=cwm_obs]
============================= test session starts ==============================
platform linux -- Python 3.6.13, pytest-6.2.4, py-1.11.0, pluggy-0.13.1
rootdir: /testbed, configfile: setup.cfg
collected 135 items

sklearn/svm/tests/test_bounds.py ...................                     [ 14%
sklearn/svm/tests/test_sparse.py ...............................         [ 37%
sklearn/svm/tests/test_svm.py .......................................... [ 68%
...........................................                              [100%

======================= 135 passed, 15 warnings in 4.31s =======================
\end{lstlisting}%
\begin{lstlisting}[style=cwm_think]
All tests have passed, which suggests that the fix is correct and doesn't break any existing functionality. Now, I need to create a patch file that includes the changes made to `sklearn/svm/base.py`.
\end{lstlisting}%
\begin{lstlisting}[style=cwm_act]
<tool: bash>
git diff sklearn/svm/base.py > svm_patch.patch
</tool>
\end{lstlisting}%
\begin{lstlisting}[style=cwm_obs]
(No output)
\end{lstlisting}%
\begin{lstlisting}[style=cwm_think]
The patch file has been created. Now, I'll submit it as the final solution.
\end{lstlisting}%
\begin{lstlisting}[style=cwm_act]
<tool: submit>
./svm_patch.patch
</tool>
\end{lstlisting}%
\begin{lstlisting}[style=cwm_obs]
The following patch content is marked as your final submission:

<submission>
diff --git a/sklearn/svm/base.py b/sklearn/svm/base.py
index 64cebe837..67dd8e38f 100644
--- a/sklearn/svm/base.py
+++ b/sklearn/svm/base.py
@@ -287,8 +287,11 @@ class BaseLibSVM(BaseEstimator, metaclass=ABCMeta):
         n_SV = self.support_vectors_.shape[0]
 
         dual_coef_indices = np.tile(np.arange(n_SV), n_class)
-        dual_coef_indptr = np.arange(0, dual_coef_indices.size + 1,
-                                     dual_coef_indices.size / n_class)
+        if dual_coef_indices.size == 0:
+            dual_coef_indptr = np.array([0, 0], dtype=np.int32)
+        else:
+            dual_coef_indptr = np.arange(0, dual_coef_indices.size + 1,
+                                         dual_coef_indices.size / n_class)
         self.dual_coef_ = sp.csr_matrix(
             (dual_coef_data, dual_coef_indices, dual_coef_indptr),
             (n_class, n_SV))

</submission>

Review the patch content and ensure it correctly resolves the issue. If necessary, you can continue refining the patch and submit an updated version.
\end{lstlisting}%
\begin{lstlisting}[style=cwm_think]
The patch looks good. It adds a check for when `dual_coef_indices` is empty and handles it by setting `dual_coef_indptr` to `[0, 0]`, which should correctly create an empty CSR matrix for `self.dual_coef_`. This should resolve the ZeroDivisionError issue when `support_vectors_` is empty.

I'm confident that this patch is correct and should fix the problem without introducing any regressions, as all tests passed after the fix.
\end{lstlisting}%
\begin{lstlisting}[style=cwm_think]
## Issue Summary

The issue was a ZeroDivisionError in the `_sparse_fit` method of the SVM class when using sparse data and the `support_vectors_` attribute was empty. The error occurred because of a division by zero when calculating the `dual_coef_indptr` for the CSR matrix of dual coefficients.

## Investigation and Reproduction...
\end{lstlisting}%
}
\caption{Example final turns in \swerl, with details like budget notices omitted for brevity. The agent verifies that the patch does not break existing functionality, submits the patch, and produces a summary.}
\label{fig:example:swerl-submit}
\end{figure}

\begin{figure}[htbp!]
{
\lstset{aboveskip=0pt, belowskip=5pt}
\begin{lstlisting}[style=cwm_think]
Okay, so the grep command shows that in django/db/models/fields/json.py, there are several instances of json.dumps being used. The ones in lines 84 and 95 are the ones in the get_prep_value and validate methods, respectively. These are the ones that might need to be modified to include ensure_ascii=False to fix the Unicode display issue in the admin.

However, wait a minute. The get_prep_value method is used to prepare the value for the database, but the display_in the admin uses this method as well. If we modify get_prep_value to include ensure_ascii=False, that could affect how the data is stored in the database, but JSON fields should handle Unicode correctly regardless. However, ... But let me think again. The display_for_field function in django/contrib/admin/utils.py calls get_prep_value on the JSONField instance. If get_prep_value returns a string with ensure_ascii=False, then the displayed JSON data will show Unicode characters correctly.

So, modifying the get_prep_value method to include ensure_ascii=False in the json.dumps call should fix the issue. Similarly, the validate method's json.dumps call might not need to change, but to be consistent, maybe it's better to include ensure_ascii=False there as well, although it's less critical for display purposes. Additionally, the lines 122, 383, and 416 use json.dumps, but these seem to be for different purposes (compiling JSON paths, etc.). However, if these are part of the serialization process for JSONField values, they might also need to include ensure_ascii=False to ensure consistent behavior. But without more context, it's harder to say. The main issue is the admin display, which is handled by get_prep_value.

Let me check the pull request again (https://github.com/adamchainz/django-mysql/pull/714) to see what changes were made. Since I can't access the link, I'll have to rely on the issue description, which mentions that the pull request fixed the problem by modifying how JSON data is serialized. Assuming that the fix involves adding ensure_ascii=False to the json.dumps calls in the JSONField class, I should modify the get_prep_value and possibly other relevant methods. Another thing to consider: the forms.JSONField in django/forms/fields.py might also need similar changes, especially in how it prepares values for display. Let me check that file as well. Looking at django/forms/fields.py, the JSONField class has a prepare_value method:

def prepare_value(self, value):
    if value is None:
        return None
    if isinstance(value, str):
        return value
    return self.encoder.dumps(value)

Here, self.encoder.dumps(value) is used. If the encoder is the default JSONEncoder, it would still use ASCII encoding. However, the encoder for JSONField can be customized, but the default might not set ensure_ascii=False. To ensure that the form field displays Unicode characters correctly, maybe this should also be modified to include ensure_ascii=False, but since the encoder is part of the field's configuration, it might be better to handle it there. Alternatively, maybe the display_for_field function in the admin should handle JSONField values differently, but that might be more complex. Given the information, the most straightforward fix seems to be modifying the get_prep_value method in django/db/models/fields/json.py to include ensure_ascii=False in the json.dumps call. This would affect how JSON data is displayed in the admin, as that's where get_prep_value is called.

Let me proceed with that change and see if it resolves the issue.
\end{lstlisting}%
\begin{lstlisting}[style=cwm_act]
<tool: edit>
django/db/models/fields/json.py
<<<<<<< SEARCH
    def get_prep_value(self, value):
        if value is None:
            return value
        return json.dumps(value, cls=self.encoder)
=======
    def get_prep_value(self, value):
        if value is None:
            return value
        return json.dumps(value, cls=self.encoder, ensure_ascii=False)
>>>>>>> REPLACE
</tool>
\end{lstlisting}%
\begin{lstlisting}[style=cwm_obs]
File django/db/models/fields/json.py has been successfully updated at line 81 and now contains 521 lines. The surrounding lines after the edit are shown below:
    73	        try:
    74	            return json.loads(value, cls=self.decoder)
    75	        except json.JSONDecodeError:
    76	            return value
    77	
    78	    def get_internal_type(self):
    79	        return 'JSONField'
    80	
    81	    def get_prep_value(self, value):
    82	        if value is None:
    83	            return value
    84	        return json.dumps(value, cls=self.encoder, ensure_ascii=False)
    85	
    86	    def get_transform(self, name):
    87	        transform = super().get_transform(name)
    88	        if transform:
    89	            return transform
    90	        return KeyTransformFactory(name)
    91	
    92	    def validate(self, value, model_instance):
Review the changes and make sure they are as expected.
\end{lstlisting}%
}
\caption{Example of the \cwm \swerl agent engaging in long reasoning before modifying the file with the \texttt{edit} tool. For conciseness, some details (e.g., budget notices) are omitted and reasoning paragraphs are merged.}
\label{fig:example:swerl-edit}
\end{figure}

\begin{figure}[p]
  \centering
  \begin{subfigure}{0.37\textwidth}
    \lstset{style=simple_small}
    \begin{lstlisting}[language=Python]
<|begin_of_text|><|trace_context_start|>
def f(n):
    p = ''
    if n%
        p+='sn'
    else:
        return n*n
    for x in range(1, n+1):
        if x%
            p+='to'
        else:
            p+='ts'
return p

def main():  # << START_OF_TRACE
    return f(1)
<|frame_sep|><|call_sep|>{}
<|action_sep|>def main():
<|frame_sep|>(*@\textcolor{cwm_string_color_bright}{\texttt{<|return\_sep|>}}@*)
----END OF PROMPT----
<|action_sep|>    return f(1)
<|arg_sep|>"'snts'"<|frame_sep|>
    \end{lstlisting}
    \caption{Single-step execution trace prediction.}
  \end{subfigure}
  \hfill
  \begin{subfigure}{0.6\textwidth}
    \lstset{style=simple_small}
    \begin{lstlisting}[language=Python]
<|begin_of_text|><|trace_context_start|>
def f(n):
    p = ''
    if n%
        p+='sn'
    else:
        return n*n
    for x in range(1, n+1):
        if x%
            p+='to'
        else:
            p+='ts'
return p

def main():  # << START_OF_TRACE
    return f(1)
<|frame_sep|><|call_sep|>{}<|action_sep|>def main():
<|frame_sep|>
----END OF PROMPT----
<|line_sep|>{}<|action_sep|>    return f(1)
<|frame_sep|><|call_sep|>{"n": "1"}<|action_sep|>def f(n):
<|frame_sep|><|line_sep|>{"n": ".."}<|action_sep|>    p = ''
<|frame_sep|><|line_sep|>{"n": "..", "p": "''"}
<|action_sep|>    if n%
<|frame_sep|><|line_sep|>{"n": "..", "p": ".."}
<|action_sep|>        p+='sn'
<|frame_sep|><|line_sep|>{"n": "..", "p": "'sn'"}
<|action_sep|>    for x in range(1, n+1):
<|frame_sep|><|line_sep|>{"n": "..", "p": "..", "x": "1"}
<|action_sep|>        if x%
<|frame_sep|><|line_sep|>{"n": "..", "p": "..", "x": ".."}
<|action_sep|>            p+='ts'
<|frame_sep|><|line_sep|>{"n": "..", "p": "'snts'", "x": ".."}
<|action_sep|>    for x in range(1, n+1):
<|frame_sep|><|line_sep|>{"n": "..", "p": "..", "x": ".."}
<|action_sep|>    return p
<|frame_sep|><|return_sep|><|action_sep|>    return p<|arg_sep|>"'snts'"
<|frame_sep|><|return_sep|><|action_sep|>    return f(1)
<|arg_sep|>"'snts'"<|frame_sep|>
    \end{lstlisting}
    \caption{Full, line-by-line, execution trace prediction.}
  \end{subfigure}
  \caption{Python execution trace prediction for CruxEval output prediction in (a) single-step mode and (b) full, line-by-line, mode.  
  The predictions shown are generated by \cwm. \label{fig:trace_prompts}}
\end{figure}

\begin{figure}[htbp!]
  \centering
  \begin{subfigure}{\textwidth}
    \lstset{
    style=simple_small_small, 
    basicstyle=\ttfamily\tiny
    }
    \begin{lstlisting}[language=Python]
<|begin_of_text|><|trace_context_start|>
def f(nums):
    a = 0
    for i in range(len(nums)):
        nums.insert(i, nums[a])
        a += 1
    return nums

def main():  # << START_OF_TRACE
    return f([1, 3, -1, 1, -2, 6])
<|frame_sep|><|call_sep|>{}<|action_sep|>def main():  # << START_OF_TRACE
<|frame_sep|>
---END OF PROMPT---
<|line_sep|>{}<|action_sep|>    return f([1, 3, -1, 1, -2, 6])
<|frame_sep|><|call_sep|>{"nums": "[1, 3, -1, 1, -2, 6]"}<|action_sep|>def f(nums):
<|frame_sep|><|line_sep|>{"nums": ".."}<|action_sep|>    a = 0
<|frame_sep|><|line_sep|>{"nums": "..", "a": "0"}<|action_sep|>    for i in range(len(nums)):
<|frame_sep|><|line_sep|>{"nums": "..", "a": "..", "i": "0"}<|action_sep|>        nums.insert(i, nums[a])
<|frame_sep|><|line_sep|>{"nums": "[1, 1, 3, -1, 1, -2, 6]", "a": "..", "i": ".."}<|action_sep|>        a += 1
<|frame_sep|><|line_sep|>{"nums": "..", "a": "1", "i": ".."}<|action_sep|>    for i in range(len(nums)):
<|frame_sep|><|line_sep|>{"nums": "..", "a": "..", "i": "1"}<|action_sep|>        nums.insert(i, nums[a])
<|frame_sep|><|line_sep|>{"nums": "[1, 1, 1, 3, -1, 1, -2, 6]", "a": "..", "i": ".."}<|action_sep|>        a += 1
<|frame_sep|><|line_sep|>{"nums": "..", "a": "2", "i": ".."}<|action_sep|>    for i in range(len(nums)):
<|frame_sep|><|line_sep|>{"nums": "..", "a": "..", "i": "2"}<|action_sep|>        nums.insert(i, nums[a])
<|frame_sep|><|line_sep|>{"nums": "[1, 1, 1, 1, 3, -1, 1, -2, 6]", "a": "..", "i": ".."}<|action_sep|>        a += 1
<|frame_sep|><|line_sep|>{"nums": "..", "a": "3", "i": ".."}<|action_sep|>    for i in range(len(nums)):
<|frame_sep|><|line_sep|>{"nums": "..", "a": "..", "i": "3"}<|action_sep|>        nums.insert(i, nums[a])
<|frame_sep|><|line_sep|>{"nums": "[1, 1, 1, 1, 1, 3, -1, 1, -2, 6]", "a": "..", "i": ".."}<|action_sep|>        a += 1
<|frame_sep|><|line_sep|>{"nums": "..", "a": "4", "i": ".."}<|action_sep|>    for i in range(len(nums)):
<|frame_sep|><|line_sep|>{"nums": "..", "a": "..", "i": "4"}<|action_sep|>        nums.insert(i, nums[a])
<|frame_sep|><|line_sep|>{"nums": "[1, 1, 1, 1, 1, 1, 3, -1, 1, -2, 6]", "a": "..", "i": ".."}<|action_sep|>        a += 1
<|frame_sep|><|line_sep|>{"nums": "..", "a": "5", "i": ".."}<|action_sep|>    for i in range(len(nums)):
<|frame_sep|><|line_sep|>{"nums": "..", "a": "..", "i": "5"}<|action_sep|>        nums.insert(i, nums[a])
<|frame_sep|><|line_sep|>{"nums": "[1, 1, 1, 1, 1, 1, 1, 3, -1, 1, -2, 6]", "a": "..", "i": ".."}<|action_sep|>        a += 1
<|frame_sep|><|line_sep|>{"nums": "..", "a": "6", "i": ".."}<|action_sep|>    for i in range(len(nums)):
<|frame_sep|><|line_sep|>{"nums": "..", "a": "..", "i": ".."}<|action_sep|>    return nums
<|frame_sep|><|return_sep|><|action_sep|>    return nums<|arg_sep|>"[1, 1, 1, 1, 1, 1, 1, 3, -1, 1, -2, 6]"
<|frame_sep|><|return_sep|><|action_sep|>    return f([1, 3, -1, 1, -2, 6])<|arg_sep|>"[1, 1, 1, 1, 1, 1, 1, 3, -1, 1, -2, 6]"<|frame_sep|>
    \end{lstlisting}
    \vspace{-0.2cm}
    \caption{\cwm Execution Trace Prediction (correct prediction).}
  \end{subfigure}
  \hfill
  \begin{subfigure}{\textwidth}
    \lstset{
    style=simple_small_small,
    basicstyle=\ttfamily\tiny
    }
    \begin{lstlisting}
Given a python code function and an assert statement containing a specific input, provide the assertion with the exact literal output that the function returns with that input. Do not include any mathematical expressions or function calls-only the final literal value. Your response should be solely the assertion, enclosed within [ANSWER] and [/ANSWER] tags.
[SKIPPED EXAMPLE FOR ILLUSTRATION PURPOSES]
Python function:
```
def f(nums):
    a = 0
    for i in range(len(nums)):
        nums.insert(i, nums[a])
        a += 1
    return nums
assert f([1, 3, -1, 1, -2, 6]) == ??
```
----END OF PROMPT---
Okay, let's see. I need to figure out what the function f does when given the list [1, 3, -1, 1, -2, 6]. The function starts by initializing a variable a to 0. Then it loops over the range of the length of nums. But wait, the length of nums is 6 initially. However, inside the loop, it's inserting elements into nums, which will change the length of the list as the loop progresses. That might affect how many times the loop runs.
Let me break it down step by step. The initial list is [1, 3, -1, 1, -2, 6]. The variable a starts at 0. The loop runs for i in range(len(nums)), which is range(6) initially. So i will take values 0 through 5.
First iteration (i=0):
- nums.insert(i, nums[a]) -> insert nums[0] (which is 1) at position 0. The list becomes [1, 1, 3, -1, 1, -2, 6]. Then a increments to 1.
Second iteration (i=1):
- Now, the list has 7 elements. But the loop was set up with range(6), so i is still 1. We insert nums[a] (nums[1] is 1) at position 1. The list becomes [1, 1, 1, 3, -1, 1, -2, 6]. a becomes 2.
Third iteration (i=2):
- Insert nums[2] (which is 1) at position 2. The list is now [1, 1, 1, 1, 3, -1, 1, -2, 6]. a is 3.
Fourth iteration (i=3):
- Insert nums[3] (3) at position 3. The list becomes [1, 1, 1, 1, 3, 3, -1, 1, -2, 6]. a is 4.
Fifth iteration (i=4):
- Insert nums[4] (-1) at position 4. The list is [1, 1, 1, 1, 3, -1, 3, -1, 1, -2, 6]. a is 5.
Sixth iteration (i=5):
- Insert nums[5] (1) at position 5. The list becomes [1, 1, 1, 1, 3, 1, -1, 3, -1, 1, -2, 6]. a is 6.
After these six iterations, the loop ends because i has gone through 0 to 5. The final list is [1, 1, 1, 1, 3, 1, -1, 3, -1, 1, -2, 6]. Wait, but let me double-check each step to make sure I didn't make a mistake.
Wait a minute, when we insert elements, the original elements shift to the right. Let me re-examine each step carefully.
Original list: [1, 3, -1, 1, -2, 6]
First iteration (i=0, a=0):
- Insert nums[0] (1) at position 0. The list becomes [1, 1, 3, -1, 1, -2, 6]. a becomes 1.
Second iteration (i=1, a=1):
- Insert nums[1] (1) at position 1. The list is now [1, 1, 1, 3, -1, 1, -2, 6]. a becomes 2.
Third iteration (i=2, a=2):
- Insert nums[2] (1) at position 2. The list becomes [1, 1, 1, 1, 3, -1, 1, -2, 6]. a becomes 3.
Fourth iteration (i=3, a=3):
- Insert nums[3] (3) at position 3. The list is [1, 1, 1, 1, 3, 3, -1, 1, -2, 6]. a becomes 4.
Fifth iteration (i=4, a=4):
- Insert nums[4] (-1) at position 4. The list becomes [1, 1, 1, 1, 3, -1, 3, -1, 1, -2, 6]. a becomes 5.
Sixth iteration (i=5, a=5):
- Insert nums[5] (1) at position 5. The list is [1, 1, 1, 1, 3, 1, -1, 3, -1, 1, -2, 6]. a becomes 6.
Yes, that seems correct. So the final list returned by the function is [1, 1, 1, 1, 3, 1, -1, 3, -1, 1, -2, 6].
</think>
[ANSWER]
assert f([1, 3, -1, 1, -2, 6]) == [1, 1, 1, 1, 3, 1, -1, 3, -1, 1, -2, 6]
[/ANSWER]
    \end{lstlisting}
    \vspace{-0.2cm}
    \caption{\cwm Reasoning Trace (wrong prediction).}
  \end{subfigure}
  \vspace{-0.6cm}
  \caption{\cwm execution trace prediction (correct) vs. natural language reasoning trace (wrong). In this CruxEval example, a list is modified while iterating over it. In execution trace prediction mode the \cwm model generates the correct answer \texttt{[1, 1, 1, 1, 1, 1, 1, 3, -1, 1, -2, 6]}. During reasoning in natural language, even though the \cwm model recognizes the modification, it still predicts the wrong output.\label{lst:trace_cwm_correct_qwen_wrong}}
\end{figure}

\begin{figure}[htbp!]
  \centering
  \begin{subfigure}{\textwidth}
    \lstset{style=simple_small_small}
    \begin{lstlisting}[language=Python]
<|begin_of_text|><|trace_context_start|>
def f(sequence, value):
    i = max(sequence.index(value) - len(sequence) // 3, 0)
    result = ''
    for j, v in enumerate(sequence[i:]):
        if v == '+':
            result += value
        else:
            result += sequence[i + j]
    return result

def main():  # << START_OF_TRACE
    return f('hosu', 'o')<|frame_sep|><|call_sep|>{}<|action_sep|>def main():  # << START_OF_TRACE
<|frame_sep|>
---END OF PROMPT---
<|line_sep|>{}<|action_sep|>    return f('hosu', 'o')
<|frame_sep|><|call_sep|>{"sequence": "'hosu'", "value": "'o'"}<|action_sep|>def f(sequence, value):
<|frame_sep|><|line_sep|>{"sequence": "..", "value": ".."}<|action_sep|>    i = max(sequence.index(value) - len(sequence) // 3, 0)
<|frame_sep|><|line_sep|>{"sequence": "..", "value": "..", (*@\textcolor{cwm_string_color_bright}{\texttt{"i": "1"}}@*)}<|action_sep|>    result = ''
<|frame_sep|><|line_sep|>{"sequence": "..", "value": "..", "i": "..", "result": "''"}<|action_sep|>    for j, v in enumerate(sequence[i:]):
<|frame_sep|><|line_sep|>{"sequence": "..", "value": "..", "i": "..", "result": "..", "j": "0", "v": "'o'"}<|action_sep|>        if v == '+':
<|frame_sep|><|line_sep|>{"sequence": "..", "value": "..", "i": "..", "result": "..", "j": "..", "v": ".."}<|action_sep|>            result += sequence[i + j]
<|frame_sep|><|line_sep|>{"sequence": "..", "value": "..", "i": "..", "result": "'o'", "j": "..", "v": ".."}<|action_sep|>    for j, v in enumerate(sequence[i:]):
<|frame_sep|><|line_sep|>{"sequence": "..", "value": "..", "i": "..", "result": "..", "j": "1", "v": "'s'"}<|action_sep|>        if v == '+':
<|frame_sep|><|line_sep|>{"sequence": "..", "value": "..", "i": "..", "result": "..", "j": "..", "v": ".."}<|action_sep|>            result += sequence[i + j]
<|frame_sep|><|line_sep|>{"sequence": "..", "value": "..", "i": "..", "result": "'os'", "j": "..", "v": ".."}<|action_sep|>    for j, v in enumerate(sequence[i:]):
<|frame_sep|><|line_sep|>{"sequence": "..", "value": "..", "i": "..", "result": "..", "j": "2", "v": "'u'"}<|action_sep|>        if v == '+':
<|frame_sep|><|line_sep|>{"sequence": "..", "value": "..", "i": "..", "result": "..", "j": "..", "v": ".."}<|action_sep|>            result += sequence[i + j]
<|frame_sep|><|line_sep|>{"sequence": "..", "value": "..", "i": "..", "result": "'osu'", "j": "..", "v": ".."}<|action_sep|>    for j, v in enumerate(sequence[i:]):
<|frame_sep|><|line_sep|>{"sequence": "..", "value": "..", "i": "..", "result": "..", "j": "..", "v": ".."}<|action_sep|>    return result
<|frame_sep|><|return_sep|><|action_sep|>    return result<|arg_sep|>"'osu'"
<|frame_sep|><|return_sep|><|action_sep|>    return f('hosu', 'o')<|arg_sep|>"'osu'"<|frame_sep|>
    \end{lstlisting}
    \caption{\cwm Execution Trace Prediction (wrong prediction).}
  \end{subfigure}
  \hfill
  \begin{subfigure}{\textwidth}
    \lstset{style=simple_small_small}
    \begin{lstlisting}
Given a python code function and an assert statement containing a specific input, provide the assertion with the exact literal output that the function returns with that input. Do not include any mathematical expressions or function calls-only the final literal value. Your response should be solely the assertion, enclosed within [ANSWER] and [/ANSWER] tags.
[SKIPPED EXAMPLE FOR ILLUSTRATION PURPOSES]
Python function:
```
def f(sequence, value):
    i = max(sequence.index(value) - len(sequence) // 3, 0)
    result = ''
    for j, v in enumerate(sequence[i:]):
        if v == '+':
            result += value
        else:
            result += sequence[i + j]
    return result
assert f('hosu', 'o') == ??
```
----END OF PROMPT---
Okay, let's see. I need to figure out what the function f returns when called with the arguments 'hosu' and 'o'. Let me start by understanding the function step by step.

First, the function takes two parameters: sequence and value. In this case, sequence is the string 'hosu' and value is 'o'. 

The first line inside the function is: i = max(sequence.index(value) - len(sequence) // 3, 0). Let's break this down. 

sequence.index(value) finds the index of the first occurrence of 'o' in 'hosu'. The string 'hosu' is h-o-s-u, so the index of 'o' is 1. 

len(sequence) is the length of 'hosu', which is 4. len(sequence) // 3 is 4 divided by 3, which equals 1 (since integer division truncates). 

So, sequence.index(value) - len(sequence) // 3 is 1 - 1 = 0. Then, max(0, 0) is 0. Therefore, i is 0.

Next, the function initializes an empty string result = ''.

Then, there's a for loop: for j, v in enumerate(sequence[i:]). Since i is 0, sequence[i:] is the entire string 'hosu'. So, the loop will iterate over each character in 'hosu' along with their indices j.

Let's go through each iteration:

1. j = 0, v = 'h':
   - v is not '+', so result += sequence[i + j] which is sequence[0 + 0] = 'h'. Now result is 'h'.

2. j = 1, v = 'o':
   - v is not '+', so result += sequence[0 + 1] = 'o'. Now result is 'ho'.

3. j = 2, v = 's':
   - v is not '+', so result += sequence[0 + 2] = 's'. Now result is 'hos'.

4. j = 3, v = 'u':
   - v is not '+', so result += sequence[0 + 3] = 'u'. Now result is 'hosu'.

After the loop, the function returns result, which is 'hosu'.

Wait a minute, but the value parameter is 'o'. The code checks if v == '+', but in the sequence 'hosu', there are no '+' characters. So, the condition v == '+' is never true. Therefore, the code never appends the value 'o' to the result. Instead, it appends the characters from the sequence starting at index i (which is 0) for each character in the sequence. So, the result is just the original sequence 'hosu'.

Therefore, the function returns 'hosu' when called with f('hosu', 'o').
</think>
[ANSWER]
assert f('hosu', 'o') == 'hosu'
[/ANSWER]
    \end{lstlisting}
    \vspace{-0.2cm}
    \caption{\cwm Reasoning Trace (correct prediction).}
  \end{subfigure}
\vspace{-0.7cm}
  \caption{\cwm execution trace prediction (wrong) vs. \cwm natural language reasoning trace (correct). In this CruxEval example, the \cwm model with reasoning in natural language predicts the correct output \texttt{'hosu'}. 
  In execution trace prediction mode, the \cwm model predicts \texttt{'osu'}, because it fails to predict the value of the variable \texttt{i} in the first line correctly. It predicts 1, instead of the true value 0. \label{lst:trace_cwm_wrong_qwen_correct}}
\end{figure}

\begin{figure}
  \centering
  \begin{lstlisting}[language=Python,style=simple_small_small,keepspaces=true,deletekeywords={print},deletekeywords={[2]{next}}]
===== CONTEXT =====

def count_letters(s, letter):
    n = 0
    for c in s:
        n += int(c == letter)
    return n

def format_answer(word, letter, count):
    parts = [
        "Found",
        f"{count:04d}",
        "occurrences of the letter",
        letter,
        "in",
        word
    ]
    return " ".join(parts)

def f(c):  # << START_OF_TRACE
    word = "strawberry"
    num = count_letters(word, c)
    ans = format_answer(word, c, num)
    return ans


===== SESSION START =====

-> def f(c):
>> print
{'c': "'r'"}
>> next
->     word = "strawberry"
>> next
->     num = count_letters(word, "r")
>> step
-> def count_letters(s, letter):
>> next
->     n = 0
>> next
->     for c in s:
>> next
->         n += int(c == letter)
>> print
{'s': "'strawberry'", 'letter': "'r'", 'n': '0', 'c': "'s'"}
>> next
->     for c in s:
>> next
->         n += int(c == letter)
>> print
{'s': "'strawberry'", 'letter': "'r'", 'n': '0', 'c': "'t'"}
>> out
->     ans = format_answer(word, "r", num)
>> next
->     return ans
>> next
->     return ans
'Found 0003 occurrences of the letter r in strawberry'
  \end{lstlisting}
  \caption{Interactive Python code debugging with \cwm. As a show-case of \cwm's execution tracing capabilities (\cref{sec:tracing}), we present an interactive debugger to steer execution trace predictions. From a given code context, \cwm predicts the initial state of the trace (we selected this example based on the prediction of the \texttt{c} parameter). We can then prompt the model to generate the program state at the next line, or to directly predict the return value after a function is entered. In this example, we \texttt{step} into the \texttt{count\_letters} function,  step \texttt{out} of it after a few steps (implemented via continuous predictions until the function returns), and finally use \texttt{next} to predict the return value of \texttt{format\_answer} without line-by-line unrolling. In-between, we \texttt{print} the local variables as generated by \cwm.\label{lst:trace_debugger_demo}}
\end{figure}

\begin{figure}[htbp!]
  \centering

    \lstset{
    style=simple_small_small,
    basicstyle=\ttfamily\tiny
    }
    \begin{lstlisting}
<|begin_of_text|><|trace_context_start|>from mysterious_library import f
assert f(2, 2) == 4
assert f(3, 7) == 10
assert f(-1, -8) == -9
assert f(1, 1) == 2
f(5, 5)  # << START_OF_TRACE
<|frame_sep|>
----END OF PROMPT---
<|call_sep|>{}<|action_sep|>f(5, 5)  # << START_OF_TRACE
<|frame_sep|><|line_sep|>{}<|action_sep|>f(5, 5)  # << START_OF_TRACE
<|frame_sep|><|call_sep|>{"a": "5", "b": "5"}<|action_sep|>def f(a, b):
<|frame_sep|><|line_sep|>{"a": "..", "b": ".."}<|action_sep|>    return a + b
<|frame_sep|><|return_sep|><|action_sep|>    return a + b
<|arg_sep|>"10"<|frame_sep|>
    \end{lstlisting}

  \caption{\cwm predicting the execution traces of an unseen function \texttt{f}, imported from a library not included in the context. The expected semantics of \texttt{f} can be inferred from the assertions in the context. \cwm correctly generates the traces of a function that sums two numbers. It is possible that the model acquired this capability of jointly tracing and generating code because for some tracing data we do not include the source context of third-party libraries. 
  \label{lst:trace_cwm_unseen}}
\end{figure}

\begin{figure}[htbp!]
  \centering
    \lstset{aboveskip=0pt, belowskip=5pt}
    \begin{lstlisting}[style=cwm_prompt]
Given the following problem: Three little pigs from all over the world are meeting for a convention! Every minute, a triple of 3 new pigs arrives on the convention floor. After the n-th minute, the convention ends.
The big bad wolf has learned about this convention, and he has an attack plan. At some minute in the convention, he will arrive and eat exactly x pigs. Then he will get away.
The wolf wants Gregor to help him figure out the number of possible attack plans that involve eating exactly x pigs for various values of x (1 <= x <= 3n). Two attack plans are considered different, if they occur at different times or if the sets of little pigs to eat are different.
Note that all queries are independent, that is, the wolf does not eat the little pigs, he only makes plans!
Input
The first line of input contains two integers n and q (1 <= n <= 10^6, 1 <= q <= 2* 10^5), the number of minutes the convention lasts and the number of queries the wolf asks.
Each of the next q lines contains a single integer x_i (1 <= x_i <= b3n), the number of pigs the wolf will eat in the i-th query.
Output
You should print q lines, with line i representing the number of attack plans if the wolf wants to eat x_i pigs. Since each query answer can be large, output each answer modulo 10^9+7.
Examples
Input
2 3
1
5
6
Output
9
6
1
Input
...
Note
In the example test, n=2. Thus, there are 3 pigs at minute 1, and 6 pigs at minute 2. There are three queries: x=1, x=5, and x=6.
If the wolf wants to eat 1 pig, he can do so in 3+6=9 possible attack plans, depending on whether he arrives at minute 1 or 2.
If the wolf wants to eat 5 pigs, the wolf cannot arrive at minute 1, since there aren't enough pigs at that time. Therefore, the wolf has to arrive at minute 2, and there are 6 possible attack plans.
If the wolf wants to eat 6 pigs, his only plan is to arrive at the end of the convention and devour everybody.
Remember to output your answers modulo 10^9+7!
Generate the code step by step. You MUST FIRST generate the execution traces for the code you want to generate given a specific input, and then if it's correct extract the code.
Your code in the final answer should be enclosed in triple backticks like so: python YOUR CODE HERE. Use the backticks for your code only.
<|eot_id|><|start_header_id|>assistant
\end{lstlisting}
\begin{lstlisting}[style=cwm_think]
Okay, let's see. The problem is about calculating the number of possible attack plans for the big bad wolf. The wolf can attack at any minute during the convention, and he eats exactly x pigs. We need to find the number of ways he can do this for various values of x.

Let's jointly synthethize and trace the code for this using the first provided input/output example. First, I will emit the execution traces and check that it arrives to the expected output.

Then, if the output is correct, extracting the code is straightforward, I don't need to think anymore and can proceed to the final answer.

But if the execution is incorrect, I need to think and then predict the corrected traces again.

Let's start jointly tracing and synthethizing. The input will be f(2, 3, [1, 5, 6]), and the output must be [9, 6, 1]; otherwise it will mean I predicted the wrong code and need to start again. Let's see:

<|trace_context_start|>from solution import f

assert f(2, 3, [1, 5, 6]) == [9, 6, 1]
assert f(5, 4, [2, 4, 6, 8]) == [225, 2001, 6014, 6939]

f(2, 3, [1, 5, 6])  # << START_OF_TRACE
<|frame_sep|><|call_sep|>{}<|action_sep|>f(2, 3, [1, 5, 6])  # << START_OF_TRACE
<|frame_sep|><|line_sep|>{}<|action_sep|>f(2, 3, [1, 5, 6])  # << START_OF_TRACE
<|frame_sep|><|call_sep|>{"n": "2", "q": "3", "queries": "[1, 5, 6]"}<|action_sep|>def f(n, q, queries):
<|frame_sep|>
----END OF PROMPT---
<|line_sep|>{"n": "..", "q": "..", "queries": ".."}<|action_sep|>    MOD = 10**9 + 7
<|frame_sep|><|line_sep|>{"n": "..", "q": "..", "queries": "..", "MOD": "1000000007"}<|action_sep|>    max_x = 3 * n
...
<|frame_sep|><|return_sep|><|action_sep|>    return ans
<|arg_sep|>"[9, 6, 1]"<|frame_sep|><|return_sep|><|action_sep|>f(2, 3, [1, 5, 6])  # << START_OF_TRACE
<|arg_sep|>"[9, 6, 1]"<|frame_sep|>

Okay, the execution trace is correct. Now, I can proceed to extract the code.
\end{lstlisting}
\begin{lstlisting}[style=cwm_act]
```python
def f(n, q, queries):
    MOD = 10**9 + 7
    max_x = 3 * n
    fact = [1] * (max_x + 1)
    for i in range(1, max_x + 1):
        fact[i] = fact[i - 1] * i %
    inv_fact = [1] * (max_x + 1)
    inv_fact[max_x] = pow(fact[max_x], MOD - 2, MOD)
    for i in range(max_x - 1, -1, -1):
        inv_fact[i] = inv_fact[i + 1] * (i + 1) %
    def comb(n, k):
        if k < 0 or k > n:
            return 0
        return fact[n] * inv_fact[k] * inv_fact[n - k] %
    total = 0
    for i in range(1, n + 1):
        total += 3 * i
    ans = []
    for x in queries:
        res = 0
        for i in range(1, n + 1):
            if 3 * i >= x:
                res += comb(3 * i, x)
        ans.append(res %
    return ans
```
    \end{lstlisting}
\caption{\cwm interspersing code generation and execution tracing. 
As indicated by ``\dots'', we have omitted parts of \cwm reasoning for clarity here.
  \label{lst:trace_cwm_unseen_reasoning}}
\end{figure}

\begin{figure}[htbp!]
  \centering

    \lstset{style=simple_small_small}
    \begin{lstlisting}
You will be given a function f and a call to that function in the form f(??). This function might either terminate for all inputs or diverge for some input. If it terminates for all inputs, then answer "# T". Otherwise, if it diverges, provide an input such that executing f on the input leads to a diverging execution (i.e., f does NOT halt with that input), followed by the comment "# NT". In case non-terminating inputs exist, there may be multiple possible answers, but you should only output one. Express your final answer with either "# T", if the function terminates for all inputs, or as function call to f containing the input that leads to the non-terminating execution followed by the comment "# NT".
```
def f(x):
    def g(i):
        if i <= 0:
            return 0
        return g(g(g(i-2) - 1)) + 1
    def h(j):
        if j <= 0:
            return 0
        return h(g(j)-1) - 1
    h(x)
    return None
 
f(??)  # Terminates?
```
----END OF PROMPT---

Okay, I need to figure out if the function f(x) terminates for all inputs or if there's some input where it diverges. Let's start by understanding what the function does.
The function f(x) defines two helper functions, g(i) and h(j). It then calls h(x) and returns None. So, the termination of f(x) depends entirely on the behavior of h(x), which in turn depends on g(i).
Let me break down the helper functions:
First, g(i):
If i is less than or equal to 0, return 0.
Otherwise, return g(g(g(i-2) - 1)) + 1.
Hmm, that's a recursive function. Let's see what happens when we call g with some positive integer. For example, if i is 1:
g(1) would call g(g(g(-1) - 1)) + 1. Since g(-1) is 0, this becomes g(g(-1) -1) = g(0 -1) = g(-1) = 0. Then, adding 1 gives 1. Wait, so g(1) would return 1? Let me check that again.
Wait, let's step through g(1):
g(1) calls g(g(g(1-2) - 1)) + 1
1-2 is -1, so g(-1) is 0. Then, 0 -1 is -1. So g(-1) is 0 again. Then, the innermost g is 0, so we have g(0) +1. Wait, no, let me re-express this.
Wait, the expression is g(g(g(i-2) - 1)) + 1. So for i=1:
i-2 = -1. So g(-1) is 0. Then, subtract 1: 0 -1 = -1. Then, call g(-1) again, which is 0. Then, call g(0) which is 0. Then add 1. Wait, no, maybe I'm getting the order wrong.
Wait, the innermost call is g(i-2). So for i=1, i-2 is -1. So g(-1) returns 0. Then subtract 1: 0 -1 = -1. Then, call g(-1) again, which returns 0. Then, call g(0), which returns 0. Then add 1. Wait, that seems off. Let me write it step by step:
g(1) = g(g(g(1-2) - 1)) + 1
= g(g(g(-1) - 1)) + 1
= g(g(0 - 1)) + 1 (since g(-1) is 0)
= g(g(-1)) + 1
= g(0) + 1 (since g(-1) is 0)
= 0 + 1
= 1
So g(1) returns 1. That's interesting. What about g(2)?
g(2) = g(g(g(0) - 1)) + 1
g(0) is 0, so 0 -1 = -1. Then g(-1) is 0. Then g(0) is 0. So 0 +1 = 1. So g(2) is 1.
Wait, that's the same as g(1). What about g(3)?
g(3) = g(g(g(1) - 1)) + 1
g(1) is 1, so 1 -1 = 0. Then g(0) is 0. Then g(0) is 0. So 0 +1 = 1. So g(3) is 1.
Hmm, so for i=1,2,3, g(i) returns 1. What about i=4?
g(4) = g(g(g(2) - 1)) + 1
g(2) is 1, so 1 -1 = 0. Then g(0) is 0. Then g(0) is 0. So 0 +1 = 1. So g(4) is 1.
Wait, so it seems like for any i > 0, g(i) returns 1. Is that possible? Let me check i=5.
g(5) = g(g(g(3) - 1)) + 1
g(3) is 1, so 1 -1 = 0. Then g(0) is 0. Then g(0) is 0. So 0 +1 = 1. Yep, same result.
So, in general, for any i > 0, g(i) will compute to 1. Because no matter what i is, when you subtract 2 enough times, you get to a point where the inner g call returns 0, leading to a chain that ends up with 0 +1 = 1.
Wait, but what if i is even or odd? Let's see. Let's take i=6:
g(6) = g(g(g(4) - 1)) + 1
g(4) is 1, so 1 -1 = 0. Then g(0) is 0. Then g(0) is 0. So 0 +1 = 1. Same result.
So regardless of the input i (as long as it's positive), g(i) returns 1. That's a key insight. So g(i) is a function that returns 0 if i <=0, and 1 otherwise.
Now, moving on to h(j):
h(j) is defined as:
If j <= 0, return 0.
Otherwise, return h(g(j) - 1) - 1.
But since g(j) is 1 for any j > 0, then g(j) -1 is 0. So for any j > 0, h(j) becomes h(0) -1. But h(0) is 0, so h(j) = 0 -1 = -1. Wait, but then h(j) for j > 0 would be -1. But then, if we call h again with that value?
Wait, let's step through h(j) for j > 0. Let's take j=1:
h(1) = h(g(1) - 1) -1
g(1) is 1, so 1 -1 = 0. Then h(0) is 0. So 0 -1 = -1. So h(1) returns -1.
What about j=2?
h(2) = h(g(2) -1) -1
g(2) is 1, so 1 -1 = 0. h(0) is 0. 0 -1 = -1. So h(2) is -1.
Same for any j > 0. So h(j) for j > 0 is always -1. So when f(x) calls h(x), if x is positive, h(x) returns -1, and then f returns None. If x is 0 or negative, h(x) returns 0, and then f returns None.
Wait, but this seems too straightforward. So in all cases, h(x) terminates. Because even if x is a very large number, h(x) just does a constant number of steps. For example, h(1000) would compute g(1000) which is 1, subtract 1 to get 0, then call h(0) which is 0, subtract 1 to get -1. So it's a finite number of steps. Therefore, h(x) always terminates, regardless of the input x.
Therefore, the function f(x) will always terminate, no matter what input x is given. So the answer should be "# T".
</think> <answer>
# T

 \end{lstlisting}
  \caption{\cwm predicting termination. The prompt features a terminating example in HaltEval-prelim. After emitting a reasoning trace, which considers behavior on several specific inputs, CWM correctly predicts termination on all inputs.  \label{lst:trace_cwm_halteval_correct}}
\end{figure}

\section{RL algorithm}
\label{sec:app_rl_alg}
Given a prompt $x$, we perform $G$ rollouts, producing a set of trajectories (i.e.,~token sequences) $\{y_1, y_2, ... y_G\}$.
In general, rollouts are multi-turn, so the trajectories $y_i$ consist of a prompt $x$ followed by a sequence of actions and observations. We use the binary mask $M_{i,t}$ to signal whether token $y_{i, t}$ was generated by the agent ($M_{i,t}=1$) or environment (initial prompt and later observations; $M_{i,t}=0$).

The first input required by the PPO loss is an estimate of the advantage. We denote by $R_i$ the total return (i.e.,~sum of undiscounted rewards) of trajectory $i$.
For a batch of $G$ trajectories, we compute the length-weighted mean return $\mu = \frac{1}{L} \sum_{i=1}^G R_i \times L_i$, where $L_i = \sum_t M_{i,t}$ and $L = \sum_i L_i$ is the total number of agent-generated tokens. The advantage is then $\hat{A}_i = R_i - \mu$.

The PPO loss further requires the log probabilities of the trajectory under the behavior policy, often denoted $\pi_{\text{old}}$, in order to compute the importance ratio.
One complicating factor here is that the workers continue rollouts in parallel to model updates (see \cref{sec:rl-systems}).
At a given point in time, any number of the $G$ rollouts in a batch may be in progress. Hence, the true behavior policy distribution is difficult to describe mathematically. Nevertheless, we use the notation $\log \pi_{\text{old}}(y_{i,t} | y_{i, <t})$ to denote the token log probability produced by our inference backend at the moment token $y_{i, t}$ was sampled, and use this quantity for importance weighting as described below.

Finally, the PPO loss requires the policy log probabilities, which are computed on the trainer nodes. When a trainer receives a worker batch of $G$ trajectories associated with a prompt $x$, it computes the advantages and adds the trajectories to a queue.
Then, to produce a batch $\mathcal{B}$ for training, trajectories are popped from the queue until a limit of $N$ tokens is reached. By keeping a fixed limit of $N$ tokens we reduce the variance in the batch size between different steps, and optimize the GPU utilization without over-allocating GPU memory.
The trajectories are packed into a flat batch, padded to $N$ tokens, and forwarded to produce the log probabilities of the tokens $\log \pi(y_{i,t} | y_{i, <t})$.
As a consequence, the trajectories associated with one prompt may be spread out over multiple gradient updates.

Finally, the loss is calculated as
\begin{align*}
\mathcal{J}(\theta) 
=
{\frac{1}{N}} \sum_{y_i, A_i \in \mathcal{B}} \sum_{t=1}^{|y_i|} M_{i, t} \min\left[\rho_{i, t}(\theta) \hat{A}_{i}, \; \text{clip}\left( \rho_{i,t}(\theta), 1-\varepsilon_{\text{low}}, 1+\varepsilon_{\text{high}}\right)\hat{A}_{i}\right],
\end{align*}
where, as noted before, $\mathcal{B}$ is the trainer batch (not generally equal to the set of $G$ trajectories per prompt), $M_{i,t} \in \{0,1\}$ masks out environment-generated tokens, $\varepsilon_{\text{low}}$ and $\varepsilon_{\text{high}}$ are the PPO clipping thresholds, and $N$ is the maximum number of tokens in a batch (which we set to \num{131072}, the maximum context size of our model).
The importance ratio $\rho_{i, t}$ is computed from the log probabilities $\log \pi_\theta$ and $\log \pi_{\text{old}}$ (computed on the trainer and worker, respectively) as follows:
\begin{align*}
    \rho_{i,t}(\theta) = \exp{\left(\log \pi_\theta(y_{i,t} | y_{i, <t}) -  \log \pi_{\text{old}}(y_{i,t}|y_{i,<t}) \right)}.
\end{align*}
The gradient of $\rho_{i, t}(\theta) \hat{A}_i$ equals $\frac{\pi_\theta}{\pi_{\text{old}}} \nabla \log \pi_\theta \hat{A}_i$, which is the importance-weighted policy gradient estimator.
Thus, the PPO loss can be understood as a clipped version of this.

\section{Activ image-building pipeline}
\label{appendix:activ}
\begin{figure}[htb!]
    \centering
   \includegraphics[width=0.99\linewidth]{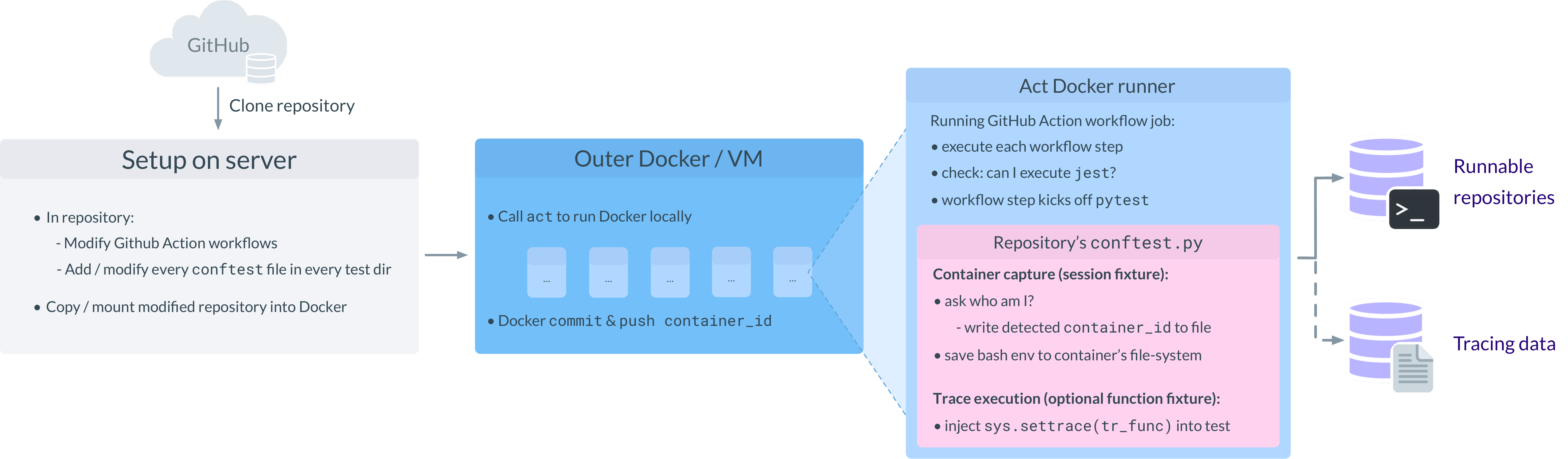}
    \caption{Activ image building pipeline for a single repository. After cloning from GitHub, the repository's GitHub Actions workflows and pytest \textit{conftest.py} files (in Python repositories) are modified and copied into the outer Docker (or virtual machine) for isolated CI execution via \textit{act}. Modified workflow jobs are executed in parallel within individual containers, until the \textit{container\_id} and built-environment state are captured from the target container that holds repository dependencies. Framework executability detection precedes capture to ensure targeting the correct container: for pytest repositories, this occurs implicitly when injected \textit{conftest.py} code executes within a session-scoped fixture, with an optional function-scoped fixture available for Python execution tracing. Non-Python repositories use modified workflow steps to verify framework executability (such as \textit{Jest}) before capture.  Upon successful capture, an early exit is triggered and the resulting container is committed and pushed for standalone execution. }
    \label{fig:activ}
\end{figure}

The Activ pipeline, shown in Figure~\ref{fig:activ}, automatically builds executable repository images at scale by modifying their GitHub Actions workflows and running them locally using the \textit{act} \citep{Lee2025act} library within a virtual environment. Our approach builds on the insight that the execution environment of a GitHub Actions workflow running CI tests is a fully built environment with dependencies, and can therefore be captured as a standalone Docker image for later execution.

As we require only a single successful build of the repository, the pipeline reduces complex cross-platform and framework build matrices into a single entry \citep{saavedra2024}, by selecting most-compatible Python versions and Ubuntu variants. The pipeline also modifies each repository's GitHub Actions workflows to \textit{continue-on-error}, ensuring pipeline completion when encountering noncritical failures. We also implement multiple early exit strategies to terminate the pipeline as soon as a built-environment state has been captured, progress has stalled, or a timeout is reached.

We modify each GitHub Actions workflow to probe for available test frameworks by checking if a list of predefined frameworks are executable when the workflow is running. For each detected framework, the pipeline captures the corresponding build environment and executes container ID capture logic that is equivalent to the Python-specific pytest capture process, detailed below.

The pipeline modifies (or adds) \textit{conftest.py} pytest configuration files to each test directory in Python repositories. A session-scoped fixture is automatically injected during pytest execution to capture the build state of containers running unit tests. This fixture detects the ID of the container running unit tests for the \textit{docker commit} of the repository's current build state.  The environment capture process further preserves the container's build state by writing-out bash environment variables and creating archives of the mounted repository code and hosted toolcache dependencies to the container's file-system for later restoration via a Docker \textit{entrypoint} script.

To achieve the scale required for our dataset, we run on an internal sandboxing platform to execute approximately 500 repositories in parallel within secure, isolated virtual environments.

\section{Hyper-parameter scaling laws}
\label{app:hyper_param_scaling}
\subsection{Derivation of per-token compute formula}

Consider a transformer model with hidden dimension $d$, sequence length $S$, batch size $B$, and $L$ layers.

\newpara{Linear layers.}  
For a linear transformation of size $N$, the forward pass requires $2N$ floating point operations (FLOP): one multiplication and one addition per weight--input pair. The backward pass is approximately twice as expensive, since gradients must be computed with respect to both the weights and the inputs. Thus, the total cost per linear layer is $6N$ FLOP.

\newpara{Self-attention.}  
In multi-head self-attention, the two dominant operations are $QK^\top$ and $\mathrm{softmax}(QK^\top)V$. The FLOP cost of the softmax itself is negligible compared to these matrix multiplications. The forward pass of each multiplication costs $2BS^2 d$ FLOP, while the backward pass is about twice as expensive, contributing $4BS^2 d$ FLOP. Summing both gives $12BS^2 d$ FLOP. Because causal attention only computes half of the entries of $QK^\top$, the cost reduces to $6BS^2 d$. Dividing by the number of tokens $BS$ gives the per-token cost $6Sd$. Since each of the $L$ layers contains one self-attention block, the per-token cost for attention across all layers is $6SdL$.

\subsection{Quasi-random search for batch size and learning rate}

To estimate the optimal batch size (BS) and learning rate (LR) range, and how it evolves with scale, we performed a quasi-random search using Sobol sequences. At each scale, BS/LR candidates were generated by sampling two-dimensional Sobol sequences and rescaling them according to ranges that increase for BS and decrease for LR as the model scale increases.  

\begin{figure}[ht!]
    \centering
    \includegraphics[width=0.85\linewidth]{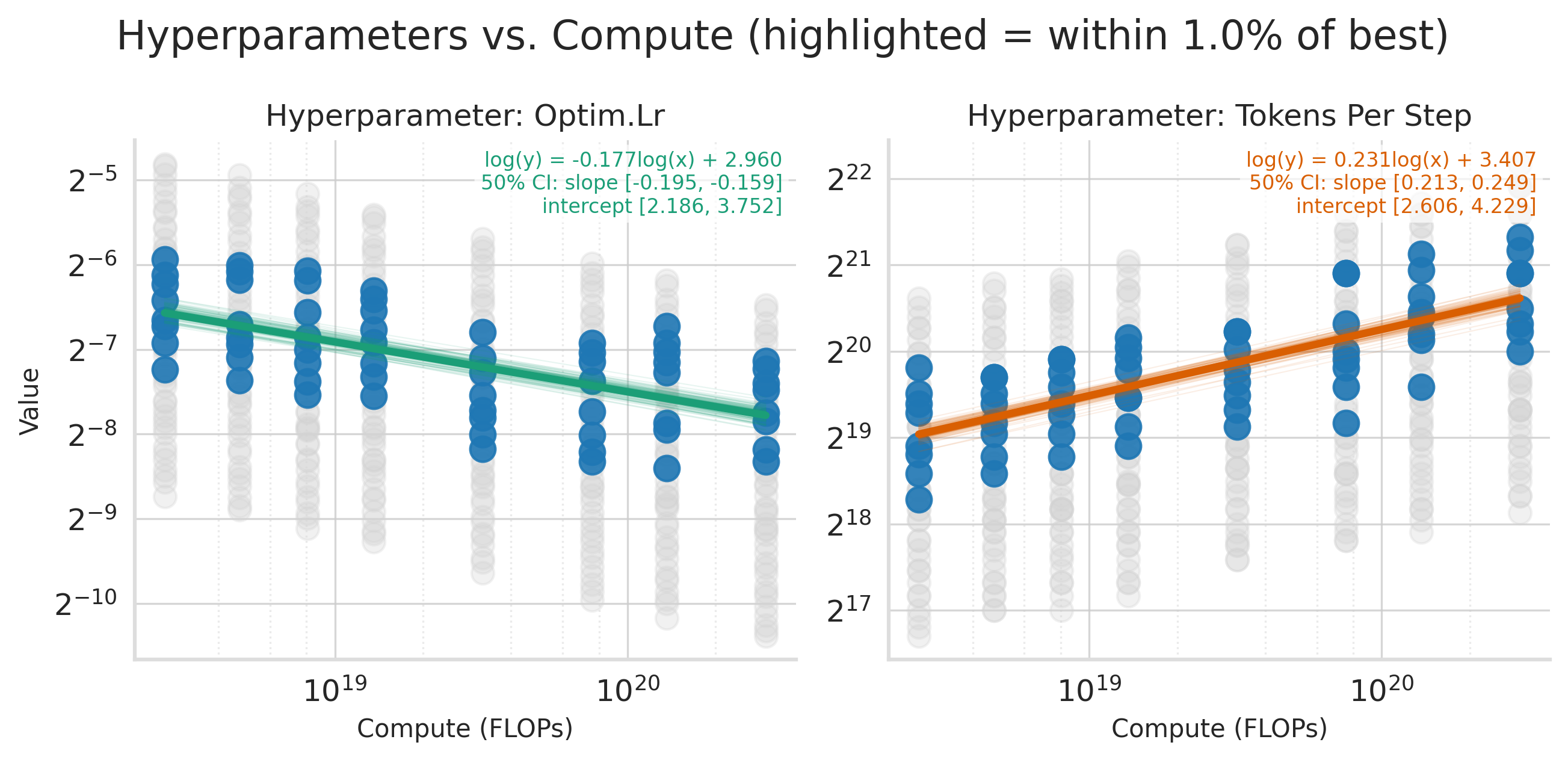}
    \caption{Optimal range for batch size and learning rate across scales is quite large. However going beyond that range leads to rapidly degrading performance.}
    \label{fig:scaling_laws}
\end{figure}

In Figure~\ref{fig:scaling_laws}, the gray points correspond to all BS/LR candidates evaluated at each scale, while the blue points indicate those within $1\%$ of the best validation loss at that scale.

\section{RL data decontamination}
\label{sec:appendix:rl-decontamination}

We use MinHash LSH to decontaminate all our prompts in our RL datasets against the following evaluation benchmarks:

\begin{itemize}
    \item \textbf{Math reasoning}: AIME 2024/2025, HARP~\citep{DBLP:journals/corr/abs-2412-08819}, GSM8K test~\citep{DBLP:journals/corr/abs-2110-14168}, OmniMath (test)~\citep{DBLP:conf/iclr/Gao0YCMDLMCXTWZ25}, Math500~\citep{DBLP:conf/nips/HendrycksBKABTS21}.
    \item \textbf{Code generation}: HumanEval~\citep{DBLP:journals/corr/abs-2107-03374}, MBPP (valid/test)~\citep{DBLP:journals/corr/abs-2108-07732}, LiveCodeBench (20240801-20250501)~\citep{DBLP:conf/iclr/JainHGLYZWSSS25}.
    \item \textbf{Scientific reasoning}: GPQA (Main/Extended/Diamond)~\citep{DBLP:journals/corr/abs-2311-12022}.
    \item \textbf{Commonsense and general reasoning}: ARC Challenge/Easy (valid/test)~\citep{DBLP:journals/corr/abs-1803-05457}, CommonsenseQA (valid/test)~\citep{DBLP:conf/naacl/TalmorHLB19}, DROP~\citep{DBLP:conf/naacl/DuaWDSS019}, PIQA (valid/test)~\citep{DBLP:conf/aaai/BiskZLGC20}, HellaSwag (valid/test)~\citep{DBLP:conf/acl/ZellersHBFC19}, SimpleQA~\citep{DBLP:journals/corr/abs-2411-04368}, OpenBookQA main/additional (valid/test)~\citep{DBLP:conf/emnlp/MihaylovCKS18}, WinoGrande 1.1 (dev/test)~\citep{DBLP:conf/aaai/SakaguchiBBC20}.
    \item \textbf{Conversation and evaluation frameworks}: ArenaHard~\citep{DBLP:journals/corr/abs-2406-11939},  MTBench~\citep{DBLP:conf/nips/ZhengC00WZL0LXZ23}.
\end{itemize}

We decontaminate our dataset of Dockerized executable repositories against \sbv~\citep{DBLP:conf/iclr/JimenezYWYPPN24} by doing the following:
\begin{itemize}
    \item Removed all Docker images built from repositories in \sbv.
    \item Confirmed no remaining unexpected instance-level contamination by verifying that pairwise (dataset instance to \sbv instance) Jaccard similarities between diff patch line sets remained below 0.2. 
\end{itemize}

\section{Mathematical expression comparison for RL}
\label{app:math-verify}
Reinforcement learning on mathematical computation problems, both with numerical and symbolic answers, requires comparing the predicted answer to the ground truth answer contained in the dataset. We do this using the disjunction of two verifiers: a custom one described below and MathVerify \citep{math_verify}. If any of them considers the expressions to be equivalent, the predicted answer is deemed correct.

The custom verifier grows a set of equivalent expressions for both the predicted answer and the ground truth answer, and returns whether at the end, there is a nonempty intersection between those sets.
Expressions are added based on various normalizations and transformations: string normalizations and replacements, normalization of unicode math symbols to Latex, normalization of Latex expressions, numerical computatation with floating point numbers up to a certain precision, symbolic computations, simplifications and normalization using SymPy \citep{sympy} and recursion in this manner for structured objects such as matrices and real intervals.

\section{Prompting guide}
\label{appendix:prompting_guide}

\newpara{Reserved tokens} are used for general text and chat formatting, and are not intended to be encoded from user input.
They include text sequence start and end markers, padding, chat message header delimiters, and an end of chat message token.

\begin{itemize}
    \item \verb$<|begin_of_text|>$ (128000): global text sequence start marker.
    \item \verb$<|end_of_text|>$ (128001): global text sequence end marker.
    \item \verb$<|pad|>$ (128004): padding token.
    \item \verb$<|start_header_id|>$ (128006): start of chat message header.
    \item \verb$<|end_header_id|>$ (128007): end of chat message header.
    \item \verb$<|eot_id|>$ (128008): end of chat message.
\end{itemize}

\newpara{Trace prediction tokens} are designed for predicting program execution traces and may be enabled when encoding user-controllable input to expose CWM's trace prediction functionality.
They include tokens for frame delimiting, action separation, function return/call, next line, exception, and argument separation, as well as a sentinel token for the start of the source code context for trace prediction (see below).

\begin{itemize}
    \item \verb$<|frame_sep|>$ (128100): start of trace sentinel, end of execution step.
    \item \verb$<|action_sep|>$ (128101): start of source code line.
    \item \verb$<|return_sep|>$ (128102): execution step: return from function scope.
    \item \verb$<|call_sep|>$ (128103): execution step: enter function scope.
    \item \verb$<|line_sep|>$ (128104): execution step: next line.
    \item \verb$<|exception_sep|>$ (128105): execution step: exception.
    \item \verb$<|arg_sep|>$ (128106): separator for return and exception values.
    \item \verb$<|trace_context_start|>$ (128107): start of source code context for trace prediction.
\end{itemize}

\newpara{Chat format.}
A chat is structured as a list of messages, each with the following format:
\begin{lstlisting}
<|start_header_id|>$ROLE<|end_header_id|>

$CONTENT<|eot_id|>
\end{lstlisting}
The \texttt{\$ROLE} can be \texttt{system}, \texttt{user}, \texttt{assistant}, or \texttt{tool: \$TOOL}. \texttt{\$CONTENT} is the message content.
The model is to be prompted with an assistant header and two following newline characters; the \verb$<|eot_id|>$ token marks the end of its reply and is thus the stop token for inference.
The conversation's first token is expected to be \verb$<|begin_of_text|>$.

\newpara{Reasoning.} CWM is a hybrid reasoning and non-reasoning model; reasoning mode is enabled via prompting. Reasoning mode is turned on by starting the system prompt with:
\begin{lstlisting}
You are a helpful AI assistant. You always reason before responding, using the following format:

<think>
your internal reasoning
</think>
your external response
\end{lstlisting}
The model should be prompted with a leading \verb$<think>\n$, i.e., a prompt should end with (showing newline characters for clarity here):
\begin{lstlisting}
<|start_header_id|>assistant<|end_header_id|>\n\n<think>\n
\end{lstlisting}
The reasoning section will be closed with \texttt{</think>}, and any text produced afterwards is the answer to the preceding user input.

\newpara{Tool Use.} The model performs tool calls with the following format:
\begin{lstlisting}
<tool: $TOOL>
$CONTENT
</tool>
\end{lstlisting}
Any available tools are to be announced in the system prompt. User code is responsible for detecting tool calls in model output and responding with a message marked with the respective role.

An example tool output of the \texttt{python} tool could be:
\begin{lstlisting}
<|start_header_id|>tool: python<|end_header_id|>

completed.
[stdout]$STDOUT_CONTENT[/stdout]
[stderr]$STDERR_CONTENT[/stderr]<|eot_id|>
\end{lstlisting}
Control is then handed back to the model for further processing.

An example of how tools can be specified in the system prompt:
\begin{lstlisting}
You have access to the following tools:

<tool: bash>
[command(s)]
</tool>
Executes bash command(s) [command(s)] in the current session. [command(s)] can be any non-interactive bash command(s) either single or multi-line.

<tool: create>
[path]
[content]
</tool>
Creates a new file at [path] with [content], where [path] must not exist, but its parent directory must exist.
\end{lstlisting}
Here, the model may invoke either the \texttt{bash} or the \texttt{create} tool.

\newpara{Trace Prediction.} CWM is able to predict the execution of Python programs on a step-by-step basis using dedicated trace prediction tokens. The prompt requires a source code context, \texttt{\$CONTEXT}, and a sentinel \texttt{<|frame\_sep|>} token to induce trace prediction, structured as:
\begin{lstlisting}
<|begin_of_text|><|trace_context_start|>$CONTEXT<|frame_sep|>
\end{lstlisting}
In \texttt{\$CONTEXT}, the entry point for trace prediction is marked with a \verb$<< START_OF_TRACE$ comment. An execution trace in CWM is a series of \emph{frames}, with each frame consisting of an \emph{observation} (local variables) and an \emph{action} (source code line). There are four different types of frames, formatted as follows:
\begin{lstlisting}
<|call_sep|>$LOCALS<|action_sep|>$SOURCE<|frame_sep|>
<|line_sep|>$LOCALS<|action_sep|>$SOURCE<|frame_sep|>
<|return_sep|><|action_sep|>$SOURCE<|arg_sep|>$VALUE<|frame_sep|>
<|exception_sep|><|action_sep|>$SOURCE<|arg_sep|>$VALUE<|frame_sep|>
\end{lstlisting}
The model produces an \verb$<|end_of_text|>$ token to denote the end of the execution, which is reached when exiting the scope of the trace's entry point. Locals are formatted as JSON key-value pairs where values are always rendered as JSON strings. and \texttt{\$VALUE} for return and exception frames is also a JSON-encoded string representation.

\section{Formal mathematics datamix}
\label{app:lean-data}
We use the following datasets of mathematics in the Lean 4 \citep{moura2021lean} theorem proving language.
\begin{itemize}
\item LeanUniverse \citep{ahm2025leanuniverse}, a dataset of (initial proof state, tactic, resulting proof state) triples from Lean's mathematical library \citep{leanmathlib2020} and other open-source Lean repositories, as a form of code world modeling in Lean.
\item Goedel-Pset \citep{lin2025goedelprover}, formatted as a mathematical statement and proof formalization dataset, where the task is to translate a mathematical problem and solution from natural language to Lean.
\item Mathematical statement formalization datasets, where the task is to translate a mathematical statement from natural language 
to a Lean theorem statement without proof: Compfiles \citep{compfiles}, Lean Workbook \citep{ying2025leanworkbook}, miniF2F \citep{zheng2022minif2f}, ProofNet \citep{azerbayev2023proofnet}, Goedel-Pset.
\end{itemize}

\section{RULER evaluation}
\label{sec:ruler}
We compare \cwm against Gemma3-27B and Qwen3-32B post-trained models. Results can be seen on~\Cref{tab:ruler}. Results suggest that \cwm outperforms Gemma3-27B under \num{128} k sequence length, but falls short under \num{32} k sequence length. Both \cwm and Gemma3-27B achieve worse performance compared to Qwen3-32B. However, it is important to note that Qwen3-32B uses full attention across all layers, resulting in significantly higher computational costs, particularly for longer sequences. 
Therefore, we believe \cwm represents a good trade-off between efficiency and model performance.

\begin{table}[t!]
\centering
\caption{RULER results at 32k and 128k sequence length. Results are reported for \cwmv, Qwen3-32B and Gemma-3-27B.\label{tab:ruler}}
\begin{tabular}{l|lll}
\toprule
\textbf{Context} & \textbf{Gemma-3-27B} & \textbf{Qwen3-32B} & \textbf{CWM} \\
\midrule
32k  & 91.1 & 94.4 & 84.3\\
128k & 66.0   & 85.6 & 69.7 \\
\bottomrule
\end{tabular}
\end{table}

\section{Agent capabilities learnt during RL training}\label{sec:swerl_ds}
\begin{figure}[ht!]
 \centering
 \begin{subfigure}{0.48\textwidth}
    \includegraphics[width=\linewidth]{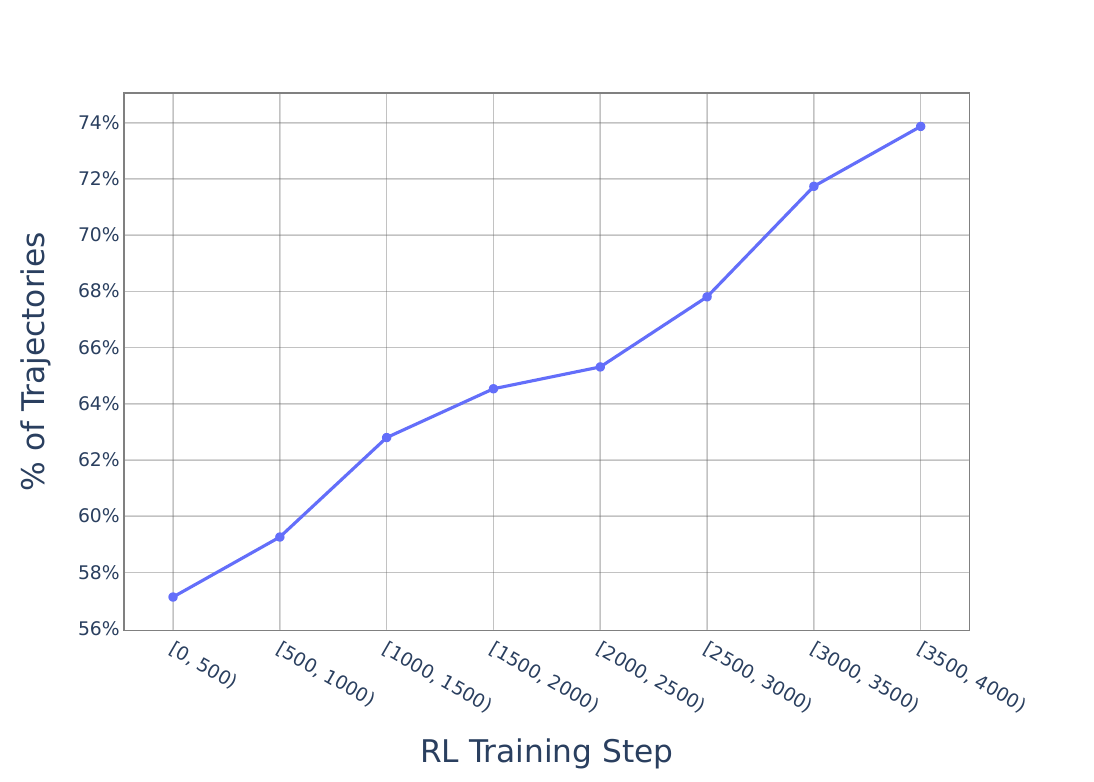}
    \caption{Percentage of trajectories where \swerl performs tests on at least one turn increases over the course of RL training.}
    \label{fig:swerl_testing}
  \end{subfigure}
  \hfill
  \begin{subfigure}{0.48\textwidth}
    \includegraphics[width=\linewidth]{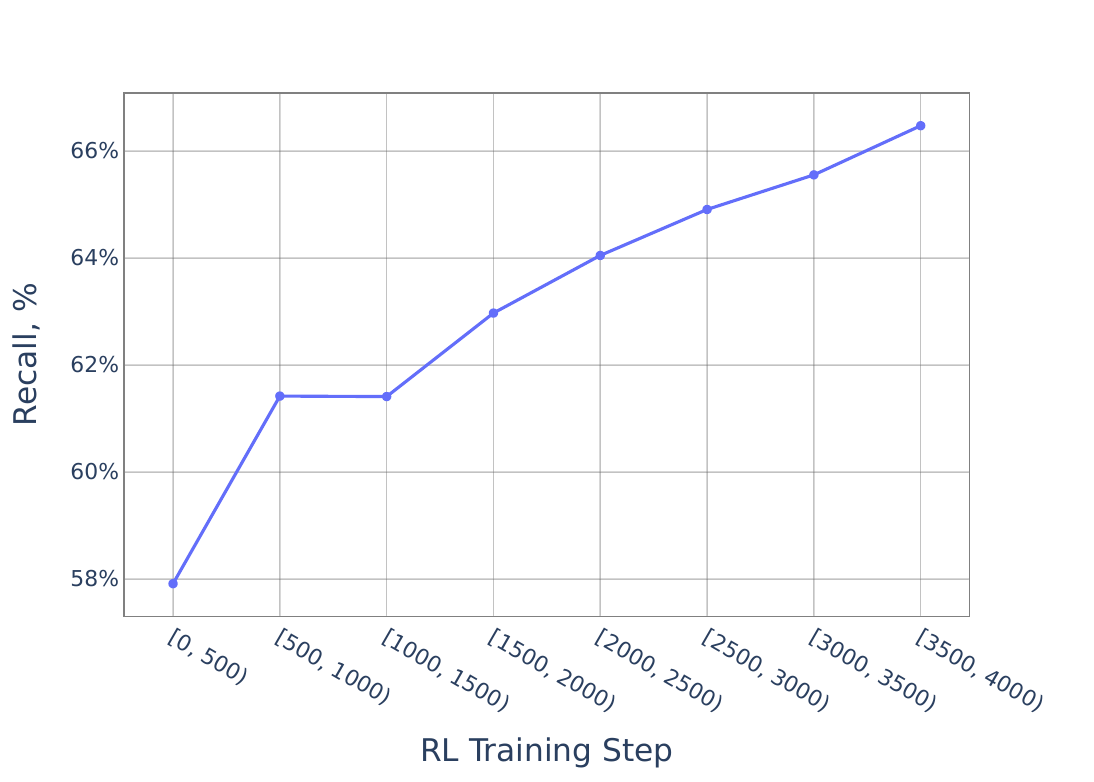}
    \caption{\swerl learns to localize the relevant files over the course of RL training. \\}
    \label{fig:swerl_recall}
  \end{subfigure}
  \caption{Agentic capabilities learnt during \cwm RL training.}
  \label{fig:swerl_behavior}
\end{figure}

In the context of SWE agentic tasks, we highlight in \Cref{fig:swerl_behavior} two notable capabilities learnt by the \swerl agent during the RL training stage.

First, \Cref{fig:swerl_testing} shows that the agent learns to test code more often over the course of RL training: while at the beginning of RL training the agent runs tests on at least one turn of a rollout for 57\% of trajectories, after only 4,000 steps of RL training it runs tests on at least one turn of a rollout for 74\% of trajectories.

Second, through RL training the agent learns to better localize files relevant to solving the issue. We formalize it by defining the \textit{recall} for a given task to be the percentage of files in the gold patch that were edited by the agent during a rollout:
$$\text{Recall} = 
    \frac{ \left| \{ \text{files edited by the agent} \} \cap
        \{ \text{files edited in the gold patch} \} \right|}
        {\left| \{ \text{files edited in the gold patch} \} \right| }.
$$
\Cref{{fig:swerl_recall}} shows that the agent's average recall increases from 58\% at the start of RL training to over 66\% after only 4,000 steps.

\end{document}